\documentclass{aa}  
\usepackage{graphicx}
\usepackage{natbib}

\newcommand{\kms}{$\rm{\,km \,s}^{-1}$}
\makeatletter
\newcommand{\rmnum}[1]{\romannumeral #1}
\newcommand{\Rmnum}[1]{\expandafter\@slowromancap\romannumeral #1@}
\makeatother
\usepackage{txfonts}
\usepackage{hyperref}
\usepackage{amstext}

\begin{document} 

\title{The MURALES survey. IV.}

\subtitle{Searching for nuclear outflows in 3C radio galaxies at
  z$<$0.3 with MUSE observations.} 
  
  \author{Giovanna Speranza\inst{1,2}
                \and Barbara Balmaverde\inst{3} 
                \and Alessandro Capetti\inst{3}
                \and Francesco Massaro\inst{4}
                \and G. Tremblay\inst{13}
                \and Alessandro Marconi\inst{5,6}
                \and Giacomo Venturi\inst{6,16}
                \and M. Chiaberge\inst{7,8}
                \and R.D. Baldi\inst{9} 
                \and S. Baum\inst{11}
                \and P. Grandi\inst{10}
                \and Eileen T. Meyer\inst{14}
                \and C. O$'$Dea\inst{11}
                \and W. Sparks\inst{15}
                \and B.A. Terrazas\inst{13}
                \and E. Torresi\inst{10} 
                }
\institute{Instituto de Astrofísica de Canarias, Calle Vía Láctea, s/n, E-38205 La Laguna, Tenerife, Spain
\and Departamento de Astrofísica, Universidad de La Laguna, E-38206, La Laguna, Tenerife, Spain
\and INAF - Osservatorio Astrofisico di Torino, Via Osservatorio 20, I-10025 Pino Torinese, Italy
\and Dipartimento di Fisica, Universit`a degli Studi di Torino, via Pietro Giuria 1, I-10125 Torino, Italy
\and Dipartimento di Fisica e Astronomia, Universit\`a di Firenze, via G. Sansone 1, 50019 Sesto Fiorentino (Firenze), Italy
 \and INAF - Osservatorio Astrofisico di Arcetri, Largo Enrico Fermi 5, I-50125 Firenze, Italy
 \and AURA for the European Space Agency (ESA), ESA Office, Space Telescope Science Institute, 3700 San Martin Drive, Baltimore, MD 21218, USA
\and Johns Hopkins University, 3400 N. Charles Street, Baltimore, MD 21218, USA
 \and  INAF- Istituto di Radioastronomia, Via Gobetti 101, I-40129 Bologna, Italy
\and INAF - Osservatorio di Astrofisica e Scienza dello Spazio di Bologna, via Gobetti 93/3, 40129 Bologna, Italy
\and Department of Physics and Astronomy, University of Manitoba, Winnipeg, MB R3T 2N2, Canada
\and Leiden Observatory, Leiden University, PO Box 9513, NL-2300 RA, Leiden, the Netherlands
\and Harvard-Smithsonian Center for Astrophysics, 60 Garden St., Cambridge, MA 02138, USA
\and University of Maryland Baltimore County, 1000 Hilltop Circle, Baltimore, MD 21250, USA
\and SETI Institute, 189 N. Bernado Ave Mountain View,CA 94043
\and Instituto de Astrof\'isica, Facultad de F\'isica, Pontificia Universidad Cat\'olica  de Chile, Casilla 306, Santiago 22, Chile
}

   \date{}

  \abstract {We analyze VLT/MUSE observations of 37 radio galaxies
    from the Third Cambridge catalogue (3C) with redshift $<$0.3
    searching for nuclear outflows of ionized gas. These observations
    are part of the MURALES project (a MUse RAdio Loud Emission line
    Snapshot survey), whose main goal is to explore the feedback
    process in the most powerful radio-loud AGN. We applied a
    nonparametric analysis to the [O~III] $\lambda$5007 emission
    line, whose asymmetries and high-velocity wings reveal
    signatures of outflows. We find evidence of nuclear outflows in
    21 sources, with velocities between $\sim$400--1000 \kms,
    outflowing masses of $\sim 10^5-10^7$ M$_\odot$, and a kinetic
    energy in the range $\sim 10^{53} - 10^{56}$ erg. In
    addition, evidence for extended outflows is found in the 2D gas
    velocity maps of 13 sources of the subclasses of high-excitation (HEG) and broad-line (BLO) radio galaxies,  with sizes between 0.4 and 20
    kpc. We estimate a mass outflow rate
    in the range 0.4--30 M$_\odot$ yr$^{-1}$ and an energy deposition rate of
    ${\dot E}_{kin} \sim 10^{42}-10^{45} $ erg s$^{-1}$. Comparing the
    jet power, the nuclear luminosity of the active galactic nucleus, and the outflow kinetic energy rate,    
 we find that outflows of HEGs and BLOs are likely radiatively powered, while jets likely only play a dominant role in galaxies with low excitation.   
The low     loading factors we measured suggest that these outflows are driven by momentum
    and not by energy. Based on the gas masses, velocities, and
    energetics involved, we conclude that the observed ionized
    outflows have a limited effect on the gas content or the star
    formation in the host. In order to obtain a complete view of the feedback process, observations exploring the complex
    multiphase structure of outflows are required.}

\maketitle

\section{Introduction}
\label{introduction}

Super massive black holes (SMBHs) hosted in radio galaxies are one
of the most energetic manifestations of active galactic nuclei
(AGN). AGN share a codependent evolutionary path with their host
galaxies through the continuous exchange of energy and matter, a
process known as AGN feedback \citep{Silk98, Fabian12}.  Outflows of
molecular, neutral, and ionized gas are thought to play a key role in
regulating both the star formation of the host galaxy and the black hole
accretion, in agreement with the observed link between black hole (BH)
mass and bulge velocity dispersion \citep{Ferrarese00,Gebhardt00,Kormendy13}
 and the evolution of the BH accretion and the star formation (SF)  history throughout cosmic time (e.g., \citealt{Aird15}).
AGN feedback also plays a crucial role in reconciling the predictions of theoretical models and observations at the high-mass end of the  luminosity function of galaxies
(e.g., \citealt{Kormendy13}  and references therein) and in explaining the bimodality of galaxies in the {\it \textup{blue}} and {\it \textup{red}} sequence (e.g., \citealt{Schawinski14}).

The feedback of AGN in radio galaxies can act on different spatial scales.  In galaxy clusters and groups, AGN radio activity
can prevent the cooling of hot, X-ray emitting gas that surrounds central galaxies \citep{McNamara12}. Radio galaxies hosted in
brightest cluster galaxies (BCGs) are able to displace the low-density
gas forming the hot phase of the interstellar medium (ISM) while expanding
\citep{McNamara00,Birzan04,Birzan12}.  So far, X-ray images of
galaxy clusters provide one of AGN feedback manifestations in the
local Universe, revealing cavities in the hot ionized medium
(with temperatures of aboutf 10$^6-10^7$ K) that are filled by the radio-emitting plasma.  This role of
feedback has been identified as kinetic (or jet) mode feedback and is associated
with radio sources characterized by radiatively inefficient accretion.
At high accretion efficiency,  the dominant mode of feedback is the radiative (or quasar) mode  (see, e.g., \citealt{Fabian12}):  When the AGN luminosity is close to the Eddington limit, the 
radiation pressure  acting on  the accreting gas generates an extended outflow of gas. In radio-loud quasars, both modes of feedback can coexist.

There is evidence of jets driving outflows on a galactic scale with typical speeds of hundreds of km s$^{-1}$ in the
interstellar medium of the host galaxy.  Most studies that reported this evidence were
focused on ionized gas and studied the interaction between the ISM and
the radio jet, which is traced by broad (often blueshifted) components of
the emission lines (see, e.g., \citealt{Axon98,Capetti99}).  The spatial
correlation of the distribution of the ionized gas and the radio
emission at low and high redshift strongly supports the role of
the jet in shaping the distribution of the gas and its ionization
state \citep{Tadhunter00}.  
Nonetheless, in some studies, the gas
masses involved in the outflows are lowl ($\lesssim$1 M$_{\odot}$)
and the effect of these outflows appears to be modest. 
For example, \citet{Mahony16}  observed an ionized mass outflow rate ranging from $\sim$0.05 to 0.17 M$_\odot$
yr$^{-1}$, corresponding to a kinetic energy of $\sim$10$^{-6}$ of the bolometric luminosity in the nearby radio galaxy, 3C293.
 Similar results were found in 3C33, in
which the estimated kinetic powers is $\sim$10$^{-5}$ of the
bolometric luminosity \citep{Couto17}.  
In  rest-frame ultraviolet spectra, some quasars show a blueshifted absorption line
(BAL quasars, \citealt{Weymann91}), indicating the presence of massive, high-velocity outflows 
(several percent of the speed of light). 
 A comparison of radio-loud and radio-quiet BAL quasars
revealed no substantial differences, and \citet{Rochais14} concluded that 
they are likely driven by similar physical phenomena. Similarly, considering a sample of radio-loud AGN, \citet{Tombesi14}
found that the fraction of sources with signatures of 
ultrafast outflows in their X-ray spectra is similar to
what is observed in the radio-quiet sample, demonstrating that relativistic jets do not preclude 
accretion-driven winds.

There are indications that in interacting, young, or recently
restarted radio galaxies the coupling between the radio jet and the
ISM might be stronger \citep{Santoro20}.  This evidence agrees with recent
numerical simulations showing that radio jets can couple strongly to the
ISM of the host galaxy: In a clumpy medium, the jet can create a
cocoon of shocked gas driving an outflow in all directions
\citep{Wagner12,Mukherjee16}.  The final effect of the jet-ISM
interaction depends on the jet power, the distribution of the
surrounding medium, and its orientation with respect to the jet.
In the merger radio galaxy 4C+29.30, \citet{Couto20} found a
 prominent outflow, with a total ionized gas mass outflow rate of 25
 M$_\odot$ yr$^{-1}$ and an energy corresponding to $\sim 6\,\times 10^{-2}$
 of the AGN bolometric luminosity.  A number of large-scale outflows
 of neutral hydrogen have been found mainly in young or restarted radio
 galaxies (identified by the blueshifted wings seen in HI absorption; see,
 e.g., \citealt{Morganti03,Morganti05,morganti18,Aditya18}. ) Other outflows have also been observed in  molecular gas and
 are characterized by velocities of between a few
 hundred and 1300 km s$^{-1}$, masses ranging from a few 10$^6$ to
 10$^7$ M$_{\odot}$ , and mass outflow rates up to 20 - 50 M$_{\odot}$
 yr$^{-1}$ \citep{Morganti20}.  Most of the mass and the energy
 carried by the outflow therefore appears to be in a neutral or
 molecular phase. The molecular outflows in radio
 galaxies appear to be less prominent than those found in
 ultra-luminous infrared galaxies (ULIRGs) and quasars (QSOs), in
 which speeds of  $\approx$1000 km s$^{-1}$ and high-mass
 outflow rates up to $\approx$1000 M$_\odot$ yr$^{-1}$ are observed
 (see, e.g., \citealt{Sturm11,Cicone12}).
 
Most AGN outflows are believed to be powered by the intense radiation field
that is produced by the accretion process onto the supermassive black hole.
Other possible drivers for the outflows are
stellar winds and supernovae; however, they are probably not 
responsible for the observed outflows because the hosts of radio-loud AGN
are typically passive early-type elliptical galaxies. In radio
galaxies, the mechanism that produces the outflow is uncertain because the radiation pressure and the radio jet can drive massive
outflows, and  it is challenging to disentangle the two contributions. 
Physical properties of outflows are important for constraining their impact on host galaxies. For example, it has been argued that 
 momentum-driven wind models (whose propagation is governed
by momentum conservation) are more successful in reproducing the size-mass relation of disk galaxies \citep{Dutton09}, the enrichment of the high-z IGM \citep{Oppenheimer06}, and the scaling relation between the black hole mass and the host galaxy bulge velocity dispersion (e.g., \citet{King15}). Conversely,
the energy-driven winds (dominated by energy conservation)  in semianalytical models have 
a  strong impact on the galaxy stellar mass function (by suppressing the star formation).
 We investigate the properties of outflows in radio galaxies in this paper.
  
We recently performed a survey of 3C radio sources in the framework of
a MUse RAdio Loud Emission lines Snapshot (MURALES). We observed 37 3C
radio sources with the integral field spectrograph MUSE at the Very
Large Telescope (VLT) at z<0.3. An extension of the project to radio
sources out to z$<$0.8 has recently been approved by ESO.  Our sample
comprises radio galaxies of different optical spectroscopic
properties: Radio galaxies have been classified on the basis of their
optical spectra \citep{Koski78, Osterbrock77} depending on the
relative ratios of the observed emission lines into high- and low-excitation galaxies and broad-line objects (HEGs, LEGs and BLOs,
e.g., \citealt{Laing94}).  We use the classification criteria
defined by \citet{Buttiglione09,Buttiglione10,Buttiglione11}.  The
differences observed in their optical spectra are linked to different
accretion modes (i.e., radiatively efficient $\text{vs}$ inefficient) and to properties of their host galaxy, including the star formation rate (see,
e.g.,\citealt{Chiaberge02,Hardcastle06,Baldi08,Balmaverde08,
  Hardcastle09,Baldi10,Best12,Hardcastle13}).  The MURALES survey
enriches the already extensive database of multifrequency observations
available for the 3C catalog, which was recently completed with the X-ray
coverage from the 3C Snapshot Survey (see, e.g.,
\citealt{Massaro10,Massaro12,Massaro15,Stuardi18,Jimenez-Gallardo20}).

 The first two papers presenting the MURALES results focused on single radio galaxies, 3C317 (hosted in the center of Abell 2052;  \citealp{Balmaverde18a}), for which we directly measured the velocity expansion of the cavities, and 3C459, a candidate  dual AGN \citep{Balmaverde18b}. 
 \citet{Balmaverde19} presented the first half of the MUSE observations, and \citet{Balmaverde20} completed the presentation of the remaining 37 radio galaxies of the sample. 
In this fourth paper of the MURALES series, we focus on the properties of the ionized gas in the
nuclear regions. Our aim is to search for signatures of high-velocity
nuclear outflows (e.g., beyond hundreds of km s$^{-1}$) and to measure their properties in an unbiased and
representative sample of low-redshift radio galaxies. Because outflows are frequently observed in the narrow-line region (NLR) of active galaxies (e.g., \citealt{Fischer13,Crenshaw15,Venturi18,Revalski18,Mingozzi19}), 
we focus on emission from the NLR in order to detect and characterize these ionized gas outflows. The low density of this region ($\text{log n}_e<
5.8\,\text{cm}^{-3}$ ; e.g., \citealt{Nesvadba06}) with respect to the BLR enables us to observe the
[O~III] emission line profile whose asymmetries (i.e., wings) can reach very high velocities and are considered signatures of outflows \citep{Spoon09, Mullaney13, Zakamska14, Brusa15}. In type 2 objects, the [O~III] emission line can be obscured by circumnuclear dust \citep{Haas05,Baum10}.
This can affect the observation of the line profile. 
For example, if the outflow is  inclined and extended beyond the potential obscuring material,
we can observe outflowing redshifted components  (e.g., \citealt{Barth08,Crenshaw10})
and complex emission line profile.
      
The paper is organized as follows: in Section~\ref{sample} we
describe the sample and present the MUSE observations
and the data analysis. Sections~\ref{nuclear} and \ref{resolved} describe the detection and properties of nuclear and extended emissions,
respectively. In Section~\ref{discussion} we discuss the physical
implications of the results, which are summarized in
Section~\ref{conclusion}. Finally, in the Appendix~\ref{Appendix~A}, we provide
a description of the individual significant sources  showing their
respective 2D velocity fields.

Unless stated otherwise, we used cgs units. We adopted a flat
cosmology using the following cosmological parameters: $H_0$= 69.6
$\text {km}\,\text{s}^{-1} \text{Mpc}^{-1} $ and $\Omega_m= 0.286$ \citep{Bennett14}.

\begin{table}
  \caption{Main properties of the 3C subsample observed with MUSE} 
  \begin{center}
\begin{tabular}{l c c c c c c c c c}
\hline
Name  &  z   & Class & L$_{\rm bol}$ & Seeing  & scale \\
      &      &       &[erg s$^{-1}$]  & $\arcsec$ & kpc/1$\arcsec$ \\
\hline
3C~015 & 0.073   & LEG & {43.39} & 0.65 & 1.39  \\
3C~017 & 0.22    & BLO & {44.77} & 0.49 & 3.55  \\
3C~018 & 0.188   & BLO & {45.34} & 0.53 & 3.14  \\
3C~029 & 0.045   & LEG & {42.88} & 0.51 & 0.88  \\
3C~033 & 0.06    & HEG & {44.97} & 0.63 & 1.16  \\
3C~040 & 0.018   & LEG & {41.99} & 0.40 & 0.37  \\
3C~063 & 0.175   & HEG & {44.42} & 0.49 & 2.97  \\
3C~076.1 & 0.032 & --  & {42.61} & 0.62 & 0.64  \\
3C~078 & 0.028   & LEG & {42.18} & 0.53 & 0.56  \\
3C~079 & 0.256   & HEG & {45.64} & 0.66 & 3.98  \\
3C~088 & 0.03    & LEG & {42.92} & 0.59 & 0.60  \\
3C~089 & 0.138   & --  & {43.29} & 0.64 & 2.44  \\
3C~098 & 0.03    & HEG & {43.78} & 0.66 & 0.60  \\
3C~105 & 0.089   & HEG & {44.24} & 0.71 & 1.66  \\
3C~135 & 0.125   & HEG & {44.83} & 0.52 & 2.24  \\
3C~180 & 0.22    & HEG & {45.12} & 1.45 & 3.55  \\
3C~196.1 & 0.198 & LEG & {44.31} & 0.48 & 3.27  \\
3C~198 & 0.081   & SF  & {43.76} & 0.78 & 1.53  \\
3C~227 & 0.086   & BLO & {44.55} & 0.91 & 1.61  \\
3C~264 & 0.021   & LEG & {41.95} & 0.85 & 0.43  \\
3C~272 & 0.003   & LEG & {40.78} & 0.39 & 0.06  \\
3C~287.1 & 0.216 & BLO & {44.53} & 0.65 & 3.50  \\
3C~296 & 0.024   & LEG & {42.52} & 1.08 & 0.48  \\
3C~300 & 0.27    & HEG & {44.79} & 0.41 & 4.14  \\
3C~327 & 0.105   & HEG & {45.03} & 0.70 & 1.93  \\
3C~348 & 0.155   & LEG & {43.20} & 1.76 & 2.69  \\
3C~353 & 0.03    & LEG & {42.92} & 1.30 & 0.60  \\
3C~403 & 0.059   & HEG & {44.54} & 0.54 & 1.14  \\
3C~403.1 & 0.055 & LEG & {42.64} & 0.80 & 1.07  \\
3C~424 & 0.127   & LEG & {43.59} & 0.98 & 2.27  \\
3C~442 & 0.026   & LEG & {42.00} & 0.61 & 0.52  \\
3C~445 & 0.056   & BLO & {45.28} & 1.48 & 1.09  \\
3C~456 & 0.233   & HEG & {45.60} & 1.27 & 3.71  \\
3C~458 & 0.289   & HEG & {44.82} & 0.50 & 4.34  \\
3C~459 & 0.22    & BLO & {44.82} & 0.43 & 3.55  \\
\hline
\end{tabular}
\end{center}
Column description: (1) source name; (2) redshift; (3) excitation
class (in some cases, the strength of the emission lines is
insufficient for a robust spectroscopic classification) from
\citet{Buttiglione10}; (4)  AGN bolometric luminosity obtained by
  applying a bolometric correction of 600 to the absorption-corrected
  nuclear [O~III] luminosity; (5) mean seeing of the observation; (6)
conversion factor from arcsec into kpc.
\label{tab:sample} 
\end{table}

\begin{figure}
\centering
\includegraphics[width=0.48\textwidth]{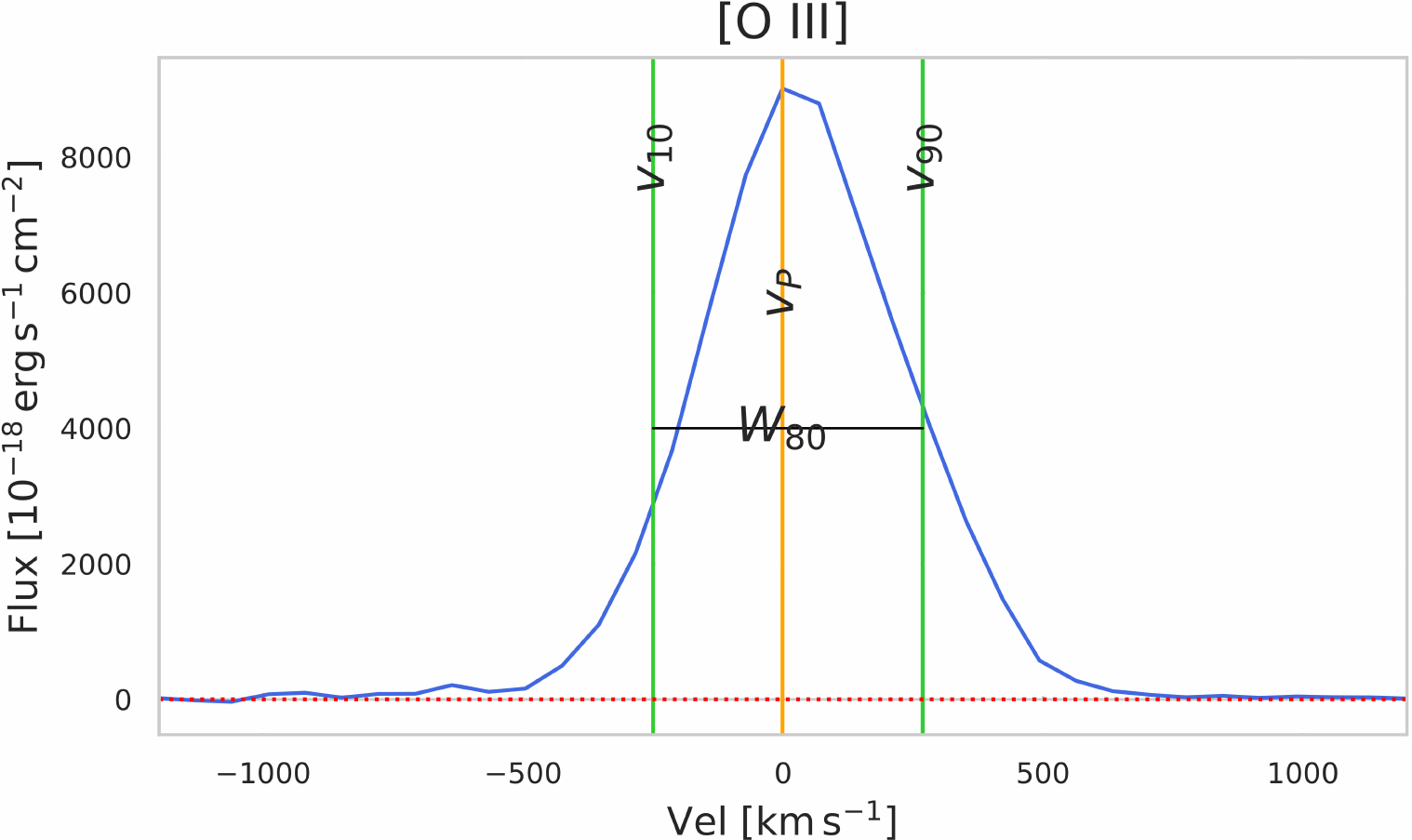}
\caption{Example of the [O~III] emission-line profile (blue curve) of
  3C~033 in the brightest spaxel of the image. The vertical orange
  line represents the velocity of the peak flux density (v$_{\text p}$), and
  the green vertical lines show the 10th (v$_{10}$) and 90th
  (v$_{90}$) velocity percentiles, which are used for the definition of
  the velocity width of the line that contains 80\% of the emission
  line flux ($W_{80}$).  In this spectrum the redshifted wing
  dominates, producing a positive value of the asymmetry
  (R > 0). The dotted red horizontal line is the continuum, linearly
  interpolated and subtracted.}
\label{fig:parameters}
\end{figure}

\section{Sample selection and data reduction}
\label{sample}

\begin{figure}
\centering
\includegraphics[width=0.49\textwidth]{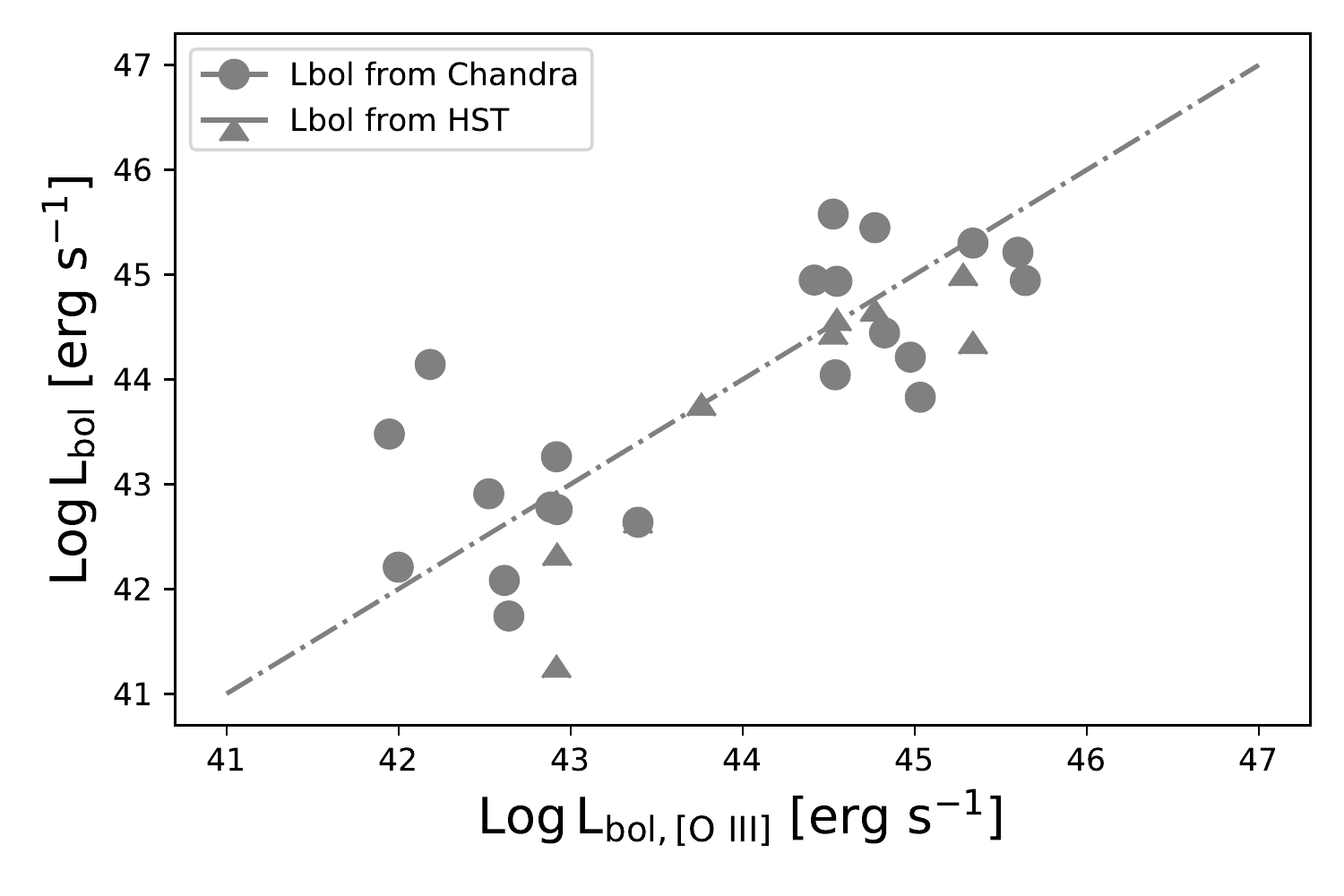}
\caption {Comparison of the bolometric values derived from the [OIII] luminosities with those obtained from 2-10 keV Chandra and 4400$\AA$ HST optical 
luminosity using  X-ray and optical bolometric corrections by \citet{Duras20}.}
\label{fig:lbol}
\end{figure}

The 3C catalog lists 298 extragalactic radio sources. We first
considered the 107 targets with good visibility from the VLT location
(i.e., $\delta < 20^{\circ} $), and we then selected the 37 sources
with a redshift below 0.3. This redshift limit allows us to
simulaneously map all the key optical emission lines from H$\beta$ at
4861\AA\ to the [S~II]$\lambda\lambda$ 6716,6731 doublet.  Our sample
comprises 14 LEGs, 12 HEGs, 6 BLOs, and one star-forming galaxy (SF),
based on the emission line ratios extracted from the nuclear
regions. In four cases, the strength of the emission lines is
insufficient for a robust spectroscopic classification. The
nuclear [O~III] luminosities of each source were measured by
integrating the nuclear emission in the first 0.6$\arcsec$ and
applying the proper aperture correction.  In two sources (namely
3C\,318.1 and 3C\,386), the [O~III] emission line is not detected, and
thus they were not considered for this analysis.  We corrected the
  fluxes for Galactic reddening using the extinction law of
  \citealt{cardelli89} and the galactic extinction listed in
  \citealt{Buttiglione09} taken from the NASA Extragalactic Database
  (NED) database. We prefer not to rely on the Balmer decrement
  because in several sources, the measurement of Hb is not sufficiently
  accurate to derive a robust estimate of the dust absorption. We
  derived the bolometric luminosity from the [O~III] luminosity by
  using the relation $L_{\rm bol}=600\times L_{\text{[O~III]}}$
  \citep{Shao13}.  Fig.~\ref{fig:lbol} compares these bolometric
  luminosities with those obtained from two alternative estimators
  based on the X-rays and optical nuclear luminosities. In particular, we considered the 2-10 keV luminosities
  from Chandra \citep{Massaro10,Massaro12,Massaro15, Massaro18} and the visible
  luminosities from HST (\citealt{chiaberge00}, scaled to 4400
  \AA\ adopting a spectral index of 1 and limiting to the unobscured
  objects), and we applied the corrections reported by
  \citet{Duras20}. The various estimates are consistent with each
  other with a spread of a factor $\sim$10. This also provides an
  estimate of their accuracies.  In Tab.~\ref{tab:sample} we list
their main properties, that is, the redshift, the optical spectroscopic
class, the bolometric luminosity ($L_{\rm bol}$), the seeing of the
observations, and the linear scale (i.e., kpc/1$\arcsec$).

These 37 radio galaxies were observed with MUSE, which is a high-throughput,
wide-field of view, image-slicing integral field unit spectrograph mounted
at the Very Large Telescope (VLT). Observations were performed using
the Wide Field Mode sampled at 0.2 arcsec/pixel.
The observations were obtained as part of programs ID
099.B-0137(A) and 0102.B-0048(A). Each source has been observed with two exposures
of 10 minutes, except for 3C~015 (2 x 13 minutes) and
3C~348 (2 x 14 minutes). The median seeing of the observations is 0\farcs65. We
used the ESO MUSE pipeline (version 1.6.2; \citealt{Weilbacher20}) to obtain fully reduced and
calibrated data cubes. The final MUSE data cube maps the entire source
between 4750 \AA < $\lambda$ < 9300 \AA. The data reduction and
analysis were described in the MURALES II paper by  \citet{Balmaverde19}, where all
details on the data reduction are reported.

 \begin{figure*}
\centering
\includegraphics[width=0.49\textwidth]{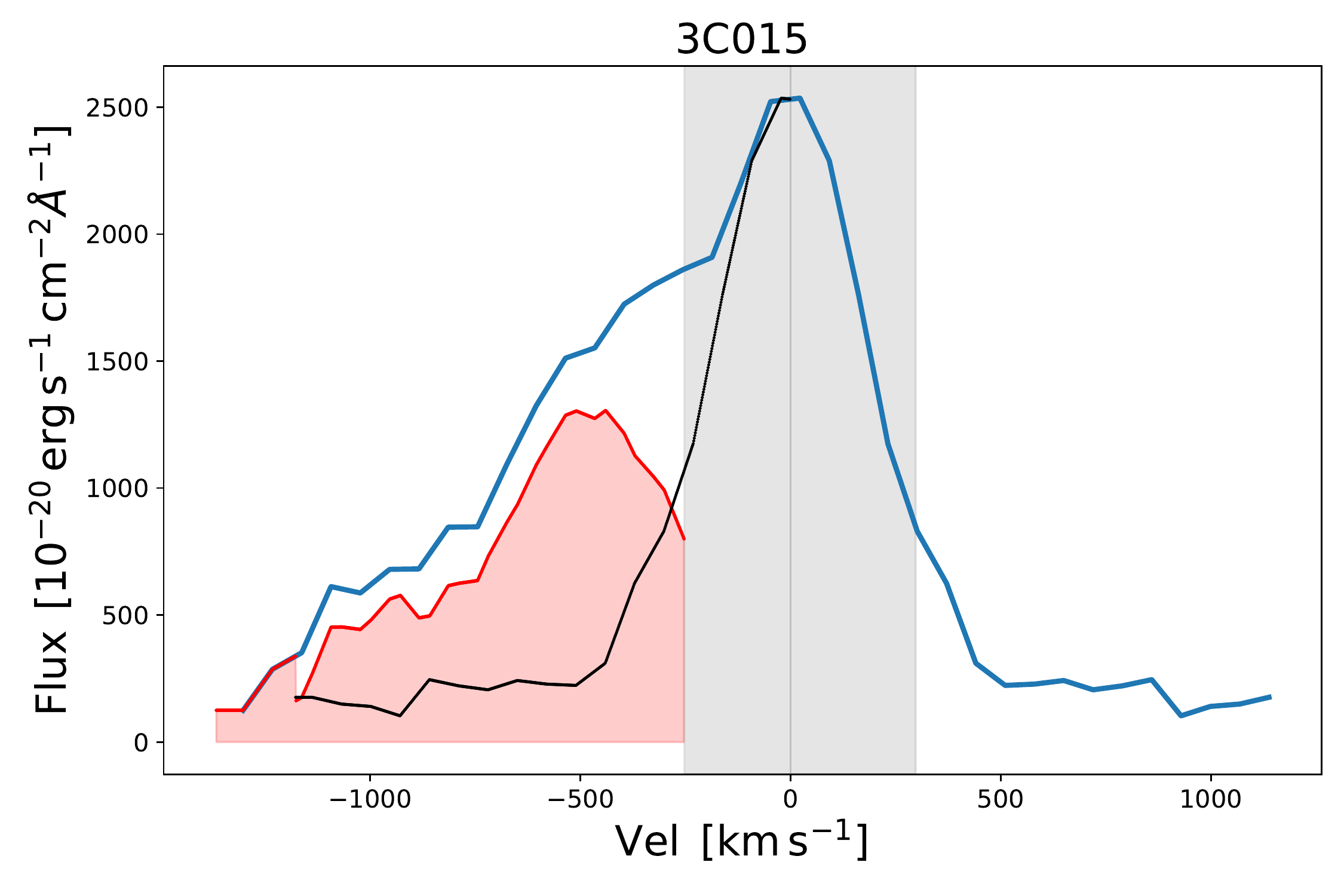}
\includegraphics[width=0.49\textwidth]{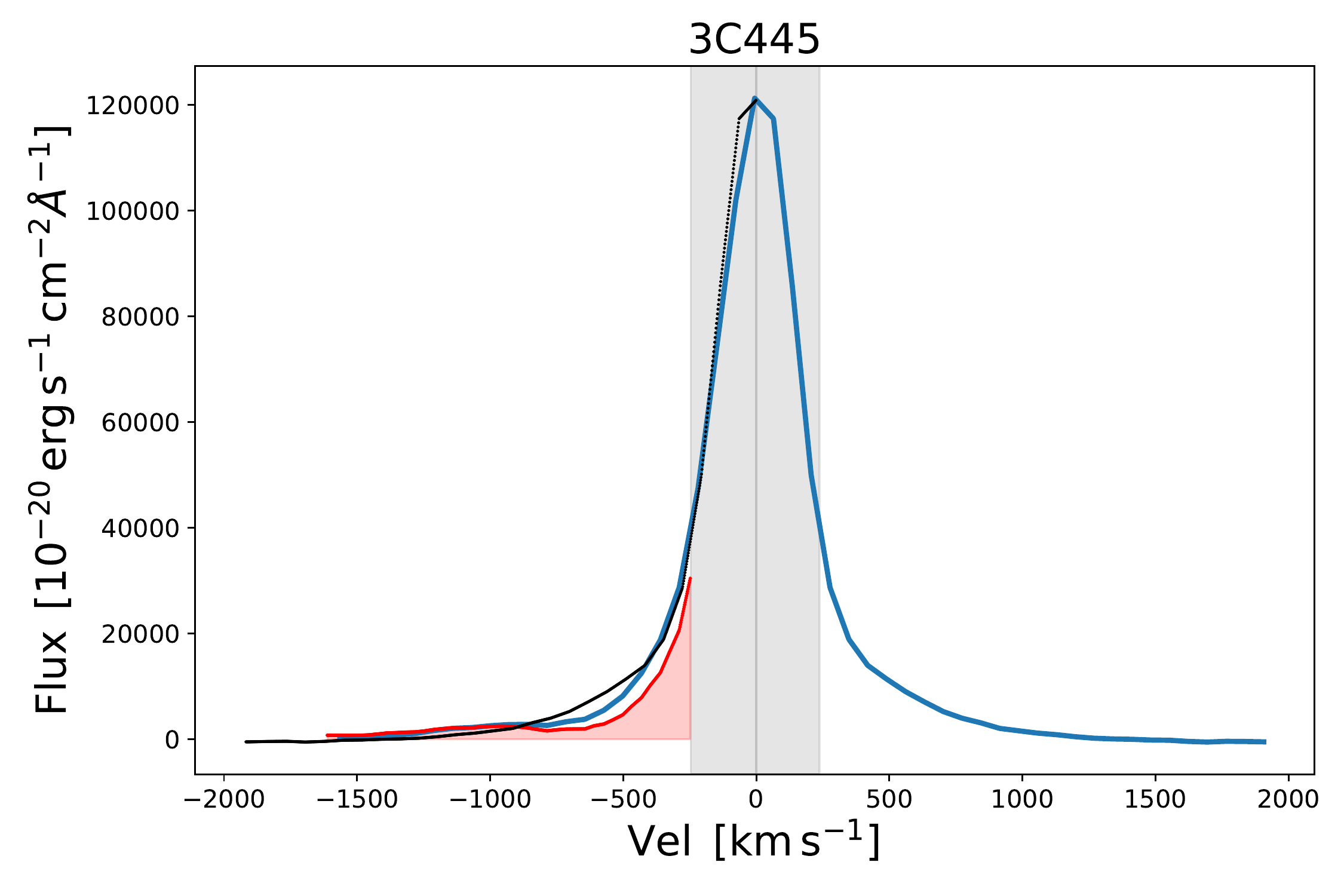}
\caption {Nuclear spectra of 3C~015 and 3C~445. Left panel: Observed [O~III]
  emission line is shown (in blue) with the
  mirror image of the redshifted line (in black). Their difference
  (in red) reveals a blueshifted wing
 in the profile, with ionized gas
  that reaches high negative velocities. We masked the central region of the [O~III] profile - at one-third
  of the maximum height of the emission line (grey region) - not considering the gas in ordered rotation. We measured the
  median velocity ($v_{50}$) of the outflow measured on the blue-wing residual
  component.}
\label{fig:ribalto}
\end{figure*}

\section{Searching for nuclear outflows}  \label{nuclear}

\subsection{Data analysis} \label{analysis}

We analyzed the nuclear profile of the [O~III] emission line to investigate the
possible presence of nuclear outflows in the selected sample of 3C
radio sources.  We extracted the nuclear spectrum of each source by coadding
   3 $\times$ 3 central spaxels (corresponding to 0\farcs6 $\times$ 0\farcs6)
   to sample a region comparable with the median seeing of the
   observations (0\farcs65).

We adopted a a nonparametric approach to perform our analysis, that is, we measured the velocity percentiles, that is,\ velocities
   corresponding to a specific percentage of total flux of the [O~III]
   emission line (e.g.,\ \citealt{Whittle85}), thus considering the line profile as a
probability distribution function. This choice was motivated by two main reasons:
 {(\rmnum{1})}  Emission line parameters are extracted directly  from the observed profile regardless of the complexity of its shape, and {(\rmnum{2})}  we can also study relatively weak [O~III] emission lines for which a fitting procedure with a multicomponent model would not yield an acceptable description.
    To
describe the kinematic of the ionized gas in the NLR, we
considered these parameters:

\begin{description}
    \item [\textbf{Velocity peak (v$_{\text p}$).}] The velocity corresponding
      to the peak of the emission line that we use as reference value for the other measurements of velocities.

    \item[\textbf{Width ($W_{80}$).}] The spectral width including
      80\% of the line flux, defined as the difference between the 90th and 10th percentile of the line profile (v$_{90}$ and v$_{10}$ , respectively). For a Gaussian profile, the value $W_{80}$ is close
      to the conventionally used full width at half maximum
      \citep{Veilleux91b}.

    \item[\textbf{Asymmetry ($R$).}] The asymmetry is defined by the combination of
      v$_{90}$, v$_{50}$ , and v$_{10}$ (the 90th,
      50th, and 10th percentile of the velocity, respectively) as follows:
\begin{equation}
R=|(v_{90} - v_{50} )| - |(v_{10} - v_{50})|
.\end{equation}
It quantifies the difference of flux between the redshifted and
blueshifted wing of the emission line. R shows negative values when
the emission line is asymmetric toward the blueshifted
side and positive values when the redshifted wing
is more prominent.
\end{description}

Fig.~\ref{fig:parameters} shows how these quantities are measured in the nuclear [OIII] emission line profile of 3C 033. The narrow
component of the line (i.e., $W_{80}$<600 km s$^{-1}$) is usually
considered a tracer of the galactic rotation and of the stellar
velocity dispersion (e.g., \citealt{Green05,Barth08}). 
$W_{80}$ provides
information about the velocity dispersion, and R  highlights the
asymmetry of the emission line profile. 

The fraction of the total ionized gas that moves at a higher velocity than the ordinary rotation is traced by the faint, broad wings of the [O~III] line.
To measure the velocity and flux of the gas in the outflow, we proceeded and adopted the following method that we illustrate in Fig. \ref{fig:ribalto} for two 
sources, namely 3C~015 and 3C~445. 
  First of all, we subtracted the continuum by  linearly fitting two continuum bands of width 40 $\AA$ , centered at 100 $\AA$
from the line peak on the blue side (to avoid the second line of
the doublet at 4959 $\AA$) and 50 $\AA$ on the red side. Then first we estimated the centroid of the emission line as the flux-weighted wavelength $v_{\text centr}$ in the inner region of
     the emission line from the peak down to one-third of the height of the
     peak (which we define as the core of the line).  For a Gaussian
     profile, this value corresponds to $\sim 90 \%$ of the
     total flux of the line, and it is produced mostly by gas moving close to
     the systemic velocity.      
Second, we obtained a mirror image of the red side of the [O~III]
      profile with respect to the $v_{\text centr}$ axis and subtracted it from the
      blue side of the line. This procedure removes the symmetric component of the emission line and only leaves the asymmetric wing.e
Finally, we masked the core of the line, which is dominated by the gas in rotation, to obtain the component
      that is exclusively associated with the outflowing gas. Moreover, in the
      central region, the subtraction of the mirror image of the red side with the blue counterpart produces strong oscillations: the possible signatures of faint outflowing gas would be
      smeared out in the core of the line. 

We also defined two additional parameters
 for the residual wing. The median velocity of the wing ($v_{50, \text{wing}})$ that corresponds to the
  50th percentile of the flux contained in the wing of emission line
  profiles and represents the median velocity of the outflow. Positive
  and negative values are associated with redshifted and blueshifted wings, respectively. The flux of the wing ($F_{\text w}$) that is the total flux
  density in the wing of the emission line.

\begin{figure}
\includegraphics[width=0.49\textwidth]{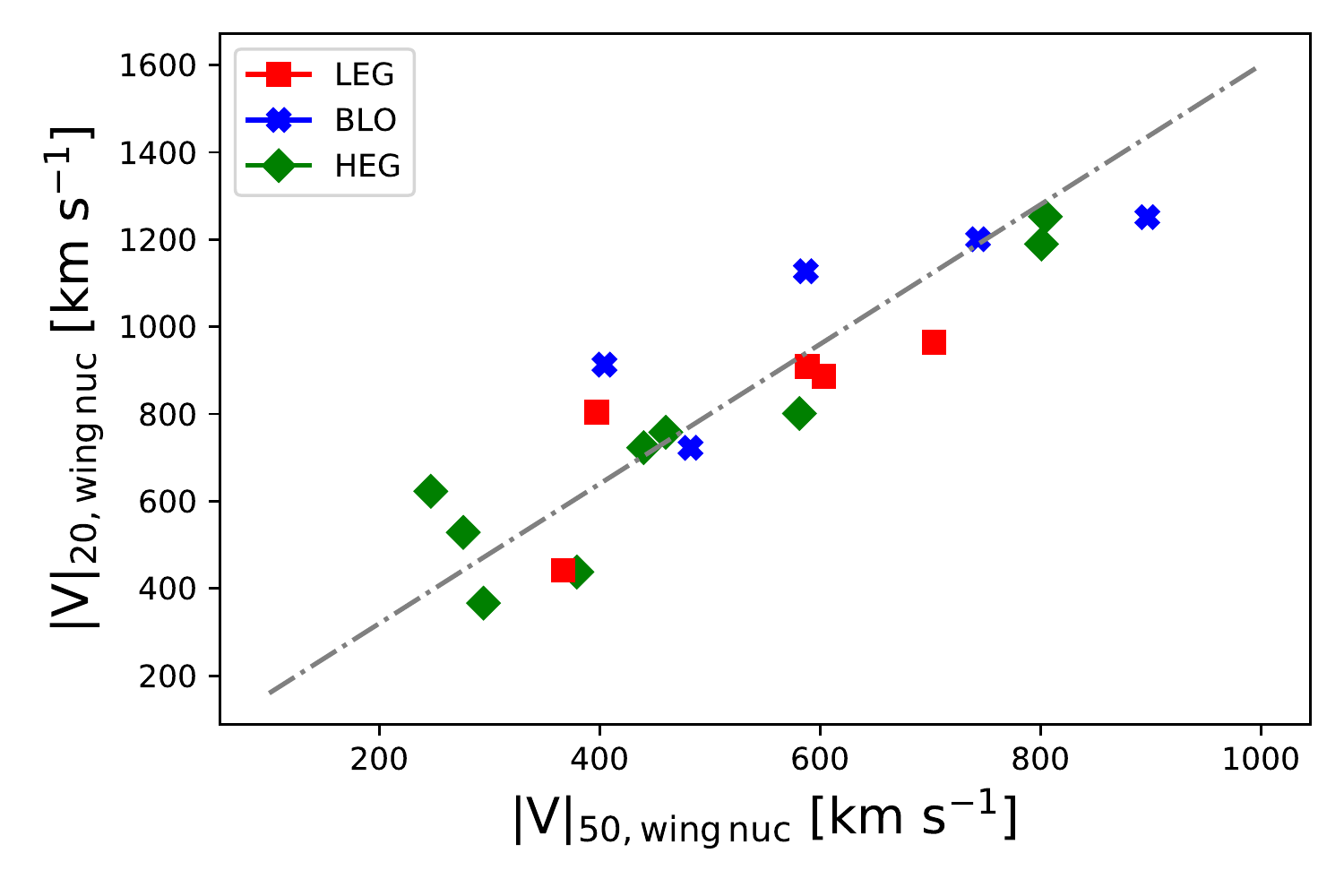}
\caption{Comparison  of the velocity measured at the 20th$^{\rm }$
  percentile of the whole [O~III] line, $v_{20}$, and of the median
  velocity ($v_{50, \text{wing}})$ of the high-velocity wing on the
  nucleus. The dash-dotted line corresponds to a constant ratio of
  1.6. This means that in our sample, v$_{20}$ would overestimate the outflow velocity.}
\label{fig:vel50vel20}
\end{figure}

In the literature, the terminal velocity of the ionized gas is often
used to explore the outflow properties. It generally corresponds to
$v_{05}$ or $v_{20}$,
that is, the flux contained in the 5th and 20th percentiles of the
emission line profiles (e.g., \citealt{Sturm11, Rupke13, Veilleux13}). However, {(\rmnum{1})} these values 
are affected by the bulk of the emission that is dominated by
rotating gas in the galaxy, therefore they may not be representative of the
outflowing gas; 
{(\rmnum{2})} they can only characterize the
blueshifted and not the redshifted component; and {(\rmnum{3})} they
are  significantly affected by the background noise because they are
measured at the faint end of the profile.
In Fig.~\ref{fig:vel50vel20} we compare our estimate
of the outflow velocity, $v_{50, \text{wing}}$, with $v_{20}$. The two
velocities are closely related, with a ratio of $\sim$1.6 (i.e., $v_{20}\,\sim1.6\,v_{50, \text{wing}}$). In the following, we use
 $v_{50, \text{wing}}$ because this velocity
is expected to be more representative of the velocity of the outflow and less affected by the galaxy 
rotational velocity field.

We estimated the uncertainties of the flux wing component $F_{\text w}$ and   $v_{50, \text{wing}}$ 
with a Monte Carlo simulation
and produced mock spectra. These were obtained by varying the flux of
each spectral element of the emission line profile by adding random
values extracted from a normal distribution with an amplitude given by
the uncertainty in each pixel. This information is stored for each
pixel in the pipeline products. The uncertainties of the parameters were
computed by taking the 1$\sigma$ of the each parameter distribution
computed from 1000 mock spectra. We consider that the outflow that is detected when
the flux wing component $F_{\text w}$ has an S/N higher than 5.

\begin{table}
  \caption{Main properties of the nuclear outflows}
  \begin{center}
\begin{tabular}{l l c c c c}
\hline
Name      &  Class  &  Log F$_w$               & S/N &  v$_{20}$  & v$_{50,wing}$ \\
          &         &   &     &  [\kms]   & [\kms]   \\
\hline
015 &   LEG & -14.35 & 14.8 & -910 & ~-590 $\pm$ 10\\
017 &   BLO & -14.57 & 12.3 & -1130 & ~-590 $\pm$ 15 \\
018 &   BLO & -13.98 & 17.4 & -1200 & ~-740 $\pm$ 10\\
029 &   LEG & -14.95 & 5.2 & -890 & ~-600 $\pm$ 25 \\
033 &   HEG & -14.63 & 6.8 & 440 &  ~340 $\pm$ 5 \\
063 &   HEG & -15.38 & 9.9 & -800 &  -580 $\pm$ 5 \\
088 &   LEG & -14.78 & 52.9 & -960 & -700 $\pm$ 5 \\
098 &   HEG & -14.25 & 8.1 & -530 & ~~-280 $\pm$  20 \\
105 &   HEG & -14.39 & 8.5 & -760 & ~~-460 $\pm$ 10 \\
135 &   HEG & -14.86 & 6.4 & -1190 & -800 $\pm$ 5 \\
196.1 & LEG & -15.31 & 5.0 & -440 & ~~-370 $\pm$ 10\\
227 &   BLO & -13.65 & 19.0 & -720 & -480 $\pm$ 5 \\
300 &   HEG & -14.95 & 6.9 & -620 & -250 $\pm$ 5\\
327 &   HEG & -14.15 & 21.7 & -370 & -290 $\pm$ 5\\
403 &   HEG & -13.56 & 71.6 & -720 & -440 $\pm$ 5 \\
442 &   LEG & -14.86 & 5.7 & -800 & ~~-400 $\pm$ 20 \\
445 &   BLO & -12.79 & 107.6 & -910 & -400 $\pm$ 5\\
456 &   HEG & -14.37 & 7.4 & -1250 & -800 $\pm$ 5\\
459 &   BLO & -14.36 & 13.8 & -1250 & ~~-900 $\pm$ 10\\
\hline
\end{tabular}
\end{center}
Column description: (1) source name; 
(2) excitation class;
(3) logarithm of the flux of the nuclear  the [O~III] emission line wing in [erg s$^{-1}$ cm$^{-2}$];
 (4) S/N of the wing flux; 
 (5 ) the 20th percentile of the velocity measured in the entire emission line; 
 (6) median velocity  of
outflows measured in the [O~III] emission line wing on the nucleus.\label{tab:nucleare}
\end{table}

 \begin{table}
  \caption{\bf{Statistical significance of the relations}}
  \begin{center}
\begin{tabular}{l l l l l}
\hline
Relations   & $\rho$ &       p & significant\\
\hline

${\rm |V|}_{\rm 50,wing\,nuc}$ vs ${\rm |V|}_{\rm 20,wing\,nuc}$                    & 0.88    & 0.0002 & yes \\
${\rm Log}\,({\rm L}_{\rm w,nuc}/{\rm L}_{\rm nuc})$ vs ${\rm Log\,L}_{\rm bol}$& -0.11  & 0.52   & no \\
${\rm Log\,{M}_{OF,nuc}}$ vs ${\rm Log\,L}_{\rm bol}$              & 0.50  & 0.0033 & yes \\
$E_{kin,tot}$  vs  ${\rm Log\,L}_{\rm bol}$                            & 0.79  & 0.0001 & yes \\   
$\dot{M}_{OF}\,[M_{\odot}/yr]$ vs ${\rm Log\,L}_{\rm bol}$                     & 0.38 & 0.08   & no \\
${\rm Log\,\dot{E}}_{\rm kin,tot}$ vs ${\rm Log\,L}_{\rm bol}$                 & 0.33  & 0.183 & no \\ 
\hline
\end{tabular}
\end{center}
 Generalized Spearman's  rank  order
 correlation  coefficient  between two variables  including censoring in the independent variable ($\rho$) and 
 probability  that  there  is  no  correlation
    between  the  variables ({\it p}).
\label{tab:stat} 
\end{table}

\subsection{Properties of the nuclear outflows}

According to the procedure and criteria described above (Section~\ref{analysis}), we
find significant nuclear outflows in about half of the sample (19 sources out of 37). Their main
properties are listed in Tab.~\ref{tab:nucleare}. Nuclear outflows are
detected in 5 out of 6 BLOs, 9 out of 12 HEGs, and 5 out of 14 LEGs, but none in the SF galaxy.  All but one source (3C~033) 
 show blueshifted wings with v$_{20} $velocities between approximately  $-1200$ and
$-400 \text {km}\,\text{s}^{-1}$ and  v$_{50,wing} $ velocities between $-900$ and
$-300 \text {km}\,\text{s}^{-1}$ . In 3C~033, the high-speed
residuals are located on the red side of the emission line (see Section~\ref{properties_outflow} for more details).

\begin{figure}
\centering
\includegraphics[width=0.5\textwidth]{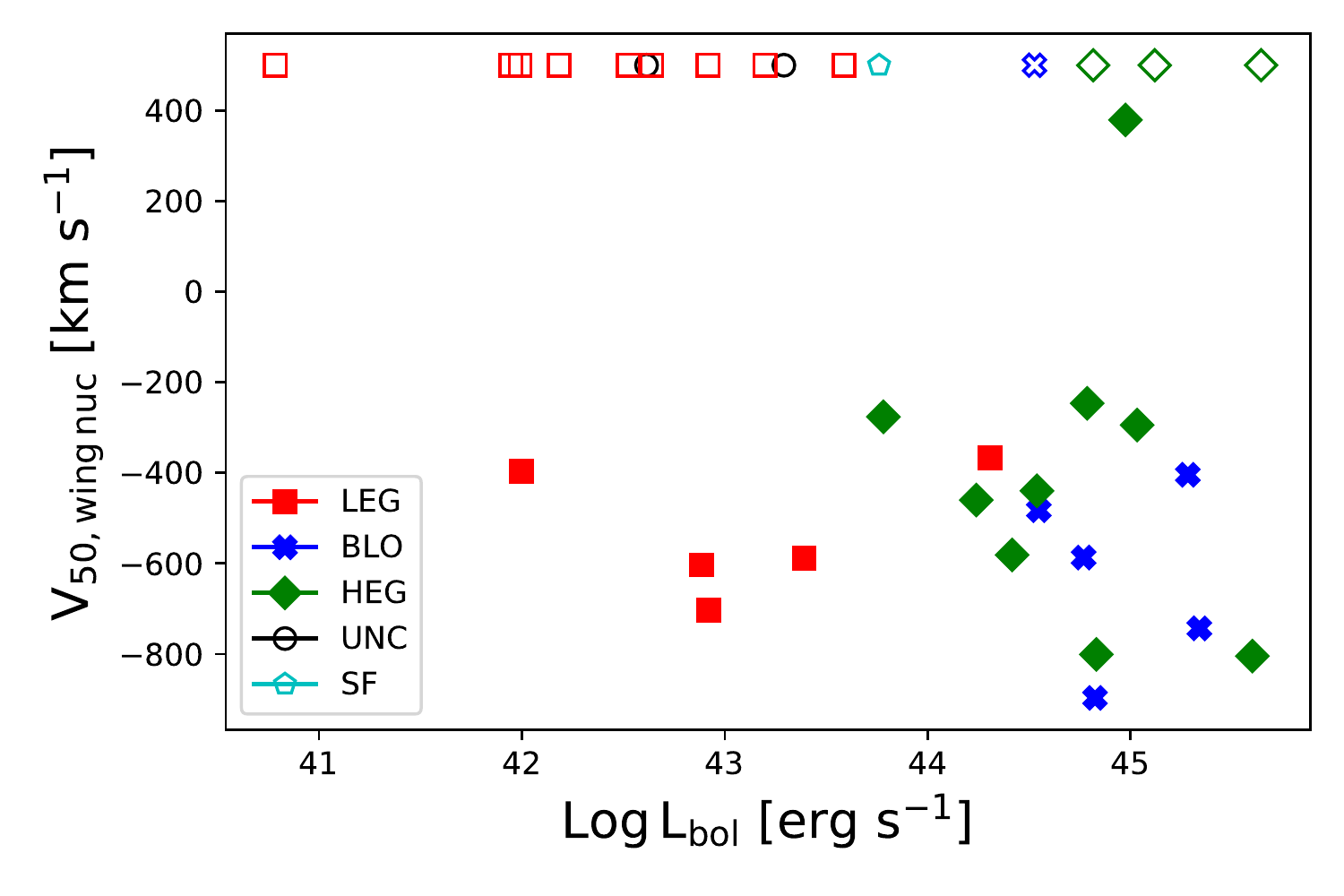}
\caption{Median velocity of the high-speed line wing (see text for
  details) as a function of the AGN bolometric luminosity. Green
  diamonds show HEGs, blue crosses show BLOs, red squares show LEGs,
  black circles show the spectroscopically unclassified sources, and the
  cyan circle represents the star forming galaxy. The empty symbols at the top
  of the figure represent sources without a detected line wing. The various classes show similar outflow velocities.}
 \label{fig:velnuc}
\end{figure}

In Fig.~\ref{fig:velnuc} we show the median velocity of the nuclear
outflow (v$_{50,\,\text{wing}}$) as function of the AGN bolometric
luminosity ($L_{\text{bol}}$) for the various spectroscopic classes of
radio galaxies.
To derive the bolometric luminosity from the [O~III] luminosity, we adopted the correction factor proposed {\bf by \citet{Shao13}} for type 1
AGN, knowing that this average correction factor cannot be
accurate for single sources. This correction is appropriate for
BLOs. HEGs show the same narrow emission line ratio properties as
BLOs \citep{Buttiglione10}. However, \citet{Baldi13} showed that their
NLR is partially obscured by the circumnulear torus, causing a deficit
of a factor $\sim 2$ in line luminosity. Finally, LEGs are characterized by a different spectral energy distribution (SED) and low efficient accretion process with respect to type 1 AGNs.
Nonetheless, althought the bolometric correction cannot be highly
accurate for these sources, the large bolometric differences obtained
between the spectroscopic classes reflect a genuinely different
accretion and efficiency level. Even with this caveat, there is no clear
connection between nuclear outflow velocities and bolometric
luminosity or spectroscopic classes.

\begin{figure}
\centering{
\includegraphics[width=0.5\textwidth]{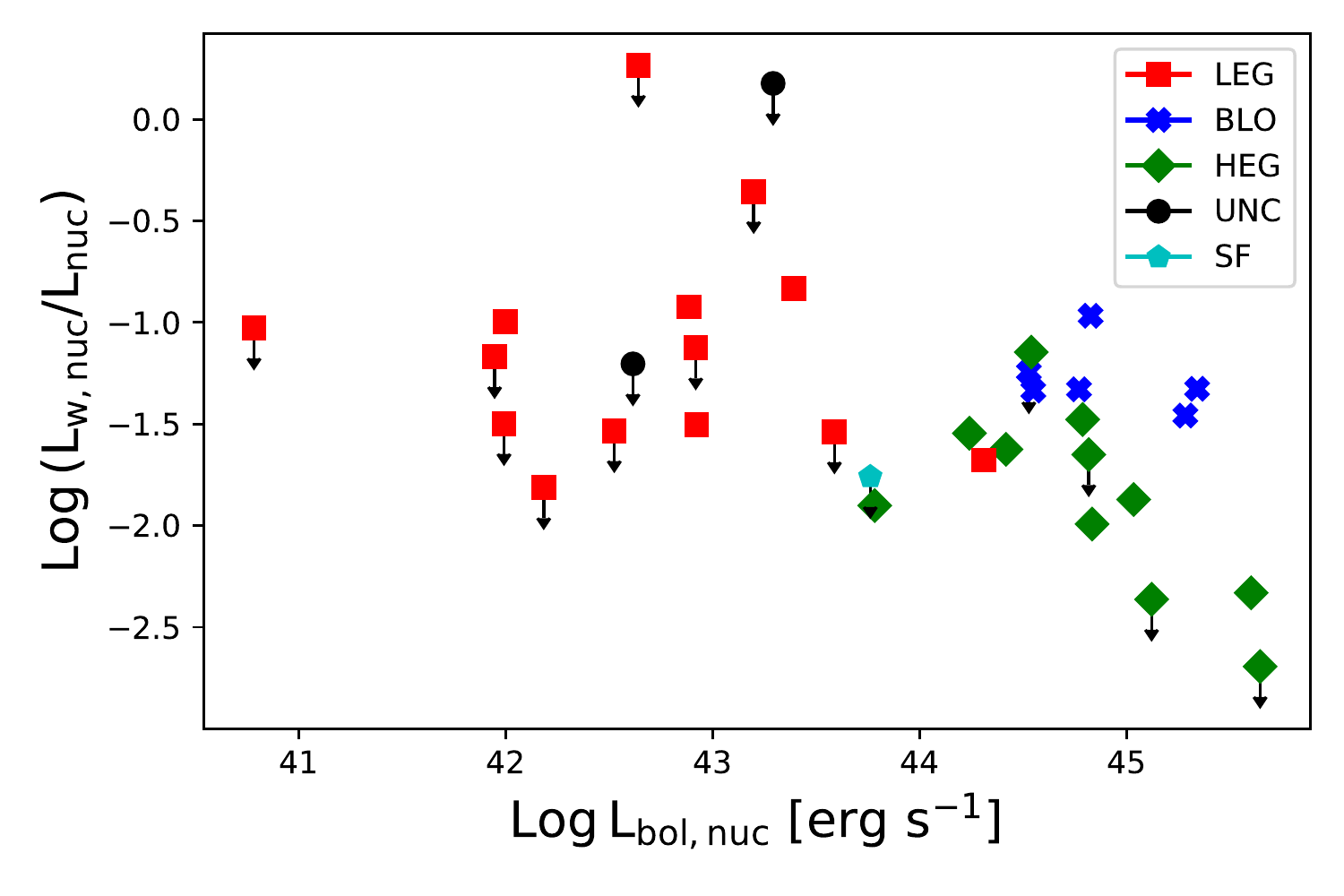}
\caption{Ratio of the [O~III] luminosity of the nuclear
  outflowing gas with respect to the whole ionized gas in the
  galaxy nucleus vs. bolometric luminosity. }
\label{fig:eff}}
\end{figure}

The mass and energy outflow rates
are important quantities for investigating the main mechanisms that drive
them and their impact on the surrounding environment. First, we
evaluated the fraction of the
gas in outflows with respect to the total. In Fig. \ref{fig:eff} we
show the $L_{\text{w,\,nuc}}/L_{\text{nuc}}$ ratio with respect to the
bolometric luminosity.  BLOs show typical ratio values of 0.05 - 0.10;
the five detected outflows in the LEGs are characterized by similar
fractions, and the HEGs extend to lower fractions, particularly
considering the three sources without a detected nuclear outflow.

We estimated the mass of the outflow using the relation discussed by
\citet{Osterbrock89},
\begin{equation} M_{of}=7.5\,\times\,10^{-3} \times
\left(\frac{10^4}{\langle n_e \rangle}\frac{L_{H
\beta}}{L_\odot}\right) M_{\odot}
.\end{equation}

 n$_e$ is the electron density and L$_{H\beta}$ is the luminosity of the H$_\beta$ emission line in erg s$^{-1}$.
 n$_e$ is commonly measured from the emission line ratio [S~II]
$\lambda$6716/$\lambda$6731 (e.g., \citealt{Osterbrock89}).  In
principle, to estimate the electron density of the outflows, we should
measure the emission line ratios only in the wings of the [S~II]
doublet, which is representative of the gas moving at high velocity. However,
it is very challenging to deblend the [S~II] doublet and take the possible presence of high-velocity wings into account, particularly in
the off-nuclear regions where these lines can be very weak. In the
literature we found various measurements of densities in the outflows:
\citet{Nesvadba06,Nesvadba08} derived $\langle n_e\rangle$ = 240--570
cm$^{-3}$ in two radio galaxies at z $\sim$2, \citet{harrison14} measured
$\langle n_e\rangle$ = 200--1000 cm$^{-3}$ in a sample of low-z AGN, and 
\citet{perna15} estimated $\langle n_e\rangle$ = 120 cm$^{-3}$ in a z
= 1.5 AGN.  Recently, the reliability of this method for measuring n$_e$
has been questioned by \citet{davies20}: they found that n$_e$ derived
from the [S~II] doublet is significantly lower than that found with
ratios of auroral and transauroral lines. However, these measurements
are obtained for compact radio galaxies that might be compressing gas
in the very central regions of their hosts, which would produce a higher
density.
Following \citet{Fiore17}, we adopted an average
gas density of $\langle n_e\rangle$ = 200  cm$^{-3}$ for all objects in the sample.
We discuss this choice in Section~\ref{discussion}. We measured the nuclear ratio [O~III]/H$\beta$
to convert the  luminosity measure in the [O~III] wing into the H$\beta$ luminosity.

We then estimated the kinetic energy of nuclear outflows as
\begin{equation}
E_{\text{kin}}=\frac{1}{2} M_{\text{of}}v_{50}^2
.\end{equation}

For the undetected sources, we estimated an outflowing mass upper limit
by adopting a fiducial value for the velocity of 500 \kms. The values of
the outflow mass and kinetic energy are collected in
Tab.~\ref{table:estese2}.

\begin{figure}
\centering{
\includegraphics[width=0.49\textwidth]{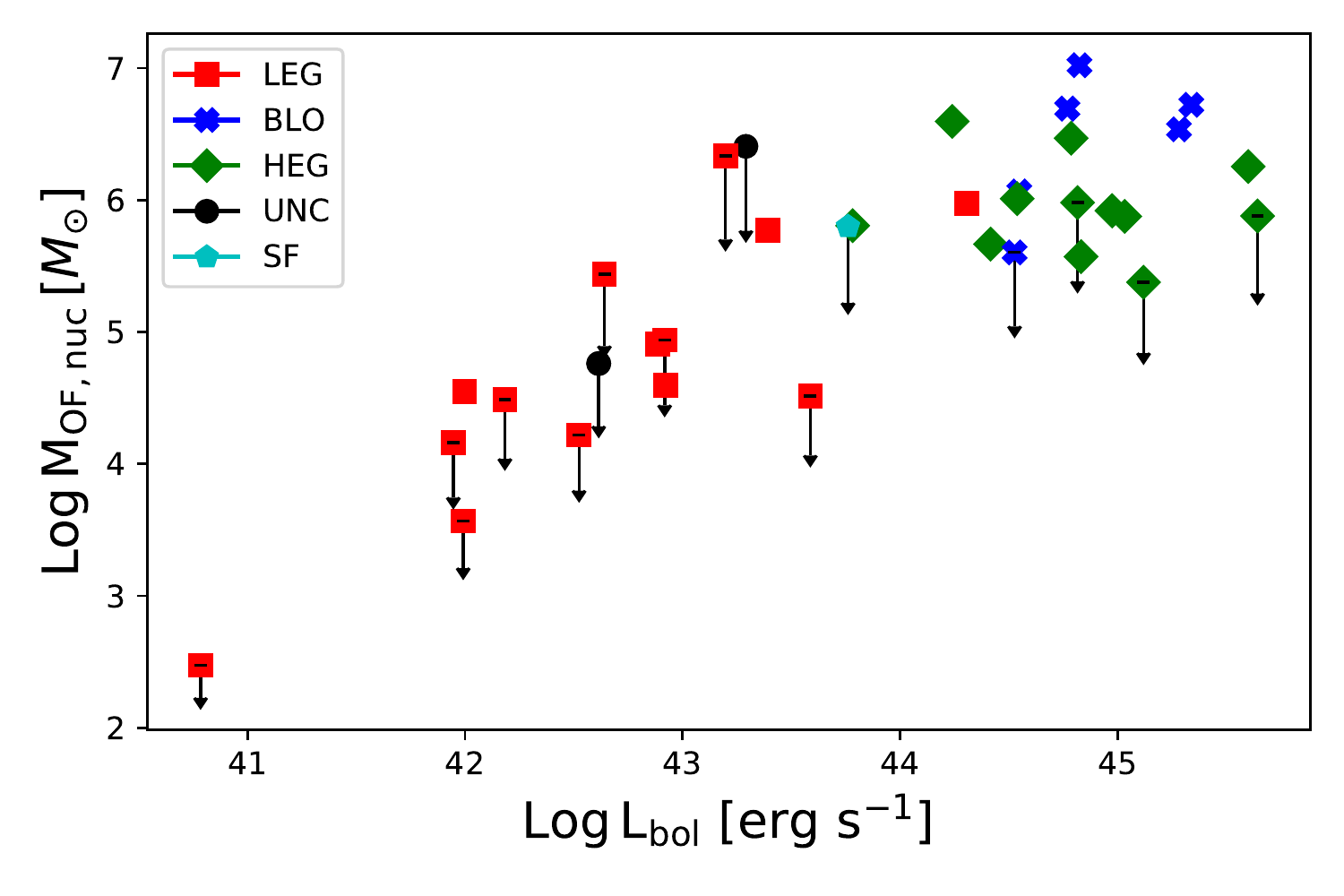}
\includegraphics[width=0.49\textwidth]{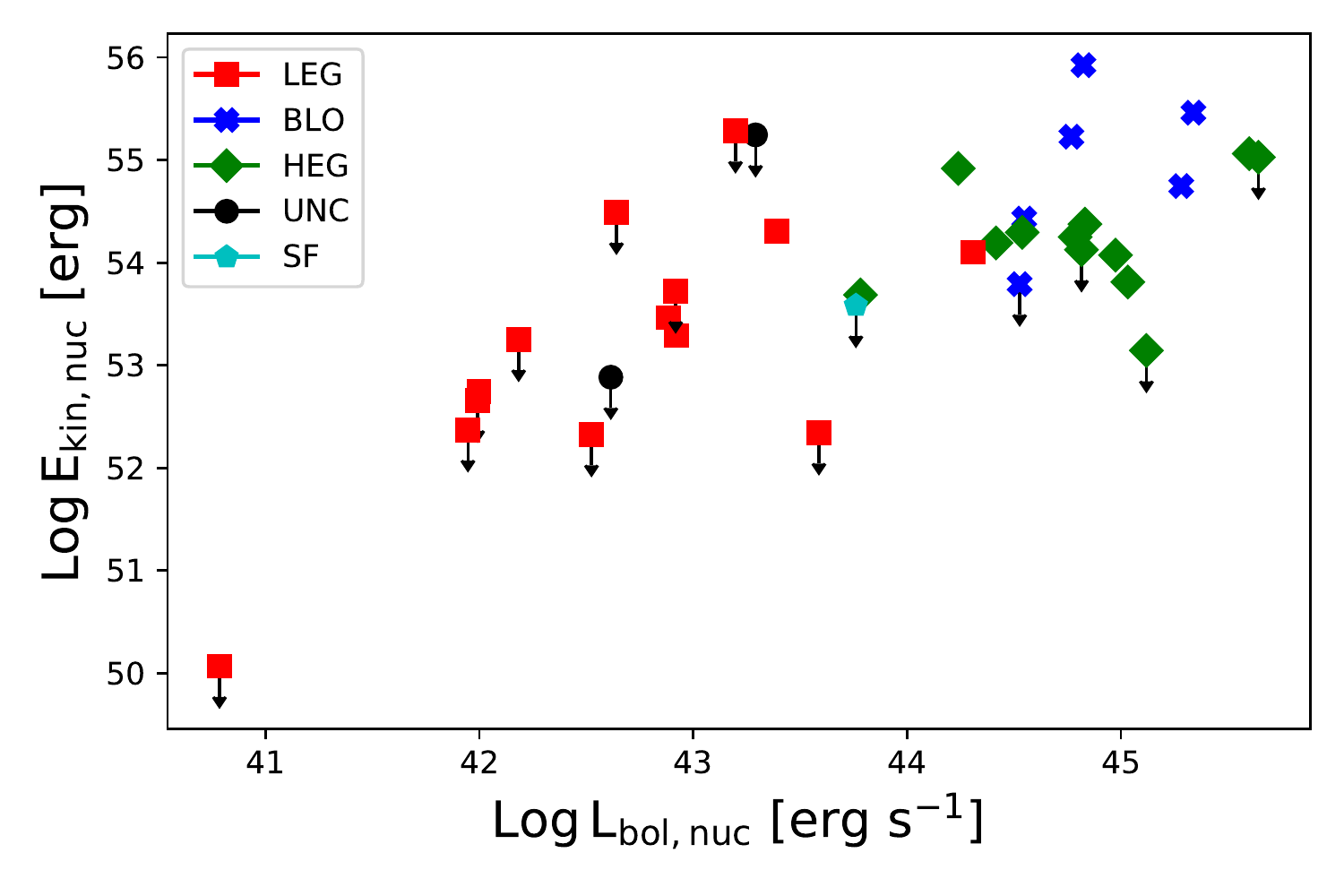}
\caption{Mass of ionized gas (top) and kinetic energy (bottom) of the
  nuclear outflows as a function of the AGN bolometric luminosity. The
  outflow mass and energy increase with the bolometric luminosity. The highest values are seen in BLOs.}
\label{ekinnuc}}
\end{figure}

The outflowing mass and
kinetic energy are plotted as a function of the AGN bolometric luminosity in
Fig.~\ref{ekinnuc}. 
Overall, both quantities increase with the
bolometric luminosity. The median outflow mass and power in BLOs is
about one order of magnitude higher than in HEGs.

\section{Searching for spatially resolved outflows}  \label{resolved}
\begin{figure*}
\centering
\includegraphics[width=1.\textwidth]{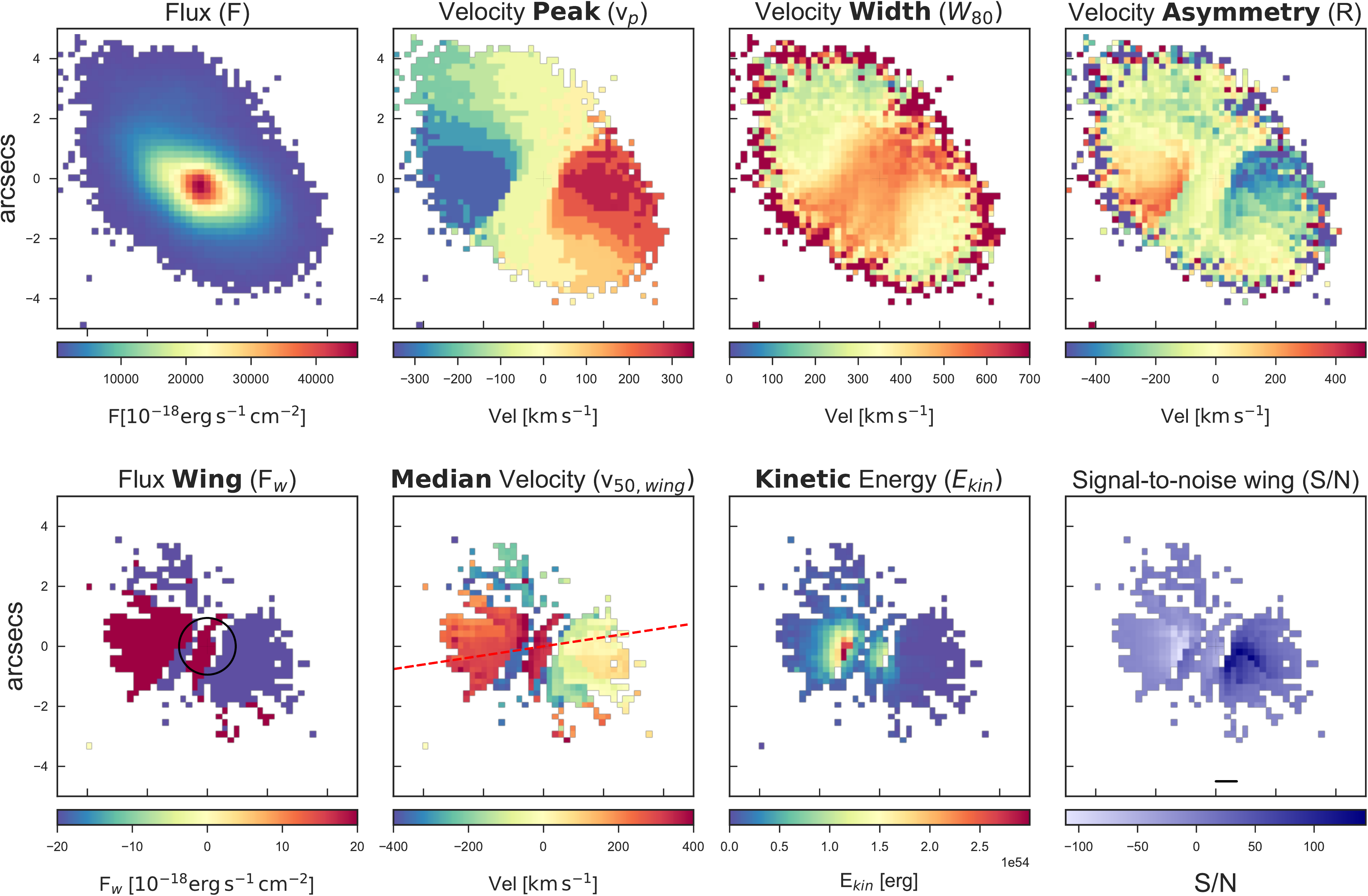}
\caption {Spatially resolved maps of the parameters defined in Section~\ref{resolved} for
  3C\,033. The maps are centered on the nucleus of the source, identified by the [O~III] peak luminosity. 
  Top panel from left to right: 1) [O~III] flux, 2) velocity peak ($v_{\text p}$) of the
  [O~III] emission line, 3) emission line width
  ($W_{80}$), and 4)  velocity asymmetry of the [O~III] profile. Bottom panel from left to right: 1)
  flux of the outflowing gas isolated by the rotational component
  (F$_{\text w}$), the black circle has a diameter of three times the seeing of
  the observations; 2) median velocity of the outflowing gas
  (v$_{50,\text{wing}}$), the dashed line marks the radio position
  angle; 3) kinetic energy of the [O~III] emitting gas; and 4)
  S/N of the [O~III] emission residing in the line wing. In all maps,
pixels with an S/N<5 have been discarded.}
\label{fig:velocitymap}
\end{figure*}

\begin{figure}
  \centering
\includegraphics[width=0.49\textwidth]{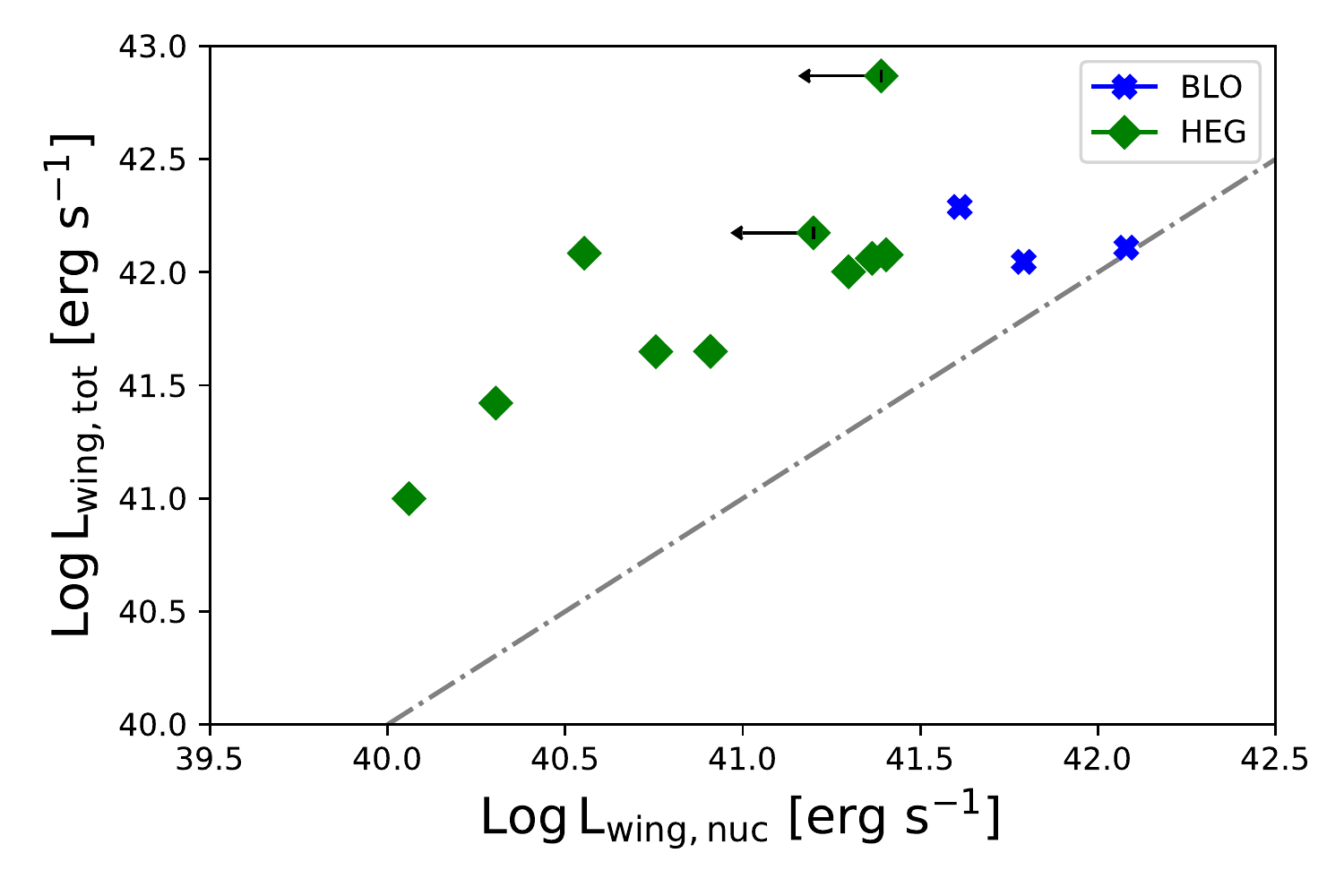}
\caption{Comparison of the [O~III] luminosity of the high-velocity
  wing, measured in the whole outflow and in the nuclear region. The three BLOs these estimates are similar, but in the HEGs the
  outflow is dominated by its extended component. The dot-dashed line represents the 1:1 relation.}
\label{nucext}
\end{figure}

 \begin{table}
  \caption{\bf Properties of the extended outflows}
  \begin{center}
\begin{tabular}{l l l l r r r}
\hline
Name    & Class &   Log F$_w$            &   |v|$_{50}$  & v$_{20}$   & PA$_{\rm jet}$ & PA$_{\rm of}$\\
\hline
033 & HEG & -13.51 & 220 & -540 & 19 & 61 \\
063 & HEG & -13.85 & 990 & -1020 & 33 & 20 \\
079 & HEG & -14.13 & 220 & -530 & 105 & 47 \\
098 & HEG & -13.32 & 300 & -770 & 25 & --- \\
105 & HEG & -13.65 & 360 & -880 & 127 & --- \\
135 & HEG & -13.96 & 220 & -680 & 75 & 40 \\
180 & HEG & -13.29 & 210 & -600 & 10 & 29 \\
227 & BLO & -12.98 & 380 & -470 & 85 & --- \\
300 & HEG & -14.28 & 740 & -920 & 132 & 110 \\
327 & HEG & -13.45 & 280 & -750 & 100 & 62 \\
403 & HEG & -12.86 & 340 & -670 & 86 & 30 \\
445 & BLO & -12.77 & 140 & -580 & 173 & 160 \\
459 & BLO & -14.11 & 540 & -790 & 97 & 110 \\
\hline
\end{tabular}
\end{center}
Column description: (1) source name; (2) excitation class; 
(3) logarithm of the gas outflow flux (in units of erg cm$^{-2}$ s$^{-1}$;  (4) median value of the
 velocity ($v_{50}$) along the the outflow; (5) 20th velocity percentile; (6 and 7) position angle of the radio
axis and of the extended outflow. 
\label{tab:estese} 
\end{table}

 \begin{table*}
  \caption{Energetical properties of the outflows}
  \begin{center}
\begin{tabular}{l c c r r r r r r r r}
\hline
Name    & Class &  & Log $M_{\text{of}}$   & Log $E_{\text{kin}}$ & R$_{out}$ & Log $\dot{E}_{\rm kin}$ &  $\dot{M}_{\rm of}$ &  Log $\dot{P}_{\rm of}$ &  Log $\dot{P}_{\rm AGN}$  &  Log $\dot{P}_{\rm jet}$         \\
        &       & & [M$_{\odot}$]   &  [erg]          & [kpc]    & [erg s$^{-1}$]          &   [M$_{\odot} yr^{-1}$] &  [dyn]]                &  (L$_{bol}$/c)         &     [erg s$^{-1}$]          \\
\hline
3C~015   & LEG  & n & 5.8    & 54.3     &$<$3.8  & $>$40.6 &  $>$0.4  & $>$33.2  &   32.9    & 44.1         \\
3C~017   & BLO  & n & 6.7    & 55.2     &$<$18.5 & $>$41.3 &  $>$1.7  & $>$33.8  &   34.3    & 45.0       \\
3C~018   & BLO  & n & 6.7    & 55.5     &$<$15.7 & $>$41.6 &  $>$2.4  & $>$34.1  &   34.9    & 44.9  \\
3C~029   & LEG  & n & 4.9    & 53.5     &$<$1.2  & $>$40.1 &  $>$0.1  & $>$32.6  &   32.4    & 43.7  \\
3C~033   & HEG  & e & 8.3    & 56.1     &3.8     & 44.5    &  33.4    & 34.7     &   34.5    & 44.4  \\
3C~040   & LEG  & n & $<$3.6 & $<$52.7  &$<$0.2  &  --     &  --      & --       &   31.5    & 43.2  \\
3C~063   & HEG  & e & 7.4    & 56.3     &4.3     & 44.4    &  15.8    & 35.0     &   33.9    & 44.9  \\
3C~076.1 & --   & n & $<$4.8 & $<$52.9  &$<$0.8  &  --     &  --      & --       &   32.1    &  43.4 \\
3C~078   & LEG  & n & $<$4.5 & $<$53.3  &$<$0.5  &  --     &  --      & --       &   31.7    & 43.4  \\
3C~079   & HEG  & e & 8.3    & 56.0     &4.6     & 44.7    &  27.6    & 34.6     &   35.2    & 45.3  \\
3C~088   & LEG  & n & 4.6    & 53.3     &$<$0.6  & $>$40.01&  $>$0.1  & $>$32.6  &   32.4    & 43.4  \\
3C~89    & --   & n & $<$6.4 & $<$55.2  &$<$11.4 &  --     &  --      & --       &   32.8    &  44.7 \\
3C~098   & HEG  & e & 5.9    & 53.6     &0.4     & 42.9    &  2.1     & 33.3     &   33.3    & 43.8  \\
3C~105   & HEG  & e & 6.6    & 54.7     &2.1     & 42.8    &  1.9     & 33.6     &   33.8    & 44.3  \\
3C~135   & HEG  & e & 7.4    & 55.1     &4.8     & 43.1    &  3.4     & 33.7     &   34.4    & 44.5  \\
3C~180   & HEG  & e & 8.0    & 55.8     &16.3    & 43.7    &  4.2     & 33.7     &   34.6    & 44.9  \\
3C~196.1 & LEG  & n & 6.0    & 54.1     &$<$15.4 & $>$40.0 &  $>$0.2  & $>$32.7  &   33.8    & 44.9  \\
3C~198   & SF   & n & $<$5.8 & $<$53.6  &$<$5.5  &  --     &  --      & --       &   33.3    & 44.0  \\
3C~227   & BLO  & e & 7.3    & 55.5     &2.8     & 44.1    &  7.7     & 34.3     &   34.1    & 44.5  \\
3C~264   & LEG  & n & $<$4.2 & $<$52.4  &$<$0.5  &  --     &  --      & --       &   31.5    & 43.4  \\
3C~272   & LEG  & n & $<$2.5 & $<$50.1  &$<$0.0  &  --     &  --      & --       &   30.3    & 41.9  \\
3C~2871  & BLO  & n & $<$5.6 & $<$53.8  &$<$23.9 &  --     &  --      & --       &   34.0    & 44.7  \\
3C~296   & LEG  & n & $<$4.2 & $<$52.3  &$<$0.8  &  --     &  --      & --       &   32.0    & 43.2  \\
3C~300   & HEG  & e & 7.9    & 56.7     &20.2    & 45.1    &  9.1     & 34.6     &   34.3    & 45.2  \\
3C~327   & HEG  & e & 7.7    & 55.7     &3.5     & 44.0    &  12.3    & 34.3     &   34.6    & 44.7  \\
3C~348   & LEG  & n & $<$6.3 & $<$55.3  &$<$38.1 &  --     &  --      & --       &   32.7    & 45.8  \\
3C~353   & LEG  & n & $<$4.9 & $<$53.7  &$<$1.4  &  --     &  --      & --       &   32.4    & 44.4  \\
3C~403   & HEG  & e & 7.0    & 55.3     &1.3     & 43.3    &  2.1     & 33.7     &   34.1    & 44.0  \\
3C~403.1 & LEG  & n & $<$5.4 & $<$54.5  &$<$2.7  &  --     &  --      & --       &   32.2    & 43.8  \\
3C~424   & LEG  & n & $<$4.4 & $<$52.3  &$<$15.2 &  --     &  --      & --       &   33.1    & 44.5  \\
3C~442   & LEG  & n & 4.5    & 52.7     &$<$0.5  & $>$39.3 &  $>$0.0  & $>$32.0  &   31.5    & 43.3  \\
3C~445   & BLO  & e & 7.6    & 55.2     &1.5     & 44.3    &  11.6     & 34.0     &   34.8    & 44.1  \\
3C~456   & HEG  & n & 6.2    & 55.2     &$<$52.5 & $>$40.8 &  $>$0.3  & $>$33.2  &   35.1    & 44.9  \\
3C~458   & HEG  & n & $<$6.0 & $<$54.1  &$<$28.3 &  --     &  --      & --       &   34.3    & 45.2  \\
3C~459   & BLO  & e & 7.8    & 56.5     &3.0     & 44.5    & 31.0     & 35.0     &   34.3    & 45.1  \\
\hline
 \label{table:estese2}
\end{tabular}
\end{center}
  Column description: (1) source name; (2) excitation class; (3) outflow properties  derived using the nuclear spectrum extracted in 3$\times$3 pixels  (n)
  or derived from the extended maps  (e);
  (4) mass of
outflowing ionized gas; (5) kinetic energy of the outflow; (6)
extension of the outflow from the nucleus; (7 and 8) energy  and mass rates
of the outflow; (9) logarithm of the  momentum outflow rate; (10) 
logarithm of the nuclear radiative momentum; (11) power of the jet.
\end{table*}

\subsection{Data analysis}

In addition to nuclear outflows, the MUSE data allow us to search for resolved, spatially extended outflows at kiloparsec scales.  We
performed the analysis described in Section~\ref{nuclear} for each
spaxel of the image in which the [O~III] emission line is detected with an
S/N$>$5 in search for evidence of high-velocity asymmetric wings in
the line profile. By extracting the parameters describing the line
profile  in each pixel, we obtained spatially resolved maps
showing the 2D kinematic properties of the ionized
gas. Fig.~\ref{fig:velocitymap} shows an example of the 2D property
maps for the line as a whole (F$_{nuc}$, $v_{\text p}$, $W_{80}$, R)
and for the high-velocity wing ($F_{\text w}$, $v_{50, \text{wing}}$,
$E_{\rm kin}$, wing S/N) for 3C~033. The analogous diagrams for the
remaining targets can be found in Appendix A.

We considered an outflow to be spatially resolved if it extends to more
than three times the seeing of the observations. We find 13 sources that show 
extended outflows.  We measured the total mass and total energy
carried by the outflow summing the values in all the significant
pixels (i.e. with S/N$>$5).  We estimated the velocity of the outflow as the median value
of the v$_{50,\,wing}$ and the outflow extension $R_{\text{of}}$ to
estimate the mass outflow rate ($\dot{M}_{\text{of}}$) and the energy rate
($\dot{E}_{\text{of}}$); we also measured the outflow momentum boost ($\dot{P_{\text{of}}}$).
We used the continuity fluid equation for a
spherical sector (see, e.g., \citealt{Fiore17}),
 
\begin{equation}
\dot{M}_{\text{of}}=3\times v_{50, mean}\times
\frac{M_{\text{of}}}{R_{\text{of}}} 
\end{equation}

\begin{equation}
\dot{E}_{\text{of}}=\frac{1}{2}\times \dot{M}_{\text{of}}\times v_{50, mean}^2
\end{equation}

\begin{equation}
\dot{P_{\text{of}}}=\dot{M}\times v_{50,wing}
.\end{equation}

For sources without a detected extended outflow, we adopted a size limit of three times the seeing and derived the
corresponding lower limits of mass and energy rates from the nuclear
emission line analysis. 
The main properties of extended outflows that are spatially resolved are reported in Table~\ref{tab:estese}, and the properties for the sources in which a lower limit has been estimated are reported in Table~\ref{table:estese2}.

\subsection{Properties of the extended outflows} \label{properties_outflow}

We detected extended outflows in 13 sources: in 3 out of 6
BLOs and 10 out of 12 HEGs, in none of the 14 LEGs, and not in the SF galaxy either. The maximum spatial extent of these outflows ranges from $\sim$ 0.4 kpc (in 3C~098) to $\sim$ 20 kpc (in 3C~300),
with a median value of $\sim$3 kpc from the nucleus. The few
sources show no clear difference between the size of the
outflows in HEGs and BLOs. The extended outflows show a variety of
morphologies, including rather symmetric circular shapes (e.g., in
3C~227), or appear as diffuse irregular regions (e.g., in 3C~098) or filamentary emission (e.g., in 3C~079), but a bipolar structure is most commonly observed (e.g., in 3C~033, 3C~135, 3C~180, and
3C~300). Generally, HEGs show more extended structures than LEGs, consistent with the result obtained with HST \citep{Baldi19}. A detailed description of the morphology of the extended
outflows is provided in the Appendix~\ref{Appendix~A}.

In Fig.~\ref{nucext} we compare the [O~III] luminosity of the high-velocity wings integrated over the whole spatial extent of the outflow with that of
the nuclear region.  While these two estimates are similar in the three BLOs, the outflow is dominated by the extended component in the HEGs. In order to achieve a comprehensive understanding of the outflows in this sources, we must take the properties of outflowing gas derived from the line emission in these extended regions into account. For the objects without detected extended
outflows, the quantities describing the ``total''
outflow properties are based on the nuclear measurements alone (as reported in Table~\ref{table:estese2} with the flag {\it n}).

The largest differences between parameters measured in the nucleus
and in the total extent of the outflow are seen in the HEGs. This is
likely due to the nuclear obscuration that in HEGs prevents
a direct line of sight toward the nucleus of the source and at
the base of the outflow, while this is visible in the BLOs. Nuclear
obscuration can be the origin of the redshifted nuclear outflow found in the 3C~033. In this HEG we do not have a clear view
of the central nuclear outflow, but we can observe the fast-moving ionized
gas in its immediate vicinity, which shows a well-defined bipolar
morphology (see Fig.~\ref{fig:velocitymap}). The redshifted component of the bipolar outflow in the central region is most likely stronger than the blueshifted component and results in a net positive outflow velocity.

\begin{figure*}
\centering{
\includegraphics[width=0.49\textwidth]{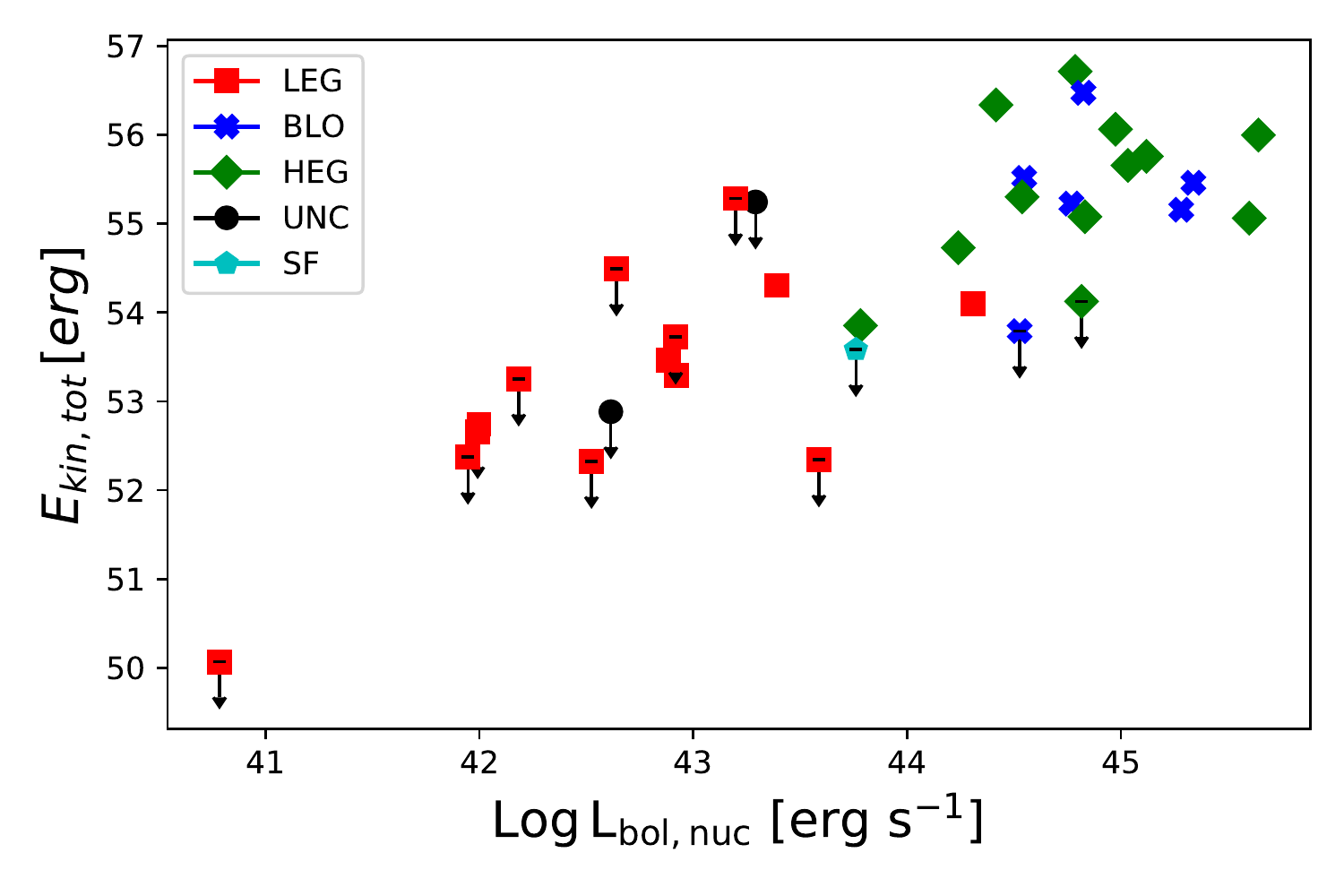}
\centering \includegraphics[width=0.49\textwidth]{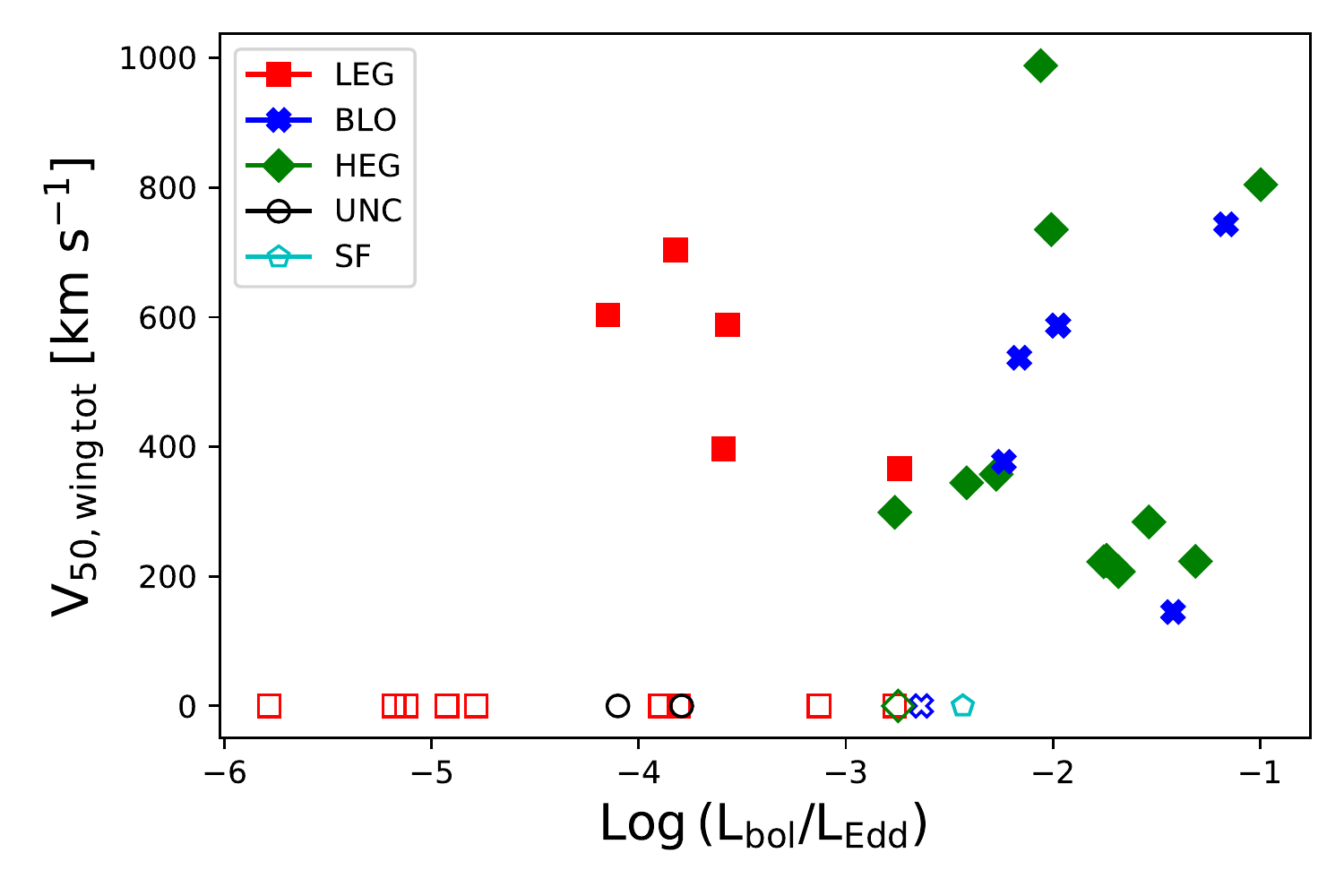}
\caption{Comparison of the outflow properties with the bolometric luminosity of the AGN (L$_{\text{bol}}$) and the Eddington ratio (L$_{\text{bol}}$/L$_{\text{edd}}$). Left: Total kinetic energy of
  the outflow vs L$_{\text{bol}}$. These quantities show a linear
dependence. 
Right: Velocity of
  the outflow vs
  L$_{\text{bol}}$/L$_{\text{edd}}$. Most sources show velocities between $\sim$ 200
  and $\sim800\, \text {km}\,\text{s}^{-1}$ with no clear trend with
  L$_{\text{bol}}$/L$_{\text{edd}}$ or differences between the various spectroscopic
  classes. The open symbols correspond to sources without detected extended outflow.}
\label{fig:ekinesteso}}
\end{figure*}
In Fig.~\ref{fig:ekinesteso} (left panel) we show the total kinetic
energy as a function of the bolometric luminosity: a clear linear
dependence is found between these quantities. An important difference emerges compared to when nuclear regions alone are observed (Fig.~\ref{ekinnuc}). In the case of the nuclei, BLOs
are found to be driving more powerful outflows, but when the
whole outflowing region is considered, HEGs show a kinetic energy that is similar to that of
BLOs. LEGs have lower bolometric luminosities than the other classes,
and the detected outflows (in five sources) show correspondingly lower
values of the kinetic energy. Most sources show outflow velocities between
$\sim$200 and $\sim$800 \kms, with no clear trend with
L$_{\text{bol}}$ or differences between the various spectroscopic
classes ( Fig.~\ref{fig:ekinesteso}, right panel). This confirms the
results from the nuclear analysis.

\begin{figure}
\centering{
\includegraphics[width=0.46\textwidth]{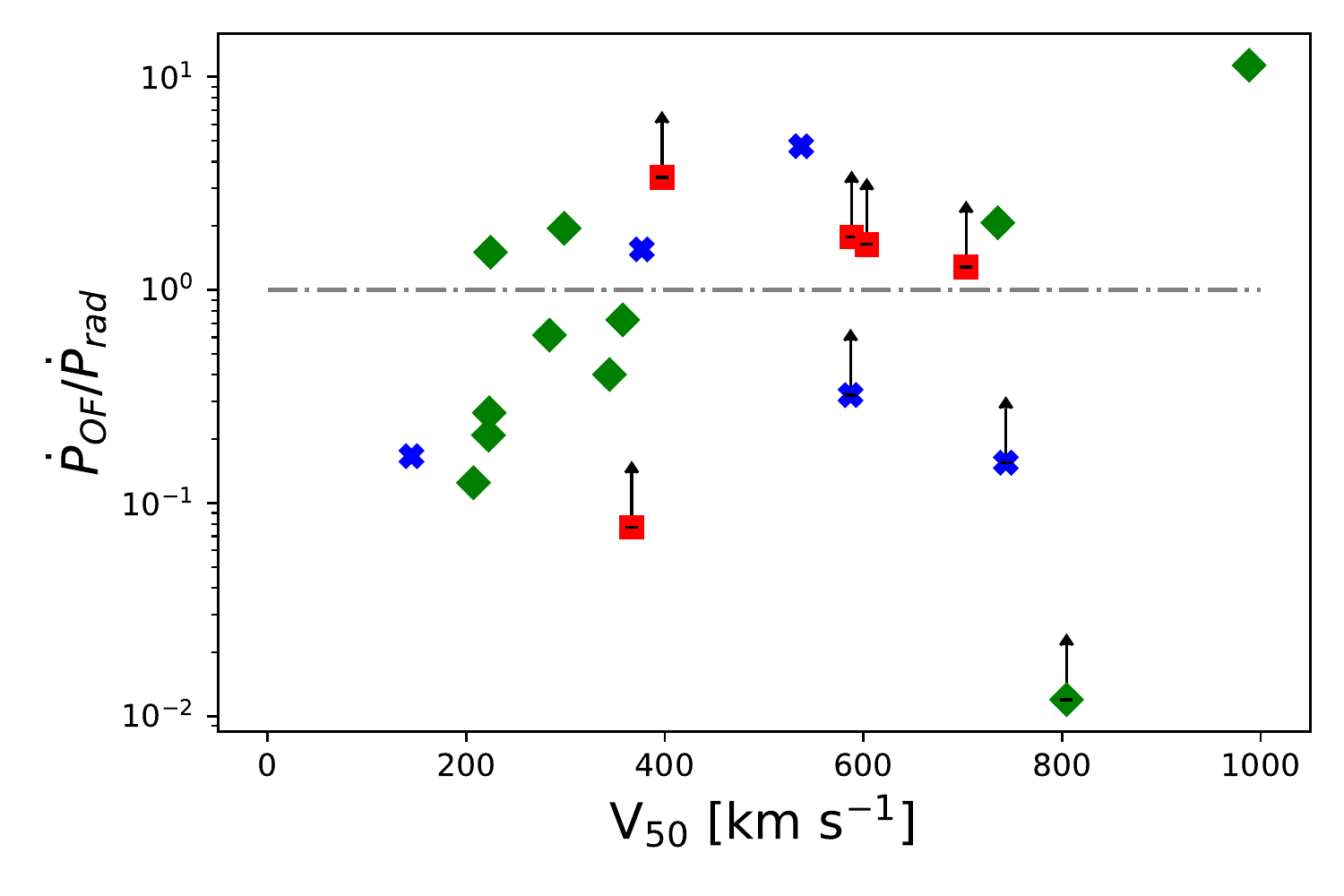}
\includegraphics[width=0.46\textwidth] {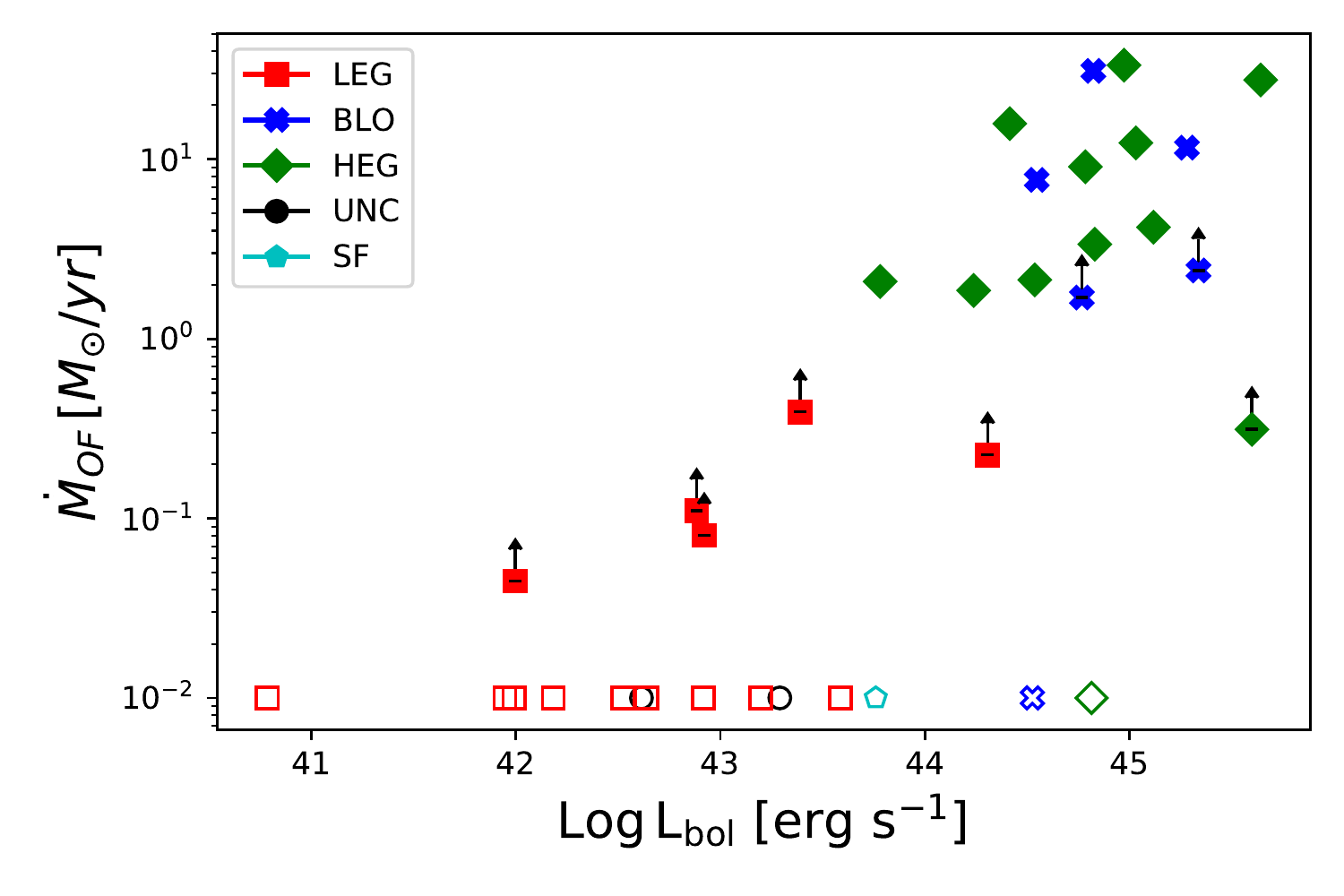}
\includegraphics[width=0.46\textwidth]{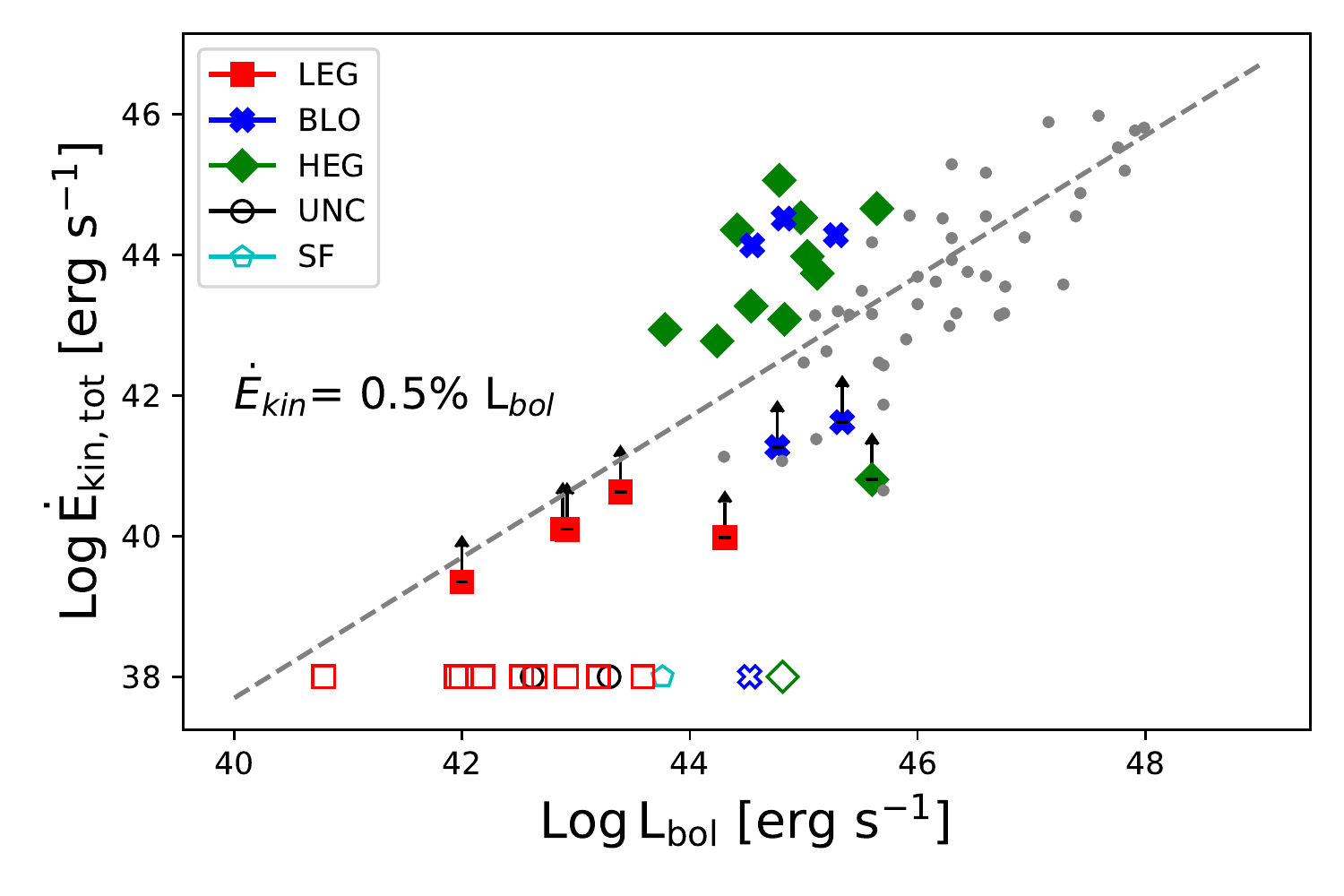}
\caption{Top panel: Outflow momentum boost divided by the AGN
  radiation momentum rate (L$_{bol}$/c) as a function of v$_{50,wings}$.  The
  dashed line marks the expectations for a momentum conserving outflow
  ($\dot{P_{OF}}/\dot{P_{rad}}=1$).  Middle panel: Wind mass
  outflow rate as a function of the AGN bolometric luminosity.  Bottom
  panel: Total outflow kinetic power vs AGN bolometric
  luminosity. 
  The grey points
  represent the data collected by \citet{Fiore17}; the dashed line
  represents the relation $\dot{E}_{\text{kin}}$= 0.5\% L$_{bol}$, above which an energy-driven outflow can substantially heat or blow the galactic gas out. The open symbols correspond to sources without detected extended outflow.}
\label{fig:dot}}
\end{figure}

The momentum flow rate $\dot{P}_{OF}$ can be compared to the radiation
pressure force ($\dot{P}_{rad}$=L$_{bol}$/c) to quantify the
efficiency of the radiative mechanism in converting the luminosity of
the AGN into the radial motion of the outflows
\citep{Zubovas12,costa14}.  In the top panel of Fig.~\ref{fig:dot}, we show
that the outflow momentum rate is between  10$^{-2}-10$  of the AGN
radiation pressure force, with no relation with the mean velocity. The
dot-dashed line marks the expectations relative to a momentum-conserving
outflow. 

Finally, in the middle and bottom panels of Fig.~\ref{fig:dot}, we compare
the total mass and energy rate of the outflows as a function of the
bolometric luminosity. The mass outflow rate of HEGs becomes similar
to that of BLOs when we consider the total outflow emission (not only
the nuclear component, as in Fig.~\ref{ekinnuc}).  The dashed line in
the bottom panel represents the threshold $\dot{E}_{\text{kin}}$=
0.005 L$_{bol}$ above which in an energy-driven (or energy-conserving)
scenario the outflow can substantially heat or blow out the galactic
gas, affecting the star formation in the host
\citep{Hopkins10,Richings17}.  We measured a median value of
$\dot{E}_{\text{kin}}$/L$_{bol}$=0.002 in the detected sources. These
low values are in line with other results in literature (e.g.,
\citealt{baron19,davies20}) and \citealt{Fiore17} (0.0016 measured at
log L$_{bol}$ = 45 erg s$^{-1}$) and are consistent with values
predicted by energy-driven models assuming realistic coupling
efficiencies (~10-20\% instead of 100\%, see,
e.g., \citealt{Costa18,Richings18}) between the accretion disk wind and
the ISM.

In Fig.~\ref{fig:dot} we also compare our results with those obtained by
\citet{Fiore17}, who collected observations of outflows in quasars from
the literature. These authors adopted the same estimate for the gas
density (200 cm$^{-3}$) as we did, but a different prescription for the outflow
velocity (they used the 20th velocity percentile of the [OIII]
emission line). Although the approach to measuring the outflow velocity
is very different, as discussed in Section 3.1, the two measurements
are closely correlated, with a median ratio of $\sim$1.6. This causes a
systematic upward offset of 0.4 dex for the Fiore et al. estimates of
$\dot{E}_{\text{kin}}$, which is much smaller than the range of measured power. We confirm that the energy rate increases with the AGN
bolometric luminosity.

\begin{figure}
\centering
\includegraphics[width=0.49\textwidth]{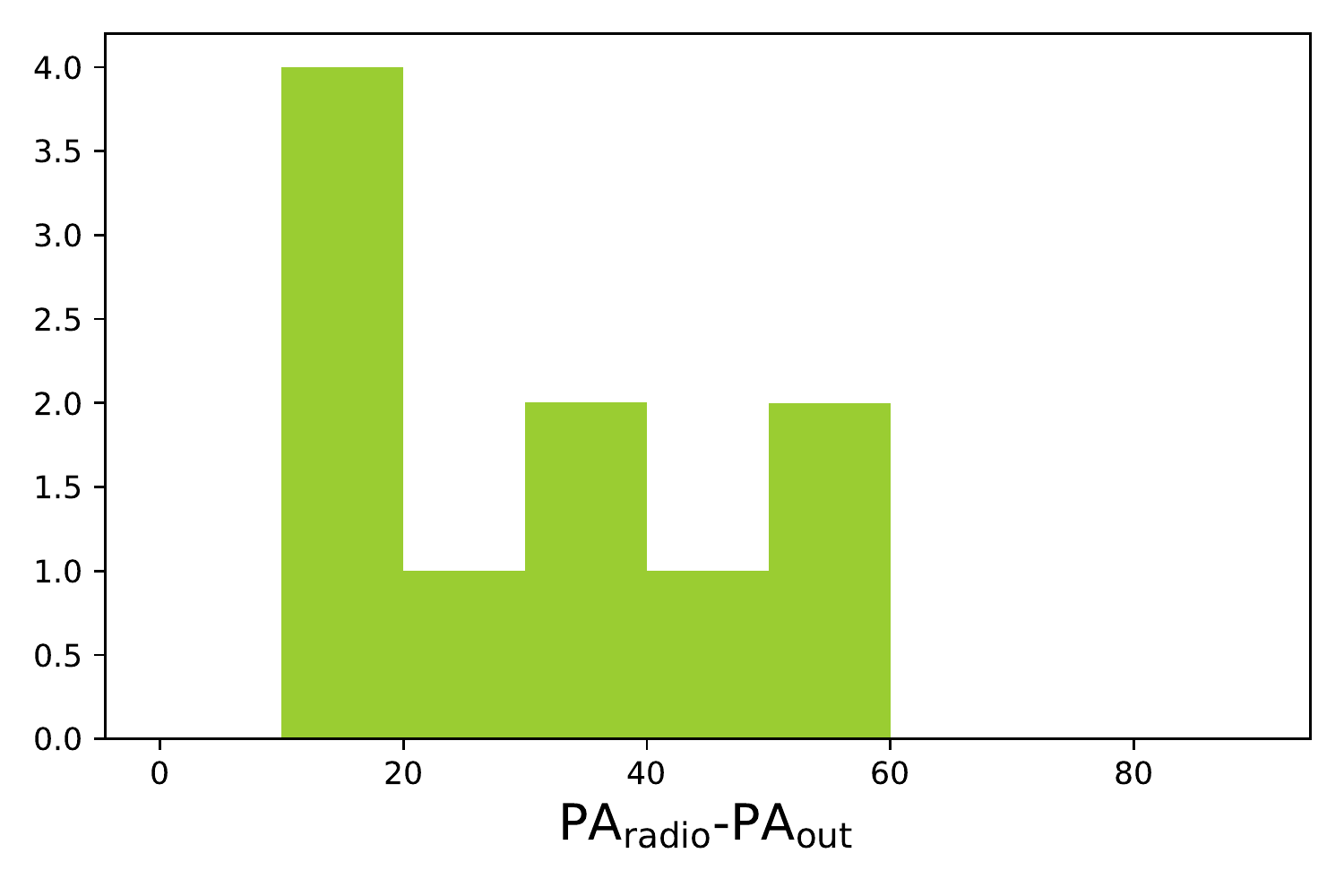}
\caption{Comparison of the position angle of the extended outflow
  and the radio structure.}
\label{fig:angle}
\end{figure}

To explore whether there is a connection between the axis of the radio
structure and of the extended outflow, we compared their
orientation. An alignment between the two axes is expected considering
the geometry of the central regions in which the radio jet emerges
perpendicular to the plane of the accretion disk and aligned with
the axis of the nuclear radiation field, consistent with what  has been observed in emission-line extended morphologies of 3C radio galaxies with the HST \citep{Baldi19}. Because several of the extended sources have circular structures with no well-defined position angle (PA), this can be successfully measured in only
ten sources, see Tab.~\ref{tab:sample}. For these sources, we show in
Fig.~\ref{fig:angle} the comparison between the jet and outflow PA.
The differences cover a wide range, between 0 and 60 degrees, with an only
marginal preference for small offsets. This result suggests that the
morphology of the outflow is mainly determined by an inhomogeneous
distribution of gas in the central regions of the host galaxy and not by nuclear illumination or the jet direction. Another possibility is that jet direction and AGN illumination are not aligned. This can occur if the accretion disk (to which the jet is perpendicular) and the obscuring torus (which determines the AGN illumination cone; \citealt{May20,Goosmann11,Fischer13}) are not aligned.

 \section{Discussion} \label{discussion}
 
  \subsection{What is the driver?}
In our sample of 37 nearby radio galaxies, we have evidence of
nuclear and/or extended outflows in 21 sources (57\% of the
sample). In radio-quiet AGN, the outflows are thought to be
powered by radiation, but in radio-loud AGN the radio jet can play an important
role in accelerating the wind up to kiloparsec scales \citep{Fabian12}.  To investigate the driver of the outflows, we compared the energy (and momentum)
from radiation pressure and from the radio jet to the kinetic energy (and
momentum) of the outflows. We have shown that a fraction of the 
0.5 - 10\%  of the bolometric luminosity is sufficient to power the observed ionized gas outflows
with radiative pressure. We can now estimate the coupling efficiency
required in a scenario in which the outflow is instead powered by the
radio jet.
 
Several studies have calibrated empirical relations of the
radio luminosity and the jet power, that is, the energy per unit time
carried from the vicinity of the supermassive black hole to the
large-scale radio source. We adopted the relation described by
\citet{daly12}\footnote{ This relation has 
been calibrated considering a sample of FRII radio galaxies, but it agrees with the relations from \citealt{Willott99,cavagnolo10}, who 
considered sample of FRI radio galaxies}: log P$_{\rm jet} =0.84 \times $log P$_{178 {\rm MHz}}$
+ 2.15, with both quantities in units of 10$^{44}$ erg s$^{-1}$.

\begin{figure}
\centering
\includegraphics[width=0.49\textwidth]{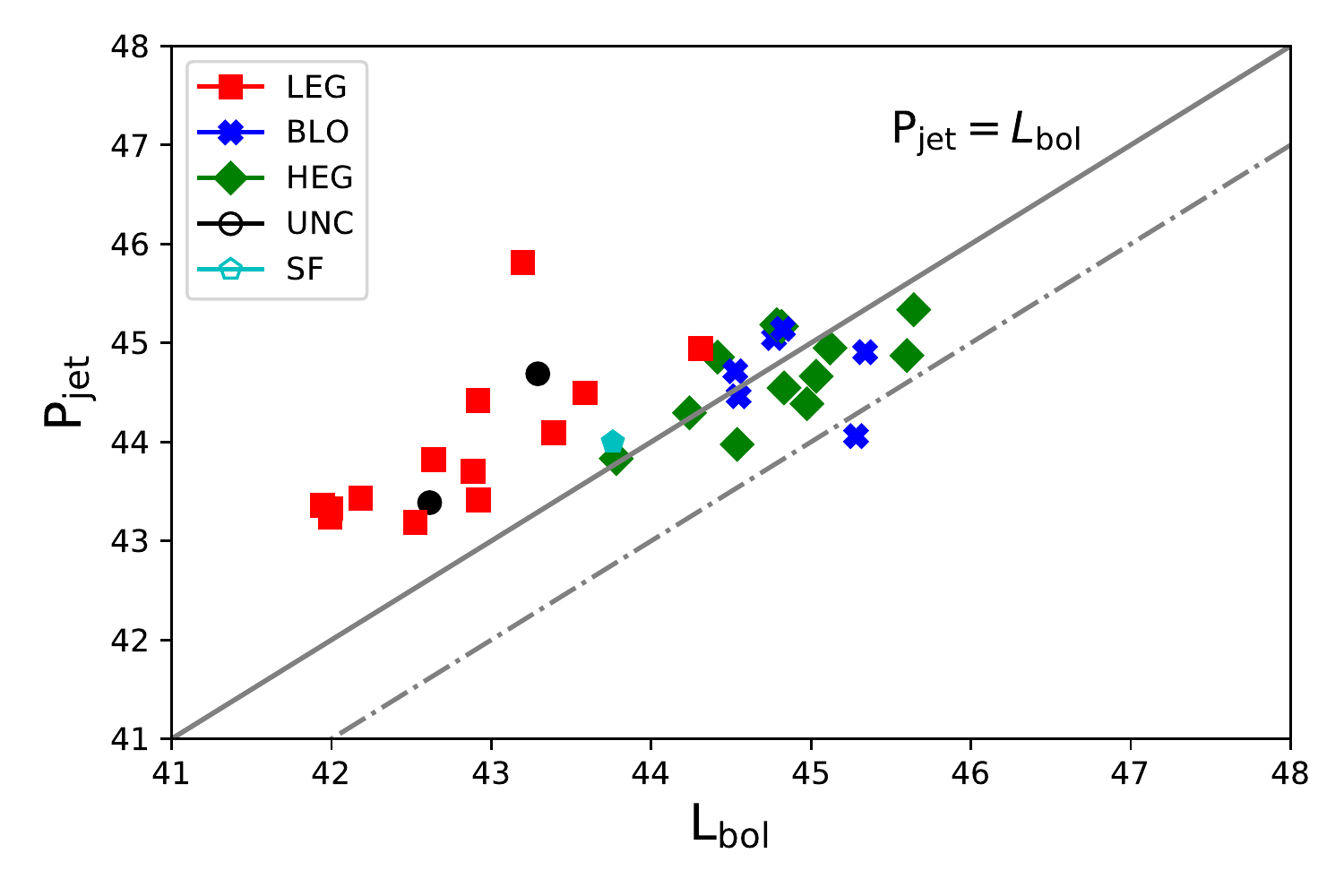}
\includegraphics[width=0.49\textwidth]{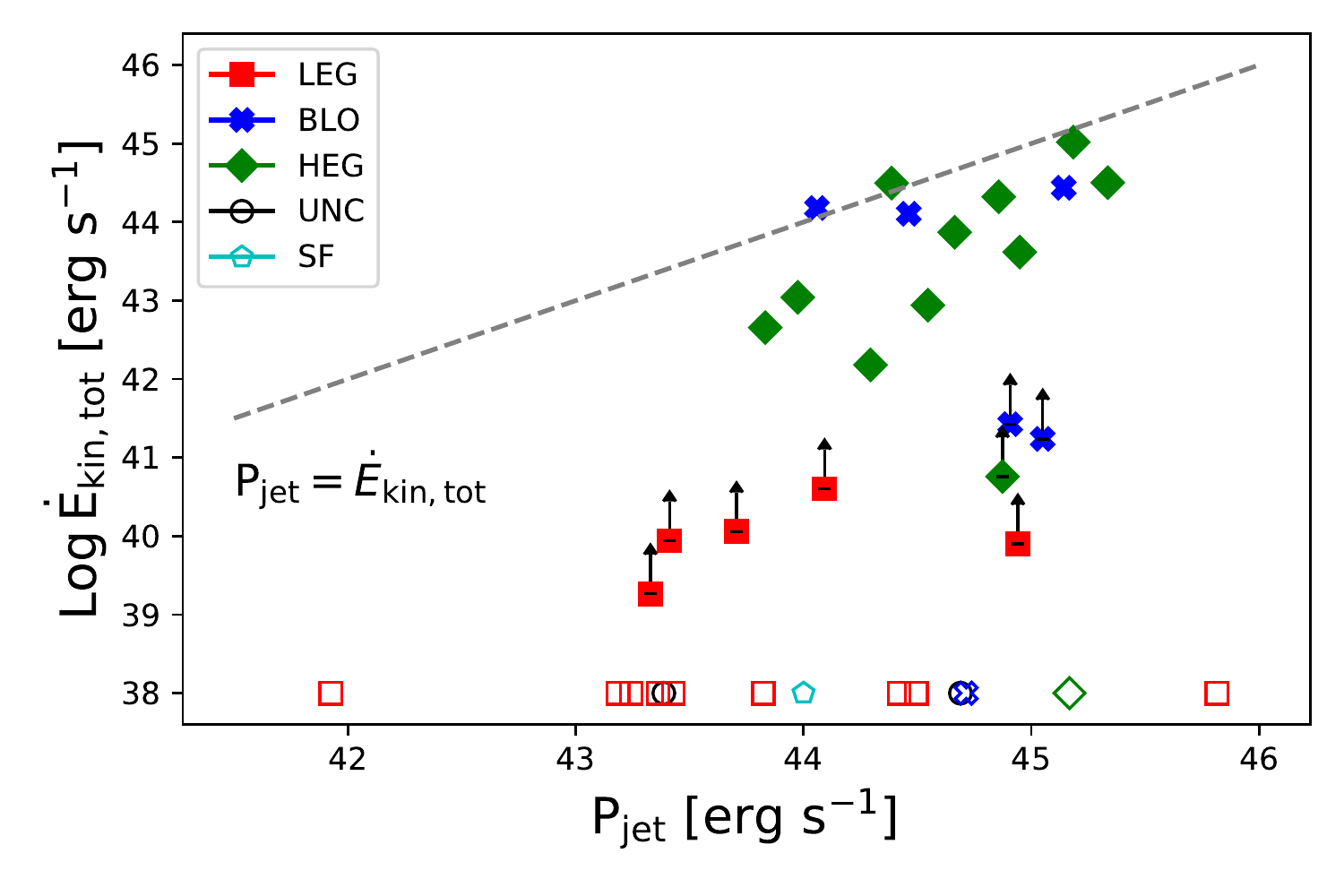}
\caption{Top panel: Jet power, i.e., the energy per unit time carried
  from the radio jet vs the nuclear bolometric luminosity.  Bottom
  panel: $\dot{E}_{\text{kin}}$ vs P$_{jet}$.   The dashed line represents the 1:1 relation. The jet 
  should have an efficiency $\eta \approx$ 0.1 - 1 to sustain the kinetic
  power of the outflow.}
\label{fig:pjet}
\end{figure}

In Fig.\ref{fig:pjet} (top panel) we compare the jet power with the
bolometric luminosity estimated from the [O~III] emission line. In
BLOs and HEGs  the bolometric luminosity is slightly higher than the jet power
 by a median ratio factor of 2, while 
in LEGs, it is lower (between 0.01 and 0.1, median ratio $\sim$0.07). Because we obtain
values of P$_{jet}$ in the same range of the $\dot{E}_{\text{kin}}$
(10$^{42}$--10$^{45}$ erg s$^{-1}$, see the bottom
panel of Fig.~\ref{fig:pjet}), an unrealistic mechanical efficiency  of $\eta \sim$ 1
would be required for the radio jet to be able to accelerate the
outflows in HEGs and BLOs. We cannot exclude a role of the jet in
accelerating the outflow in LEGs, but this energetic argument rules out the radio jet as the main
driver of the most powerful and extended outflows we observed in HEGs
and BLOs. This is consistent with the fact that HEGs, LEGs, and BLOs
share the same range of radio powers, but we observe more powerful and
extended outflows in the most luminous sources (HEGs and BLOs) and not
in the less luminous LEGs.  Therefore the radiation released by the
accretion process onto the SMBH seems to be the most viable mechanism to
ionize and accelerate the gas to the observed velocities.

\subsection{Energy or momentum driven?}
 
In a momentum-driven outflow, the relativistic winds originating from
the accretion disk shocks the ambient medium and cools
efficiently \citep{Ciotti97,King03}, transferring only its momentum
flux to the ISM. Conversely, in the energy-driven or the
energy-conserving outflow, the cooling of the shocked outflow from the
SMBH is inefficient and generates a hot-wind bubble that expands from the
center of the galaxy, transferring most of its kinetic
luminosity. Hydrodynamical simulations predict that energy-driven
outflows can reach momentum fluxes exceeding 10 L$_{\rm Edd}$/c within
the innermost 10 kpc of the galaxy \citep{costa14} and are the most
massive and extended outflows.

We estimated the amount of momentum and energy transferred from
the active nucleus to the ISM. These quantities assess the relative
importance of AGN-driven winds in the context of galaxy evolution.
One of main source of uncertainty in the measurement of the outflow
energetics is the density of the outflowing gas because the ionized
gas masses, and hence the kinetic energies and mass outflow rates,
scale linearly with the electron density (n$_e$).  Following \citet{Fiore17}, we adopted $\langle n_e\rangle$ =
200 cm$^{-3}$ for all objects in the sample. Our estimates might need
to be rescaled for different values of the density, but the
relative comparison between different sources is robust because
we do not expect large variation in the gas density of the outflows
from source to source.

We obtained mass outflow rates in the range of $0.4-30$ M$_\odot$
yr$^{-1}$ and kinetic energies of about 10$^{42}$--10$^{45}$
erg $^{-1}$ and $\dot P_{OF}/\dot P{_{rad}}$ between $\sim$0.01 and 10,
with observed v$_{20}$ velocities in the range 400--1500 km s$^{-1}$.
These values are similar to what is observed in other types of AGNs
within the rather large uncertainties.  For example,
\citet{harrison14} analyzed IFU observations of 16 type 2 AGN below
redshift 0.2 with a similar bolometric luminosity range
($\sim$10$^{44}$--10$^{46}$ erg s$^{-1}$), and found high velocities
in the range 510-1100 km s$^{-1}$ in all their targets. In 70\% of
the sample, the outflow is resolved and extended on kiloparsec scales. They
estimated an energy outflow rate of about
(0.15--45)$\times$10$^{44}$ erg s$^{-1}$, which matches our results.
\citet{Fiore17} found that about half of the ionized winds have
$\dot{P_{OF}}/\dot{P_{rad}}\lesssim$ 1, and the other half has
$\gtrsim$ 1 and a few $>$10, suggesting that these latter ionized
winds may also be energy conserving.  Energy-conserving winds are
characterized by values of $\dot{P_{OF}}/\dot{P_{rad}}$ of 15-20
(e.g., \citealt{Faucher12,Zubovas12}).  Conversely, values of
$\dot{P}_{OF}/\dot{P}_{rad} \lesssim$ 10 suggest that these ionized
winds are momentum conserving, as predicted by the \citet{King03}
model.  The low outflow momentum boost we obtained (see
Fig.~\ref{fig:dot}, top panel) suggests a momentum-conserving
(radiative) regime, as observed in BAL winds and fast X-ray winds.
However, we recall that the contribution of molecular
and neutral atomic phase might be significant and increase the momentum
boost: the ionized gas is usually the less important part in
terms of mass and energies when multiphase outflow studies are
available (see,
e.g., \citealt{Carniani17,Fiore17,Fluetsch19}).

\subsection{Connection between outflow and nuclear properties}

The existence of a link between the bolometric luminosity of
the AGN and the velocity of the outflow is controversial. There are
claims for a positive correlation (e.g., \citealt{Mullaney13}),
but other authors using different samples find, in agreement with our
results for the 3C radio galaxies, no clear trend between
these quantities \citep{harrison14}. 
Using a different sample of objects, \citet{bischetti17} reported a correlation between the maximum outflow velocity and AGN luminosity that is absent when the individual samples are considered separately.
 We note that this trend is expected for energy-conserving winds (L$_{bol}\propto v_{max}^5$, \citealt{costa14}) and may explain why we do not see this relation in our sample, in which the data suggest that the outflows are produced by momentum-conserving winds.

Instead, we find a positive trend between the kinetic energy of the
outflow and the bolometric luminosity. Standard measurements of
the outflow velocity based on percentile velocities of the whole line,
that is, v$_{20}$, might be subject to a bias because stronger
emission lines have more luminous wings, inducing the observed
correlation. However, our approach for measuring the outflow velocity accounts for this bias, and we argue that this positive trend is a robust result.
  
\subsection{Differences between LEGs, HEGs and BLOs}

The clearest differences between different spectroscopic types is
that all LEGs appear to be compact (we estimated an average upper limit of 1
kpc) and only HEGs and BLOs show extended outflows. The average radius
is 4 kpc for HEGs and 3 kpc for BLOs.
Moreover, LEGs accelerate a larger fraction of the gas: about 10\% of the total ionized gas is part
of the outflow, compared to $\sim$1\% in HEGs and BLOs.  The
outflow velocities are comparable, but in LEGs, we detect lower mass rate and
kinetic energy in the outflows.

The observed differences of the outflow in HEG and LEGs might be due
to different environments (e.g., \citealt{Macconi20}).  LEGs are more frequently found in
clusters than BLOs or HEGs \citep{Ramos13}, and their
supermassive black holes are thought to accrete mainly hot gas.  Many
radio galaxies of LEG type are able to displace the hot, X-ray
emitting gas with radio jets and lobes.  
HEGs are thought to have larger amounts of cold gas, supported by the presence of a dusty torus and efficient accretion.
  In theoretical models, it is relatively easy to displace the hot,
diffuse interstellar medium, but it is more difficult to hydrodynamically accelerate cold clouds.  Recent simulations have shown that a
combination of instabilities and simple pressure gradients will drive
the cloud material to effectively expand in the direction
perpendicular to the incident outflow. In this case, the outflow can
effectively expand in the direction perpendicular to the incident
outflow and propagate out to large radii.  We speculate that the large extent of the ionized outflows present in HEGs and BLOs might be related to the large amount of cold gas with respect to LEGs
 and to a more powerful ionizing nucleus.

%

\subsection{Effect of outflows} \label{impact}

To evaluate whether the ionized outflowing gas can permanently escape
the galaxy, we estimate the escape velocity v$_{esc}$ from the host.
We used the prescription by \citet{Desai04}, which links the circular
velocities v$_c$ to the line-of-sight stellar velocity dispersion
$\sigma$ (v$_c \approx 1.54 \sigma$) and the relation by
\citet{Rupke02}, who adopted a simple gravitational model based on an
isothermal sphere, which predicts v$_{esc}(r)=\sqrt2
v_c[1+ln(r_{max}/r)]^{(1/2)}$.  For a galaxy with a velocity
dispersion of 300 km s$^{-1}$ at a radius of r=10 kpc and assuming
r$_{max}$=100 kpc, we estimate a typical escape velocity of $\sim$1170
km s$^{-1}$. We measured outflow velocities well below this threshold
and outflowing gas masses of about a few solar masses per year.

We infer that the outflowing gas cannot leave the
gravitational potential well of the host galaxy, even if the radio jet facilitates pushing this gas out  \citep{Fromm18, Best05}. This
result, combined with the relatively low values of conversion between
bolometric luminosity and outflow energy rate,
$\dot{E}_{\text{kin}}$/L$_{bol}$=0.2\%, suggests that the observed
outflows are not able to significantly affect the gas content or the
star formation in the host. This could be consistent with an auto-regulating  feedback-feeding mechanism proposed by \citet{Gaspari20}, for example.

 \section{Summary and conclusions} \label{conclusion}

We have observed 37 3C radio galaxies at z$<$0.3 with the
VLT/MUSE integral field spectrograph, obtained in the framework of the MURALES
project. 
 The main aim of
this paper was to explore  ionized outflows and the
relation with the radio jet. This sample includes all
spectroscopic subtypes of radio galaxies, LEGs, HEGs, and BLOs.

We performed a nonparametric analysis of the [O~III] emission line,
exploring the asymmetries in the wings of the line profile. We measured
v$_{50,wing}$, the mean velocity of the wings residuals. We consider this definition more representative of the outflow velocity than the percentile velocities that are commonly used in literature, which are typically
affected by the entire flux emission, that is, are dominated by
gravitational motions. However, for a comparison with other studies in
literature, we also measured v$_{20}$ and found a good correlation
with v$_{50,wing}$, with v$_{20} $ larger than v$_{50,wing}$ by a
median factor 1.6.

We found evidence of nuclear outflows in 21 sources ( 5 out of 6 BLOs, 9 out of 12
HEGs, and 5 out of 14 LEGs) with blueshifted v$_{20} $velocities between
$\sim$ 1200 and $\sim$ 400 $\text {km}\,\text{s}^{-1}$ and
v$_{50,wing} $ velocities between $\sim$ 300 and $\sim$ 900 $\text
{km}\,\text{s}^{-1}$ (still blueshifted). We observed only one redshifted wing in 3C 033,
likely because of obscuration in the nuclear region.  We measured the
luminosity of the wing and the kinetic energy of the outflow.

The same analysis, performed in all  spaxels, revealed
extended outflows in 13 sources with sizes between 0.4 and 20
kpc. Comparing the luminosity and kinetic energy of these extended outflows with
those obtained from the nuclear analysis, we found large
differences between parameters measured on the nucleus and on the
total extent of the outflow in HEGs, likely due to a combination of
prominent extended outflows and nuclear obscuration. To provide a more comprehensive view of the outflows in these
sources, we therefore included measurements of these extended regions. For the objects without detected extended outflows, the
quantities describing the total outflow properties are based on the
nuclear measurements alone. We measured the mass and kinetic energy
outflow rates assuming an outflow radius corresponding to the maximum
observed extension of the outflows (for the compact sources, we assumed
a radius equal to three times the seeing of the observations, and we
provided lower limits) and as velocity the mean value of v$_{50,wing}$ in the outflow region.
  
We detected nuclear and/or extended outflows in 21 out of 37 radio
sources we analyzed. We observed different outflow morphologies, but they were
mainly circular or bipolar. In the extended outflows where we are able
to estimate the orientation, this is not closely correlated with the
radio jet direction. This is likely due to an inhomogeneous
distribution of the gas in the host galaxy possibly produced by the jet
expansion that drives away gas along its path. We found mass outflow
rates and kinetic energy rates in line with the values obtained
by studying other samples of quasars ( M$_{of}\sim$ 0.4--30 M$_\odot$
yr$^{-1}$, ${\dot E}_{kin} \sim 10^{42}-10^{45}$ erg s$^{-1}$), but with
a low momentum loading factor $\dot{P_{OF}}/\dot{P_{rad}}<$ 10. This
result suggests that our outflows probably are momentum driven and not
energy driven, as observed in many X-ray fast winds and BALs and also in many galactic-scale ionized outflows; see, for example, the right panel of Fig. 2 in \citet{Fiore17}.

Comparing the nuclear bolometric luminosity and the jet power, we
speculate that these winds are likely accelerated by the radiation
released by the accretion process and not by the radio jet, at least
in HEGs and BLOs.  In these sources, the radio power is even a factor 10 lower than the bolometric luminosity, and a mechanical efficiency of about 1 would be required by the radio jet to accelerate the
outflows.  In LEGs, the bolometric luminosity is higher than the
jet power, and we cannot exclude a strong contribution of the jet
to the acceleration process.

 We conclude that nuclear outflows are common in radio galaxies, and
 that they are extended in almost all the HEGs and BLOs, in
 which the nuclear radiation powers the outflow.  The velocities, gas
 mass, and kinetic energy rate are probably too low to have a
 significant effect on the host galaxy. However, the ionized gas is
 just one of the outflow phases, which also include atomic and molecular
 gas. In order to obtain a full characterization of the properties
 and effect of the AGN outflows in the population of radio galaxies, multiphase observations of our sample are required.

\begin{acknowledgements}
We acknowledge the anonymous referee for her/his report. Based on observations collected at the European Organisation for Astronomical Research in the Southern Hemisphere under ESO programmes 0102.B-0048(A) 099.B-0137(A).
	This research has made use of the NASA/IPAC Extragalactic Database (NED), which is operated by the Jet Propulsion Laboratory, California Institute of Technology, under contract with the National Aeronautics and Space Administration.
This work is supported by the ``Departments of Excellence 2018 - 2022’’ Grant awarded by the Italian Ministry of Education, University and Research (MIUR) (L. 232/2016).
This investigation is supported by the National Aeronautics and Space Administration (NASA) grants GO9-20083X and GO0-21110X.
GS acknowledges financial support  from
the European Union's Horizon
2020 research and innovation programme under Marie Sk\l odowska-Curie
grant agreement No 860744 (BID4BEST), from the State
Research Agency (AEI-MCINN) of the Spanish MCIU under grant
"Feeding and feedback in active galaxies" with reference
PID2019-106027GB-C42, and from IAC project
P/301404, financed by the Ministry of Science and Innovation, through
the State Budget and by the Canary Islands Department of Economy,
Knowledge and Employment, through the Regional Budget of the Autonomous
Community.
GV acknowledges support from ANID program FONDECYT Postdoctorado 3200802.
SB and CO acknowledge support from the Natural Sciences and Engineering Research Council (NSERC) of Canada.
BAT was supported by the Harvard Future Faculty Leaders Postdoctoral Fellowship.
\end{acknowledgements}



\begin{appendix}
\section{Nuclear spectra and spatially resolved maps of kinematical parameters.}
\label{Appendix~A}
In Fig.~A.1 we present for all objects analogs of
Fig. \ref{fig:ribalto}, that is, the nuclear spectrum extracted from the
central 0\farcs6 $\times$ 0\farcs6, showing the decomposition of the
[O~III] profile into rotating and outflowing gas.  In six sources
(3C\,089, 3C\,318.1, 3C\,386, and 3C\,403.1), the [O~III]
emission line is not detected, and they were not considered for this
analysis.

Finally, in Fig. A.2 through A.26 we show the maps of the eight
parameters defined in Section~\ref{resolved}, that is, the velocity peak
($v_{\text p}$) of the [O~III] emissio -line, the emission line width
($W_{80}$), the line asymmetry, the S/N peak of the whole [O~III]
line, the flux of the outflowing gas isolated by the rotational
component (F$_{\text w}$), the median velocity of the outflowing gas
(v$_{50,\text{wing}}$), the kinetic energy of the [O~III] emitting
gas, and the S/N of the [O~III] emission residing in the line
wing. The last four parameters are estimated only in pixels with an
S/N$>5$. In the panel showing the line wing flux, we draw a circle with
a diameter of three times the seeing of the observations: the sources
in which the wing is observed at larger distances are considered as
those associated with an extended outflow. In the panel showing the wing
velocity, the dashed line marks the radio position angle.

\begin{figure*}
  \label{fignuc}
  \centering
\includegraphics[width=0.3\textwidth]{fignuc/3C015nuc.pdf}
\includegraphics[width=0.3\textwidth]{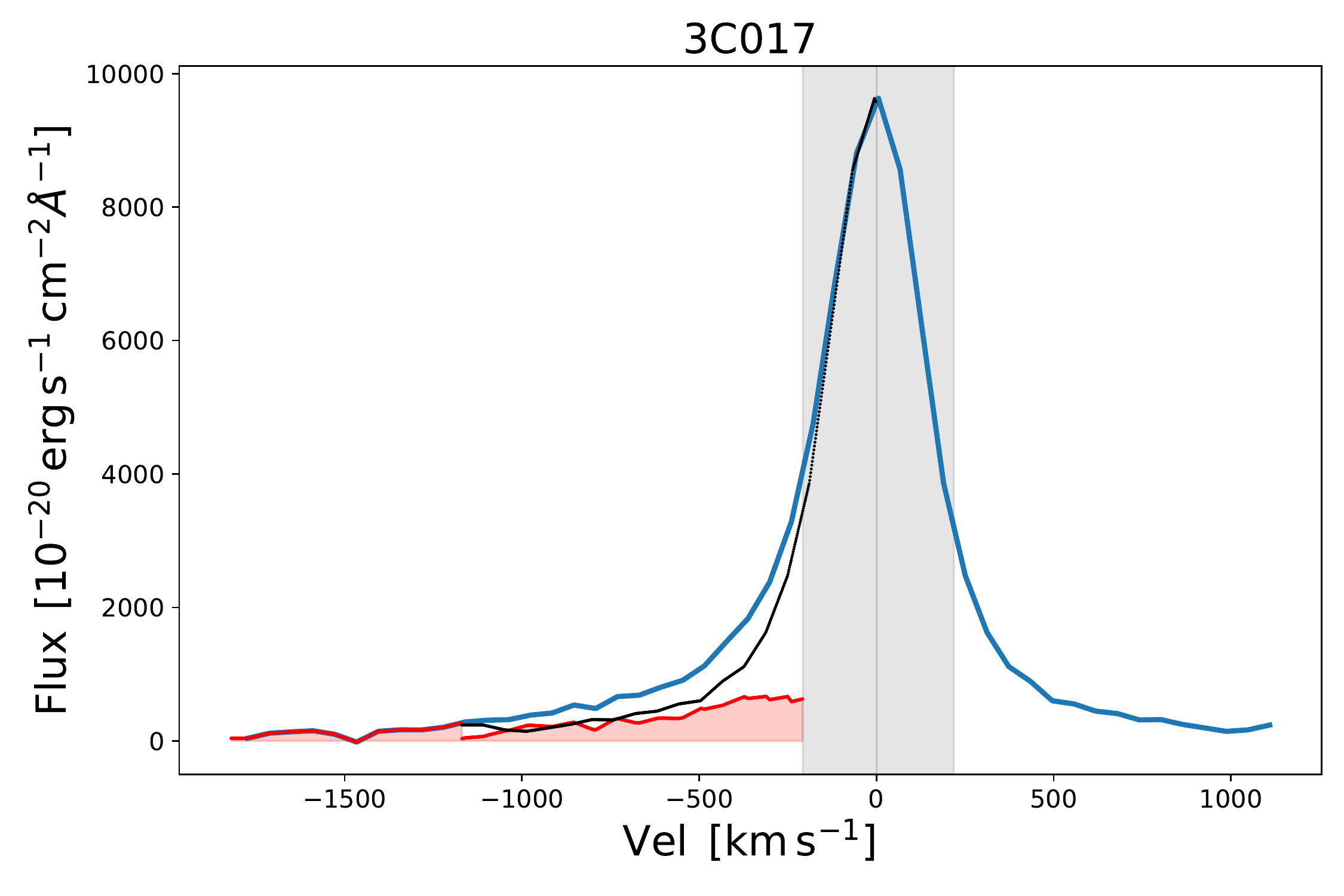}
\includegraphics[width=0.3\textwidth]{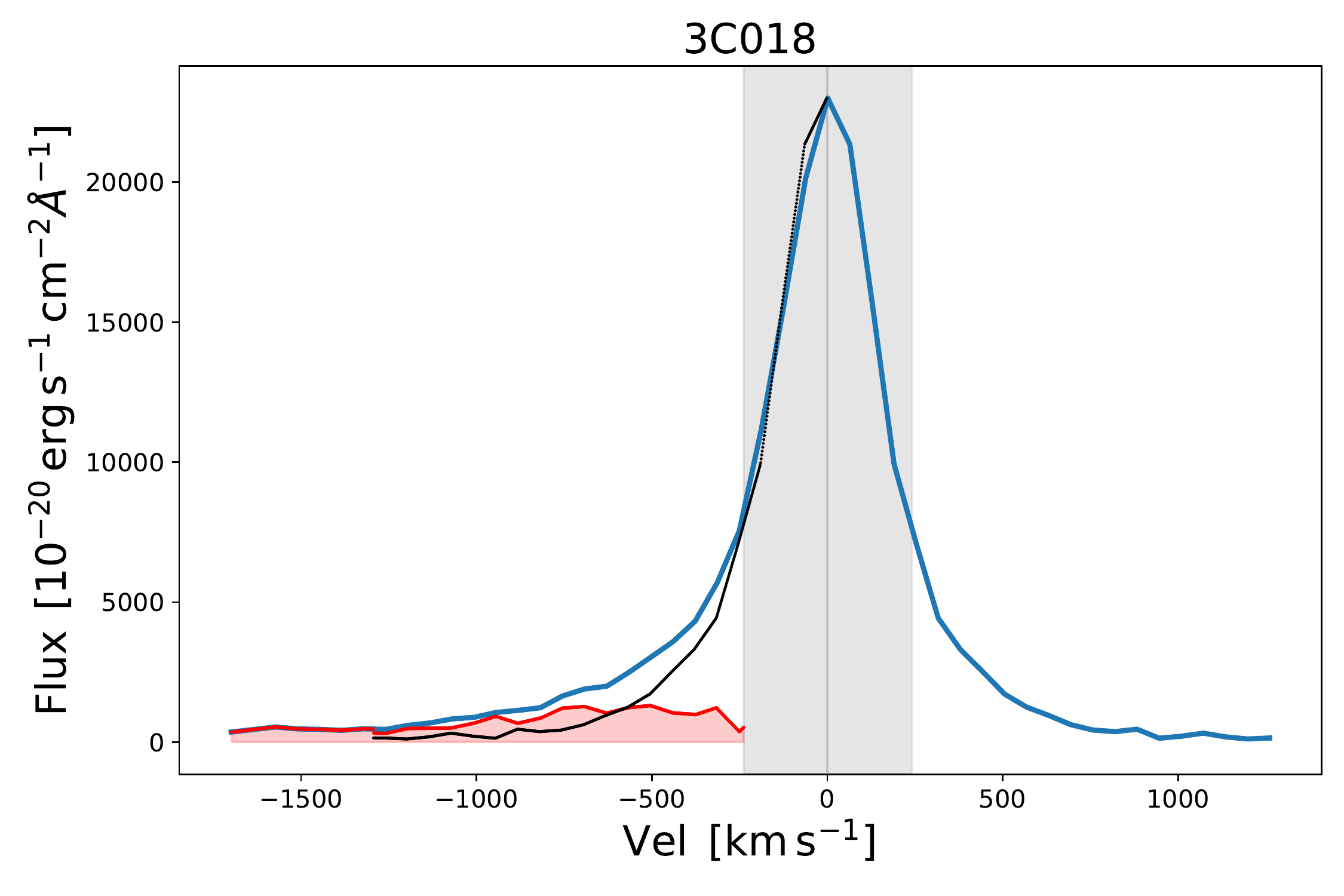}
\includegraphics[width=0.3\textwidth]{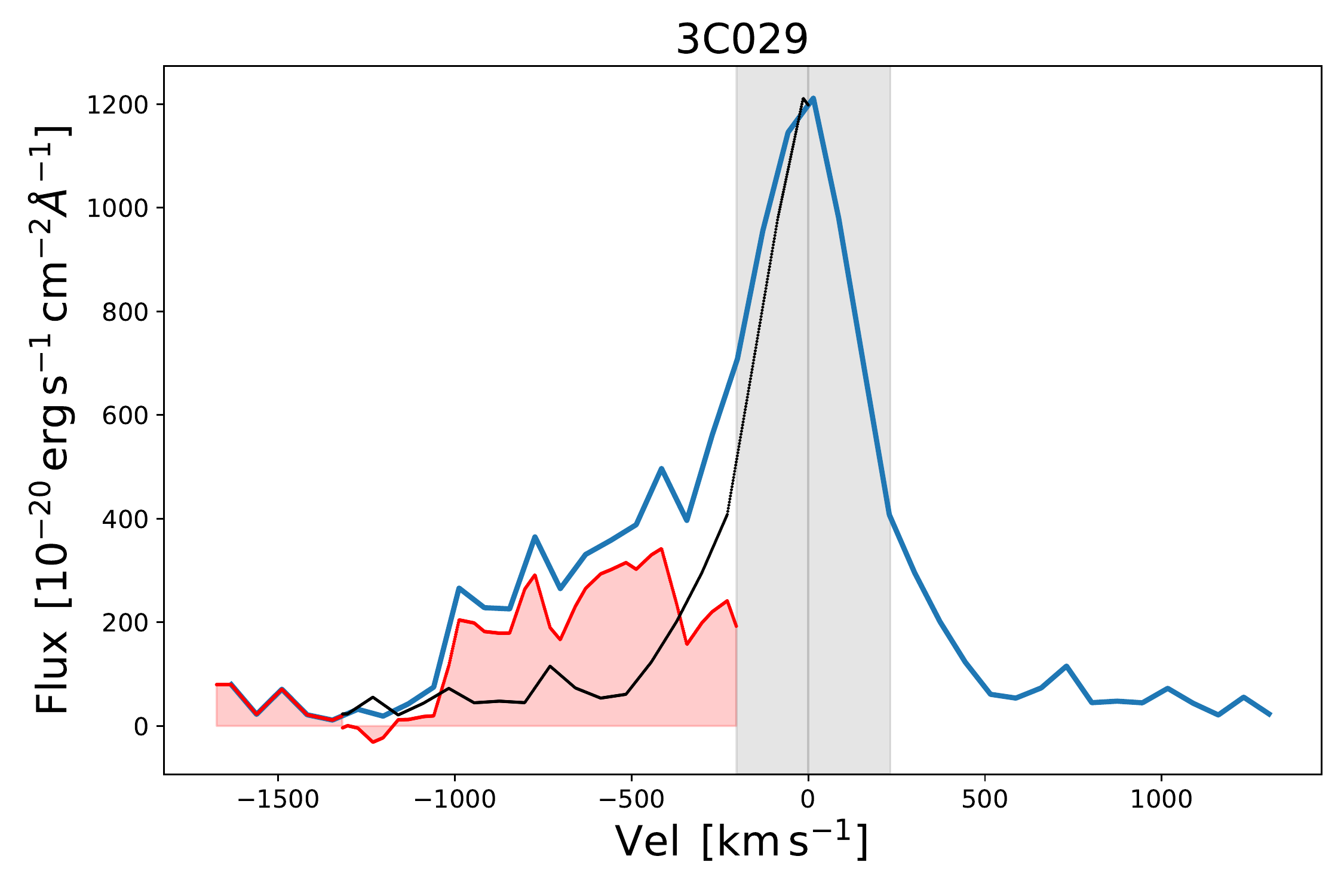}
\includegraphics[width=0.3\textwidth]{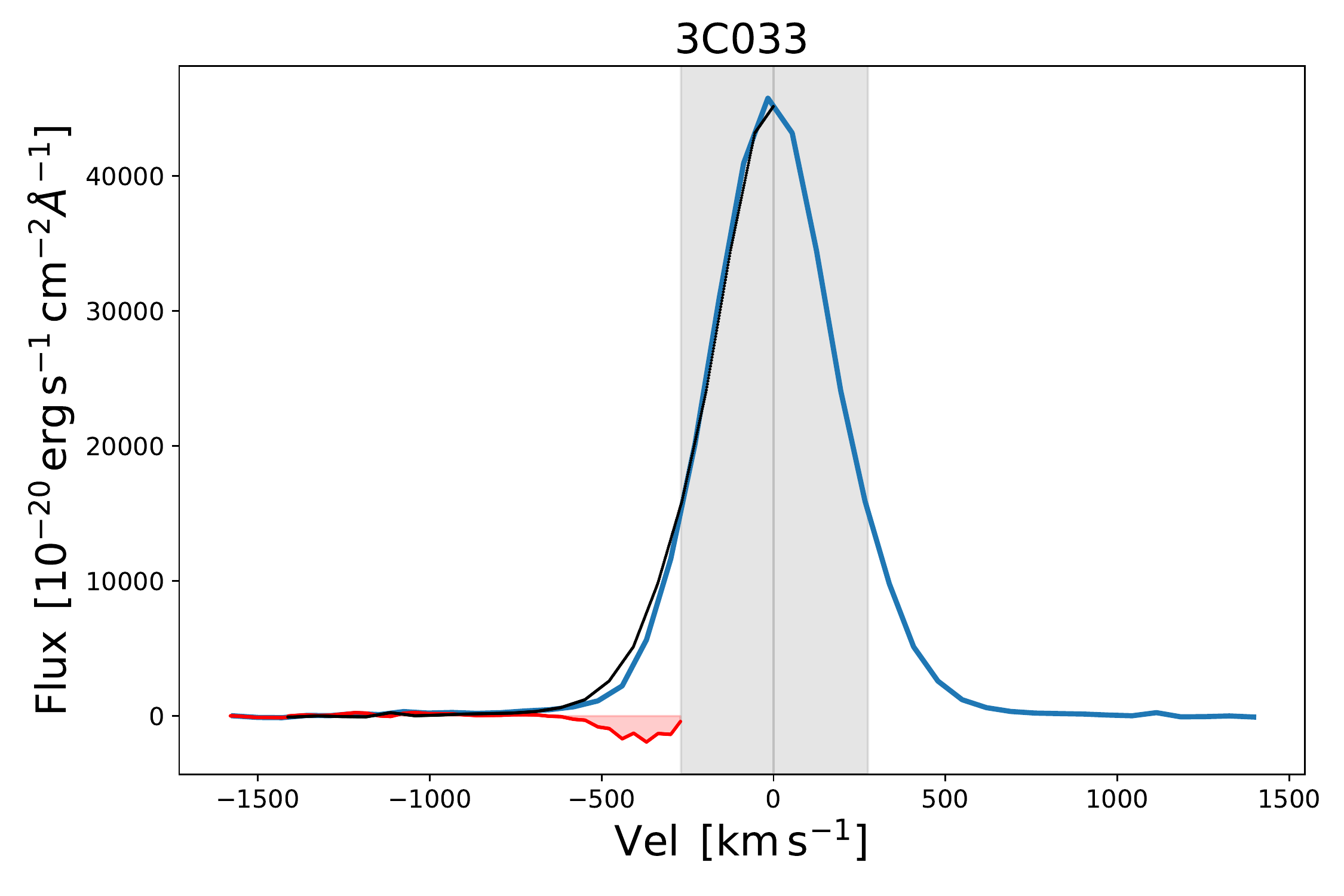}
\includegraphics[width=0.3\textwidth]{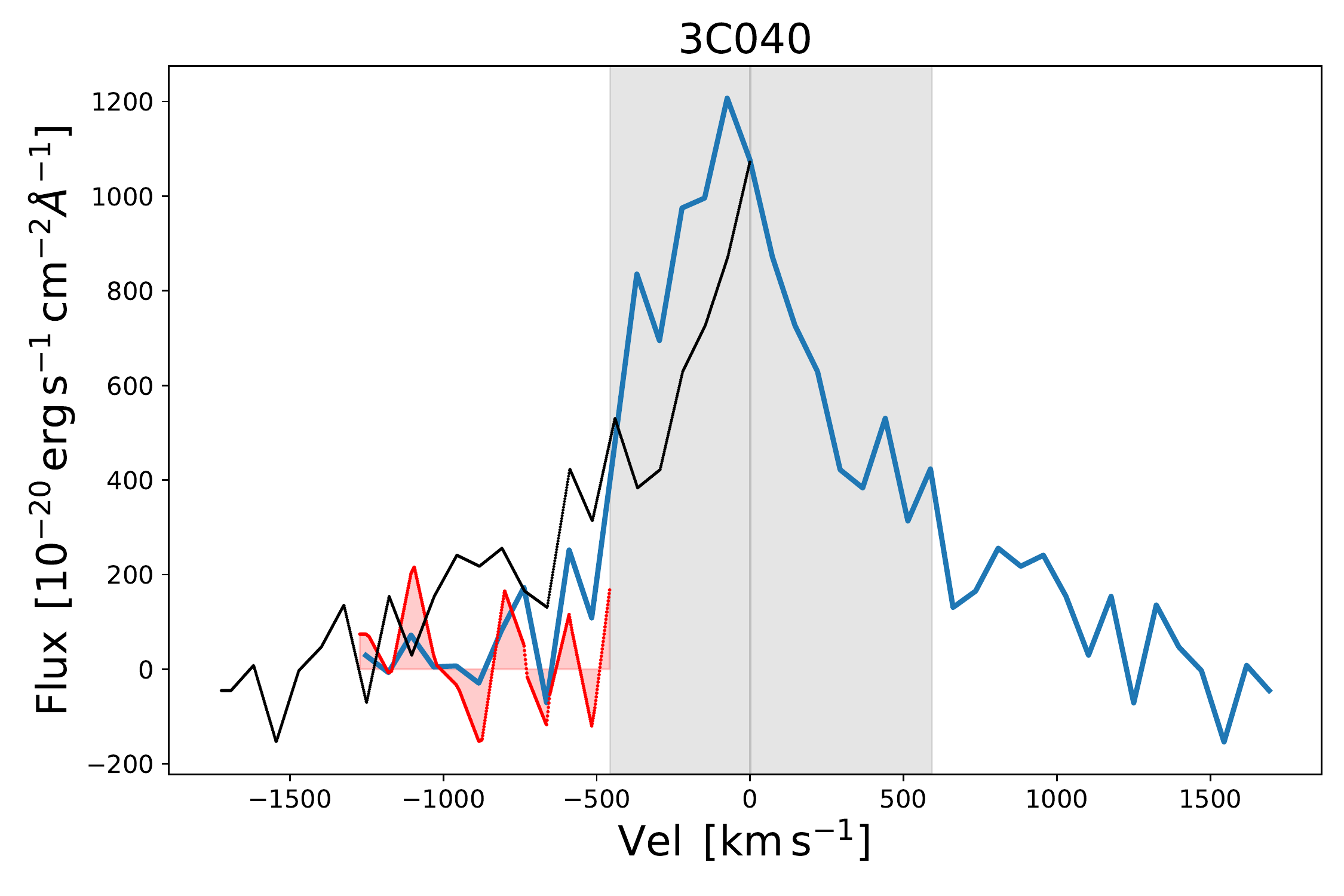}
\includegraphics[width=0.3\textwidth]{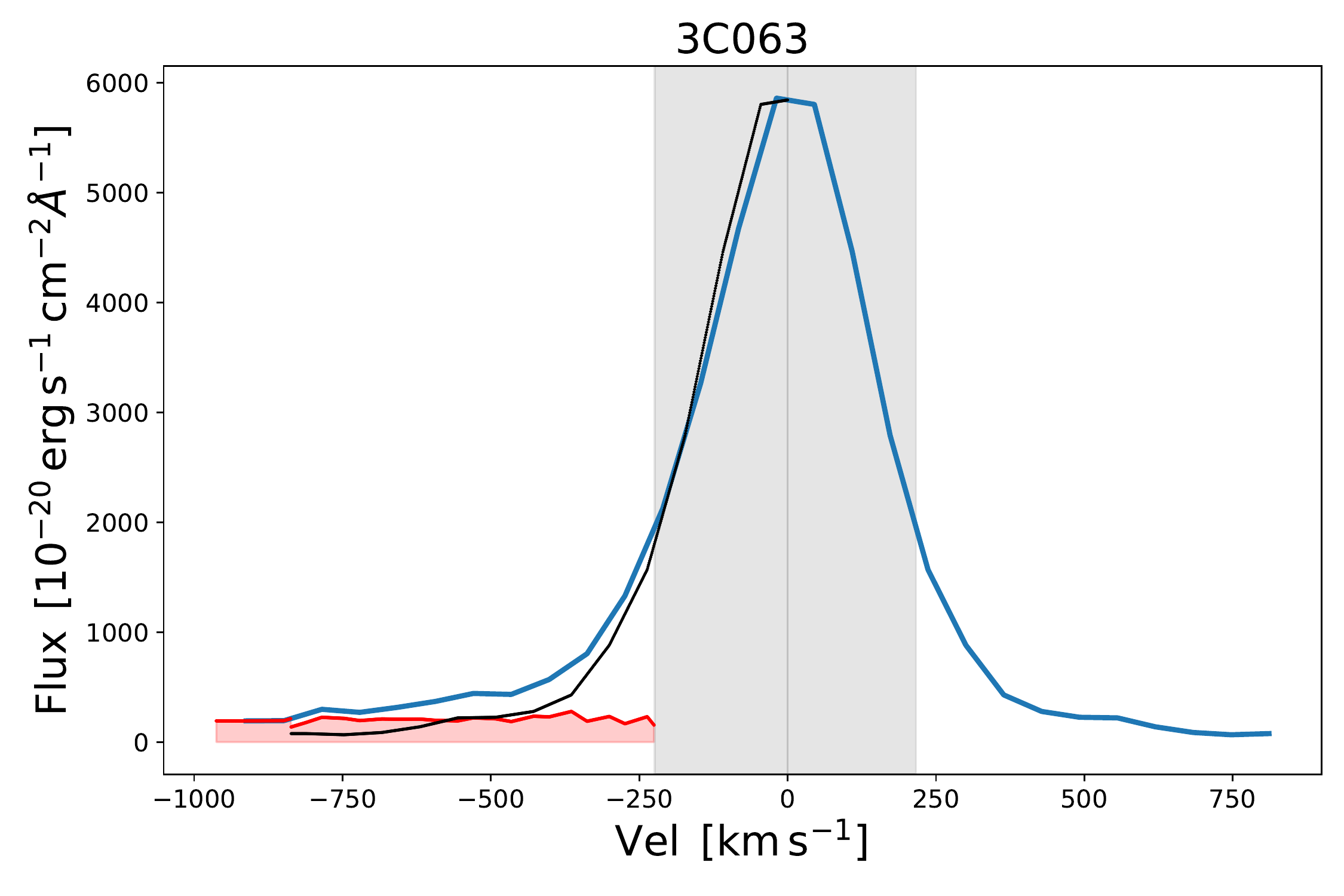}
\includegraphics[width=0.3\textwidth]{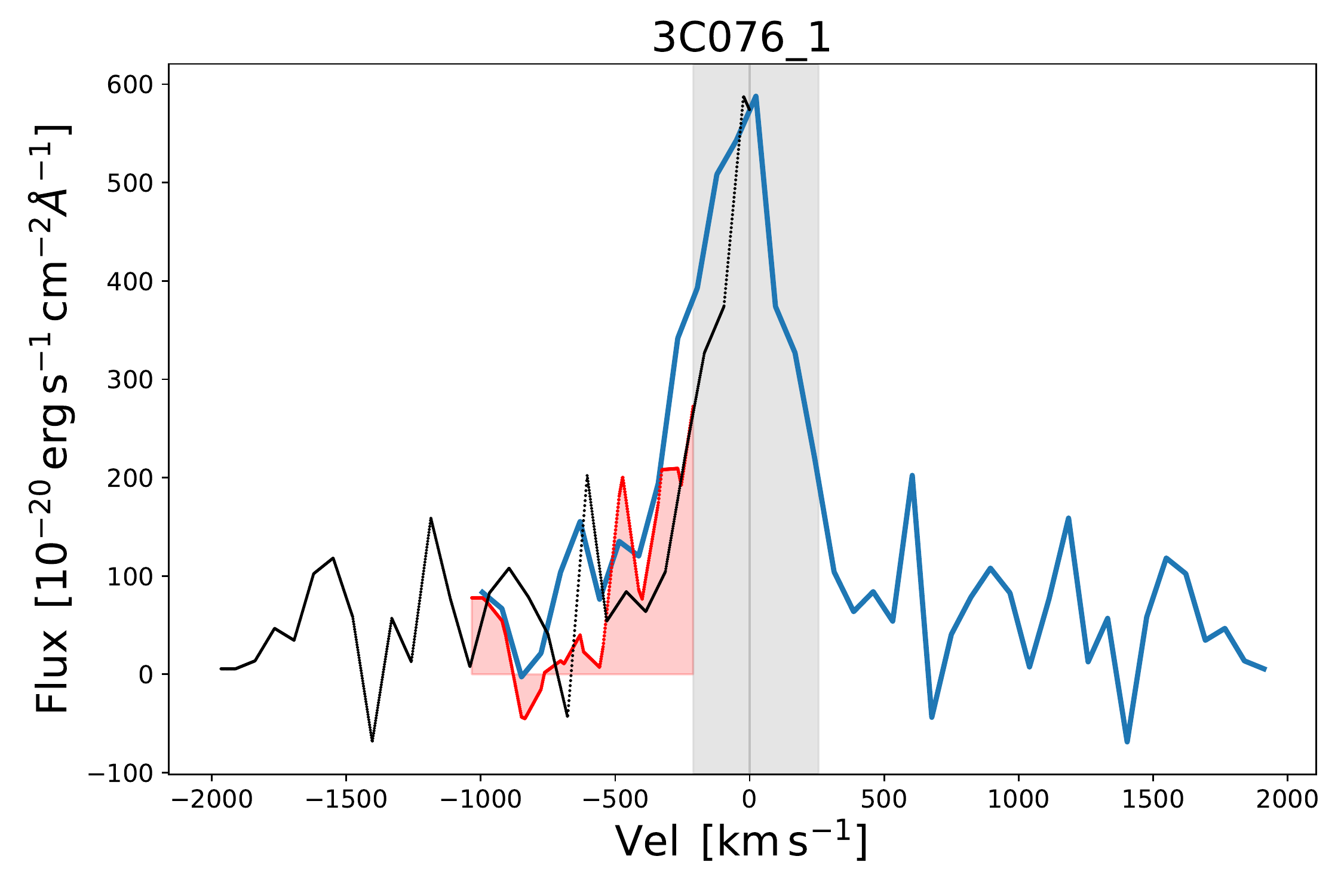}
\includegraphics[width=0.3\textwidth]{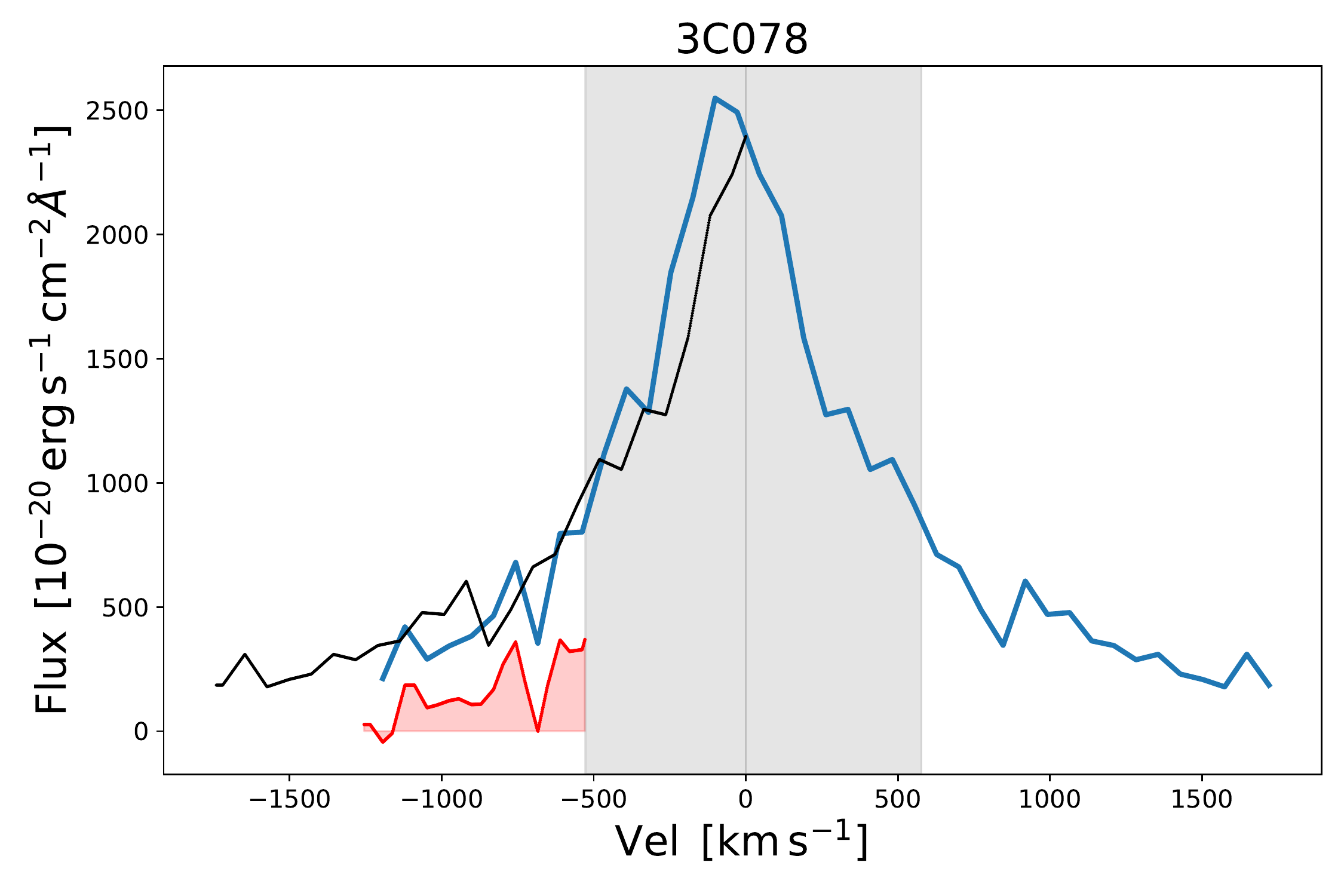}
\includegraphics[width=0.3\textwidth]{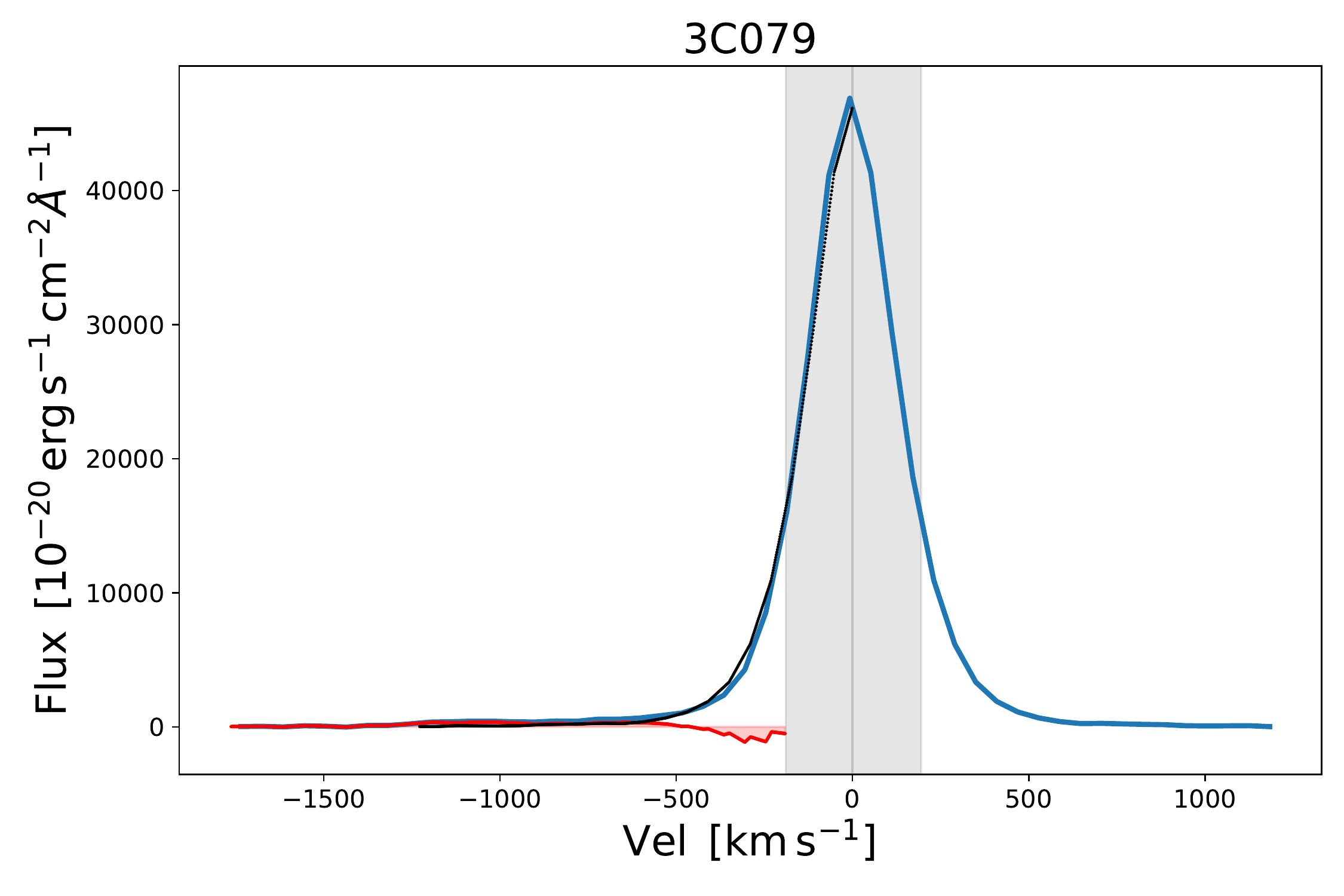}
\includegraphics[width=0.3\textwidth]{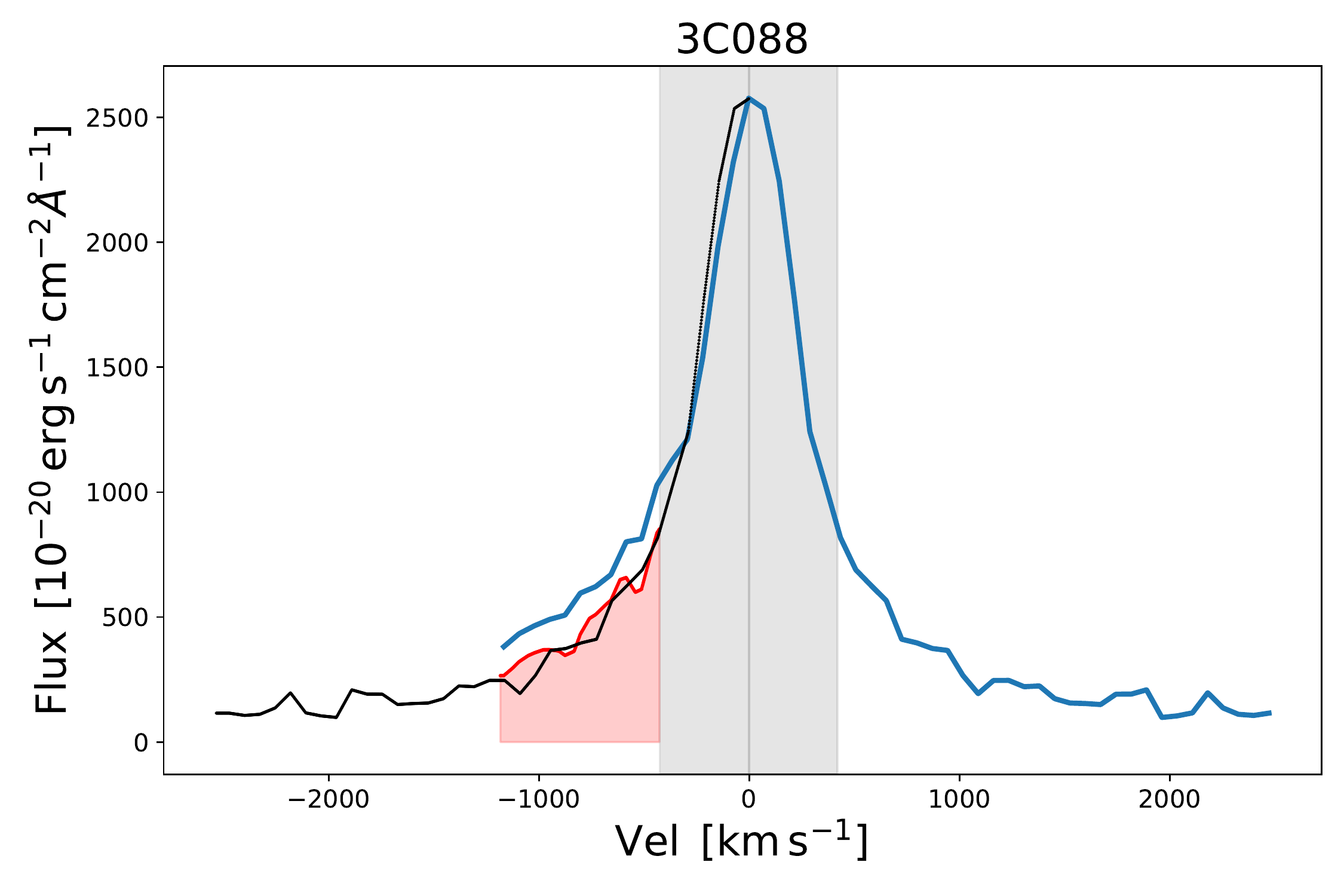}
\includegraphics[width=0.3\textwidth]{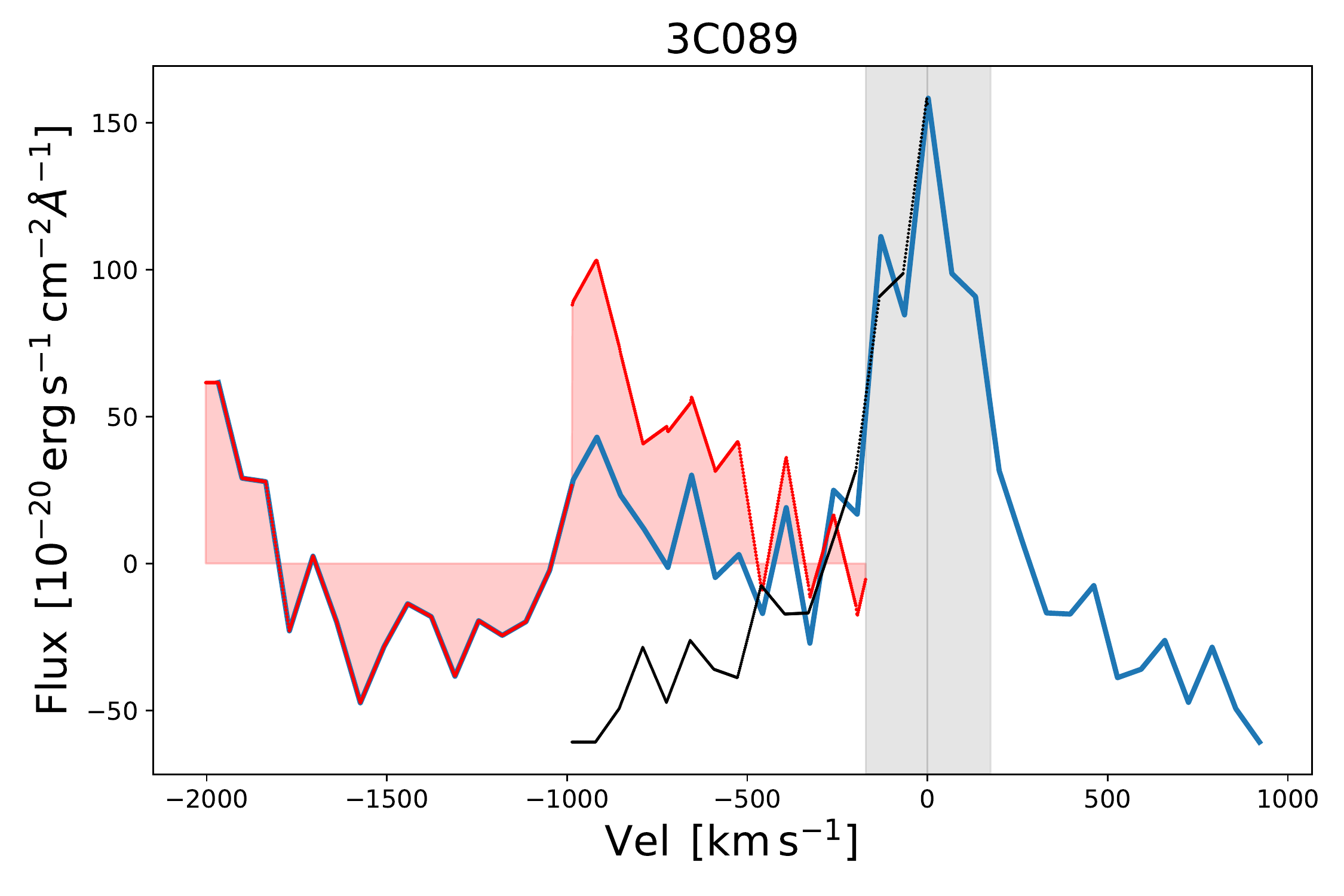}
\includegraphics[width=0.3\textwidth]{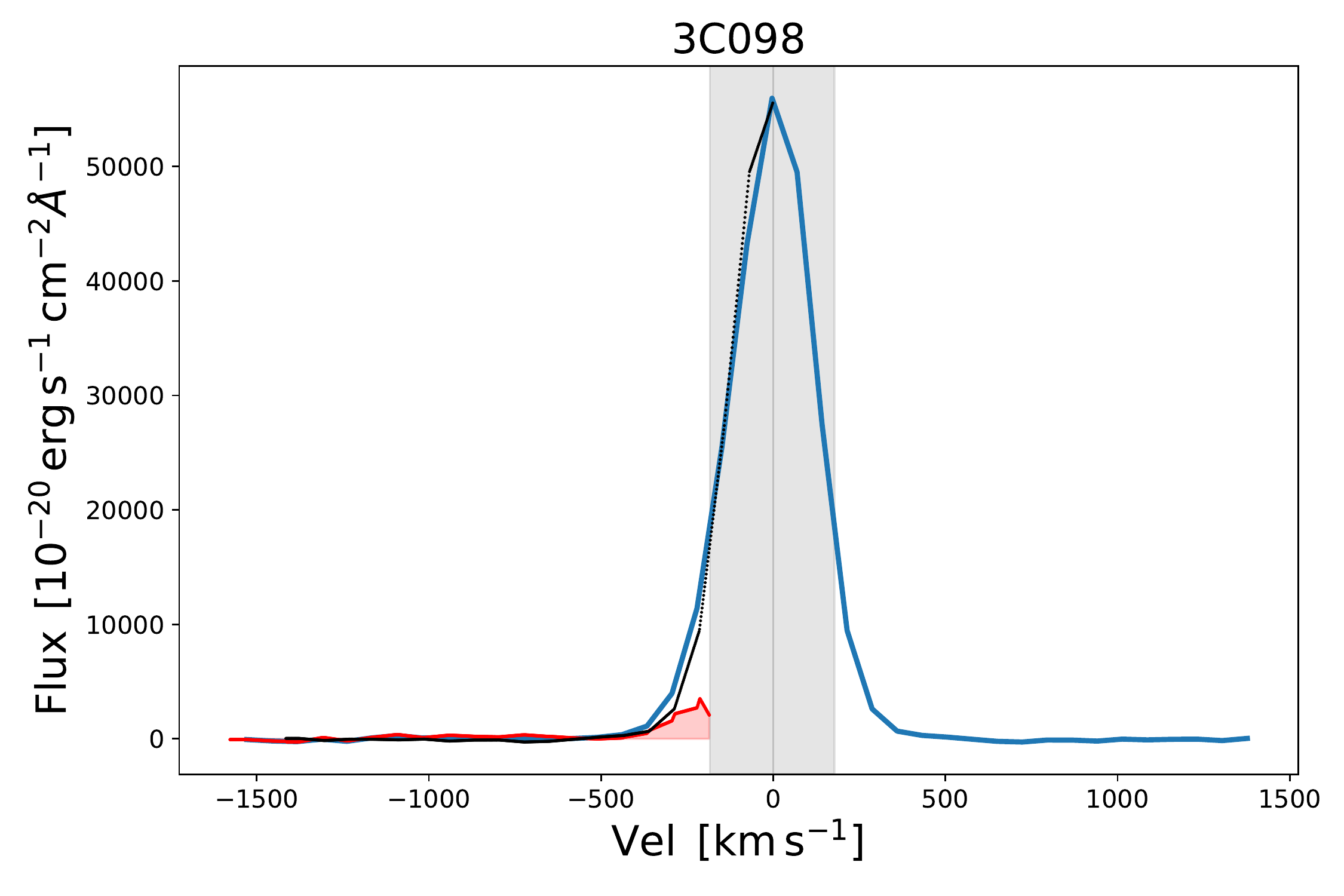}
\includegraphics[width=0.3\textwidth]{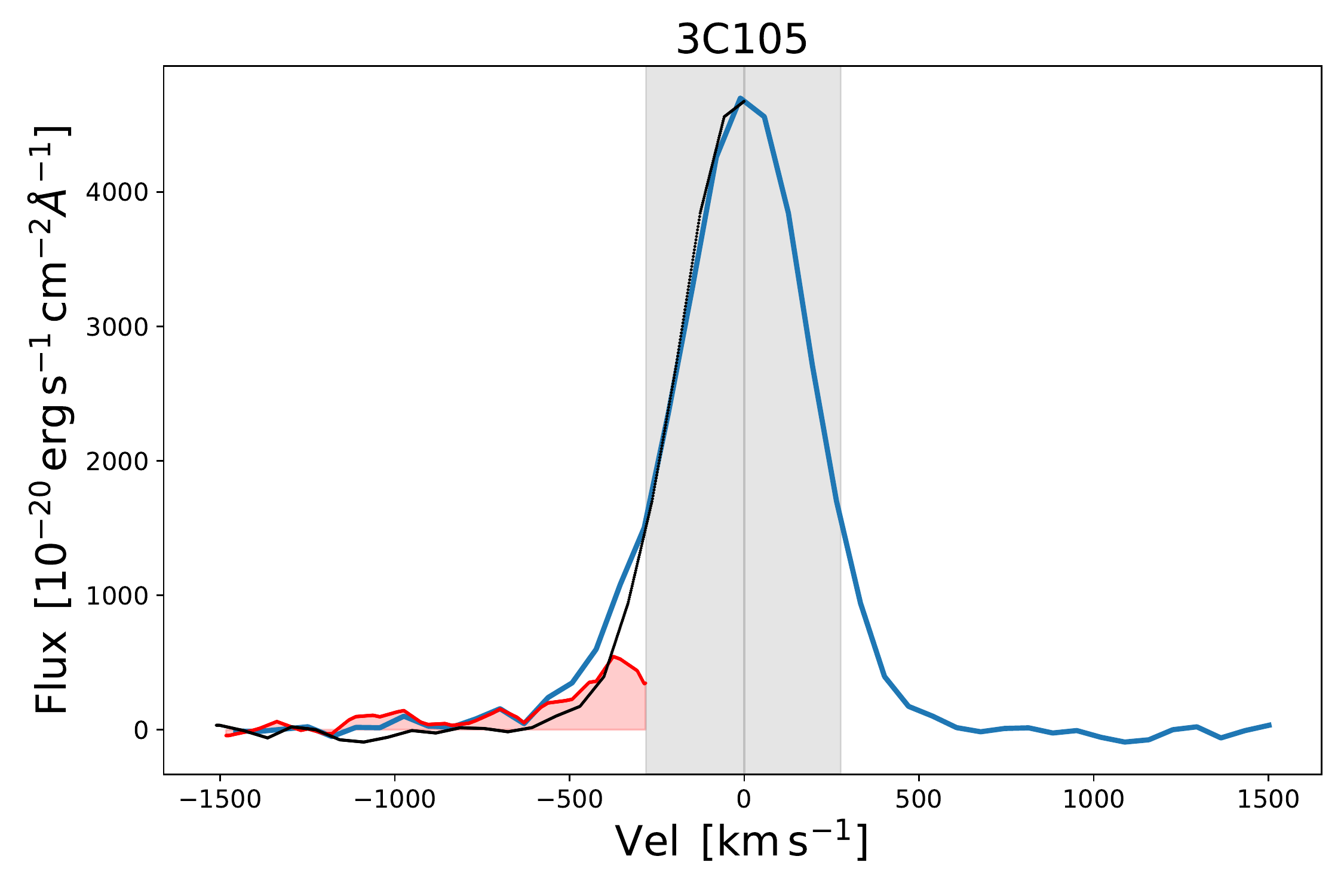}
\includegraphics[width=0.3\textwidth]{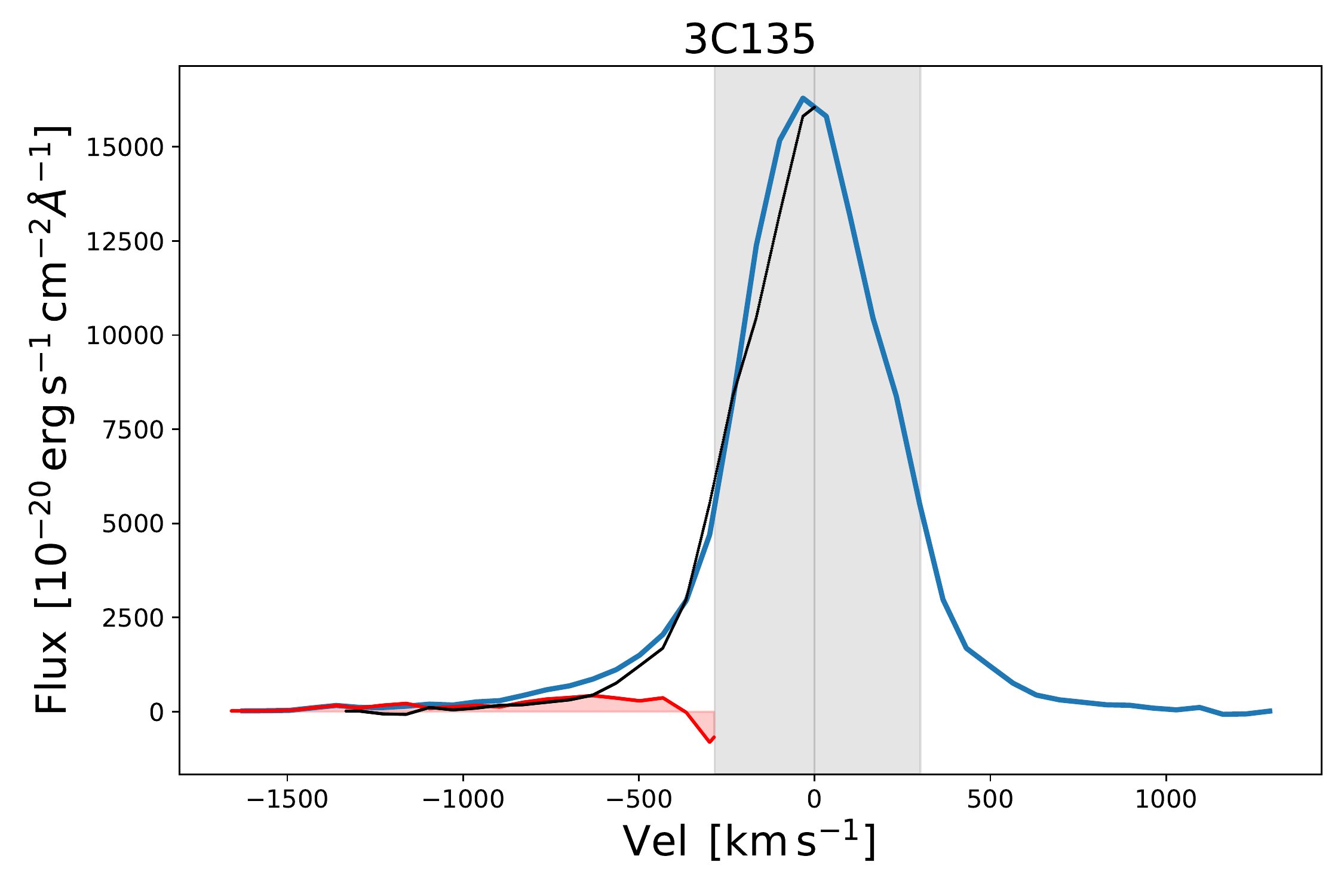}
\includegraphics[width=0.3\textwidth]{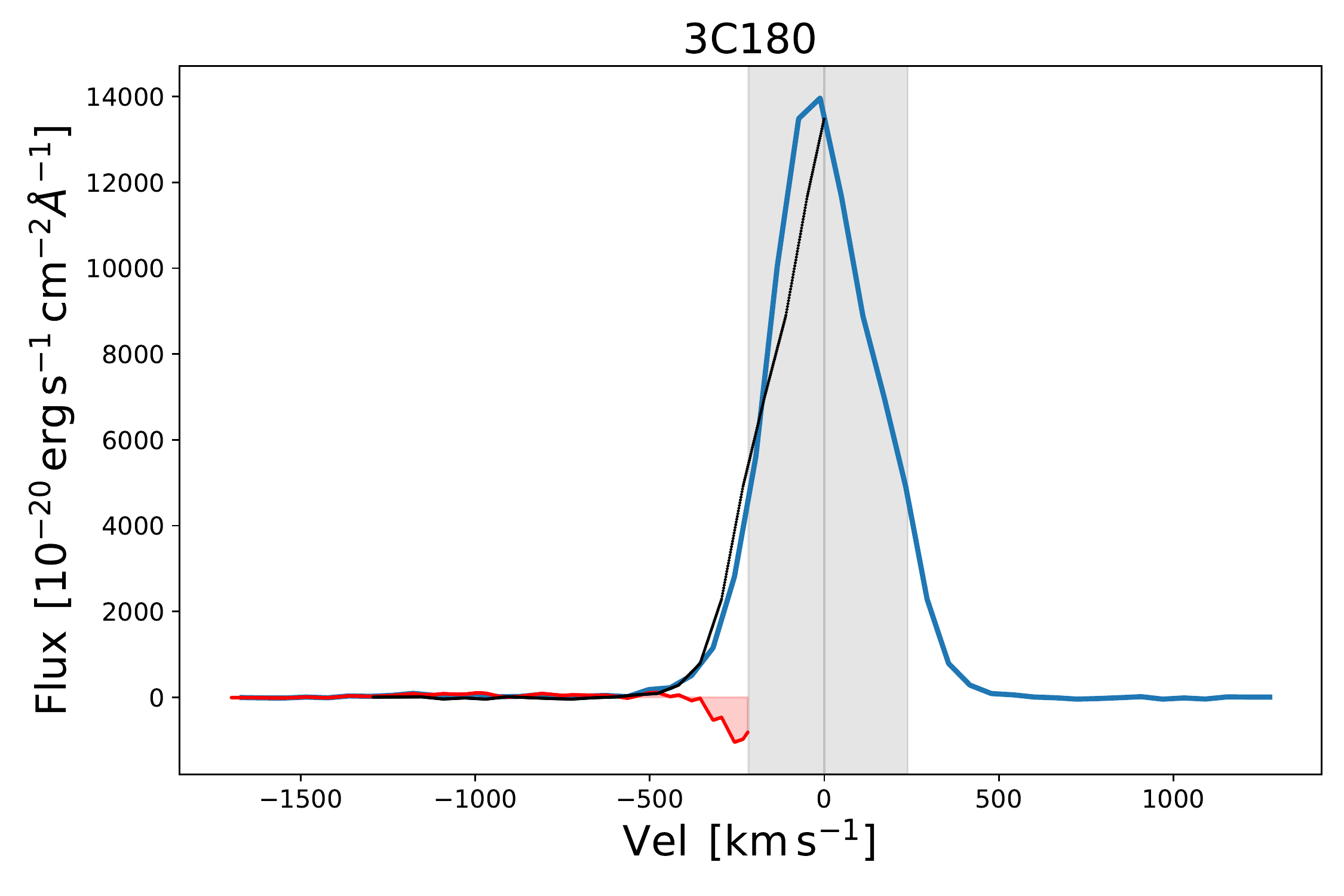}
\includegraphics[width=0.3\textwidth]{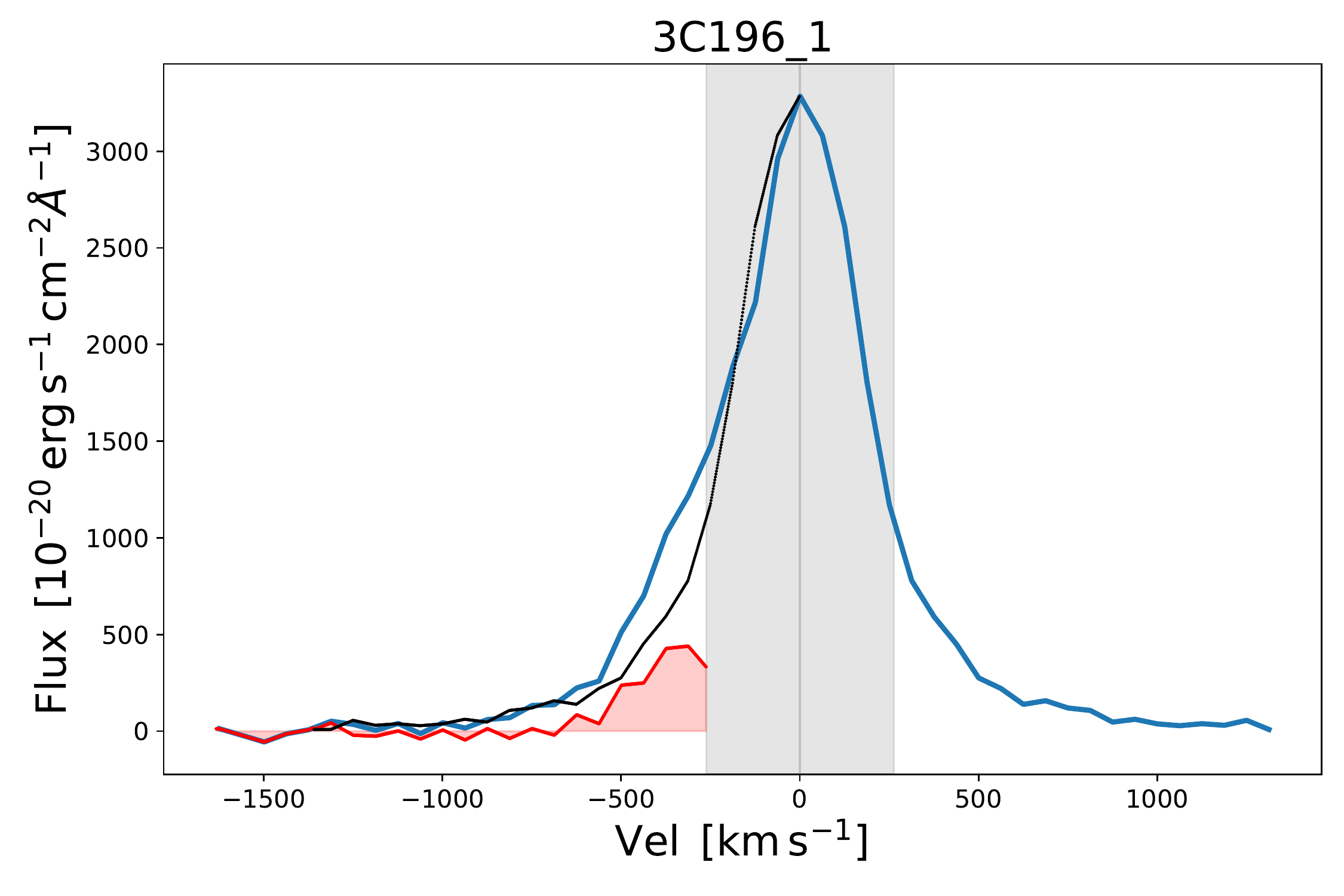}
\includegraphics[width=0.3\textwidth]{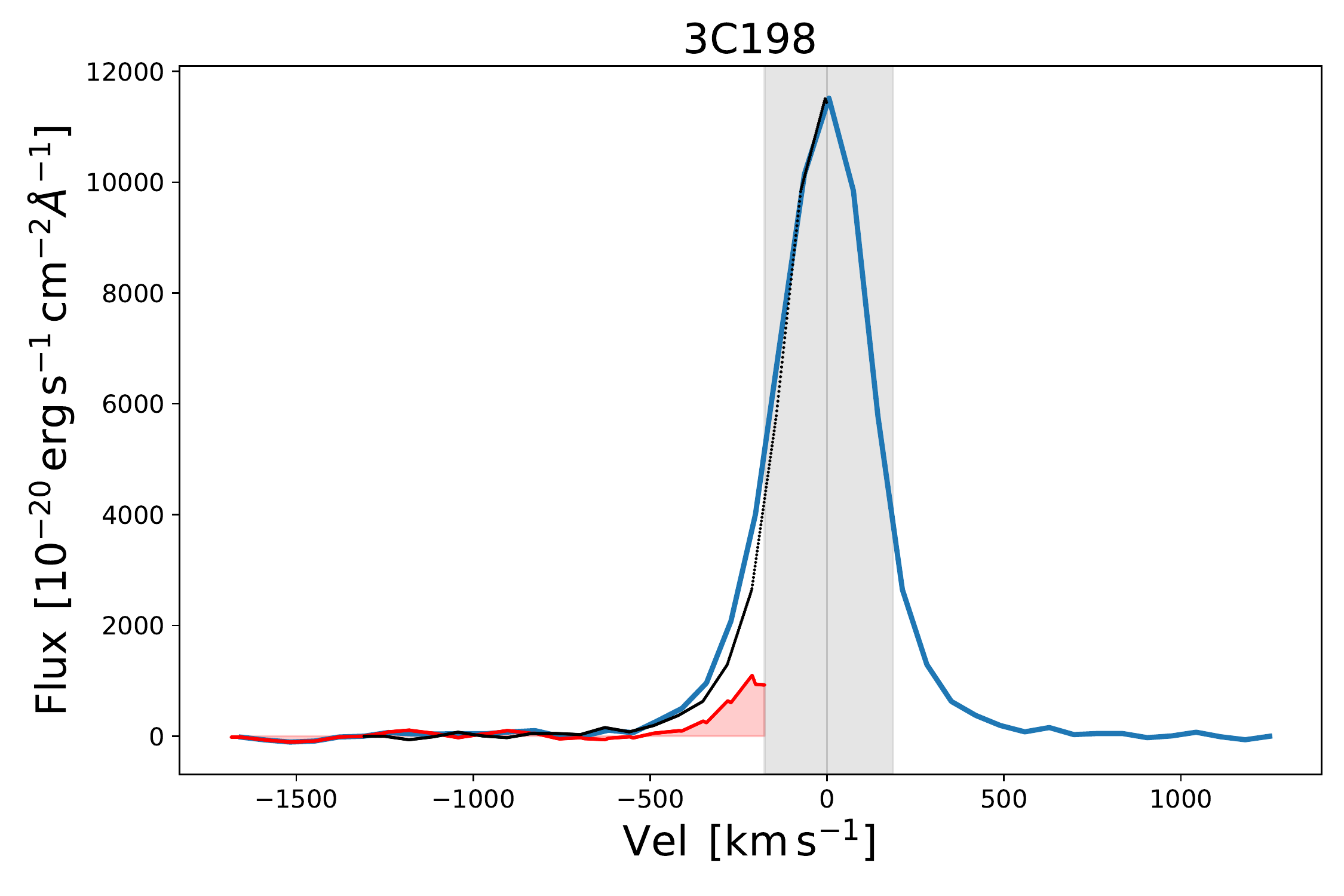}
\includegraphics[width=0.3\textwidth]{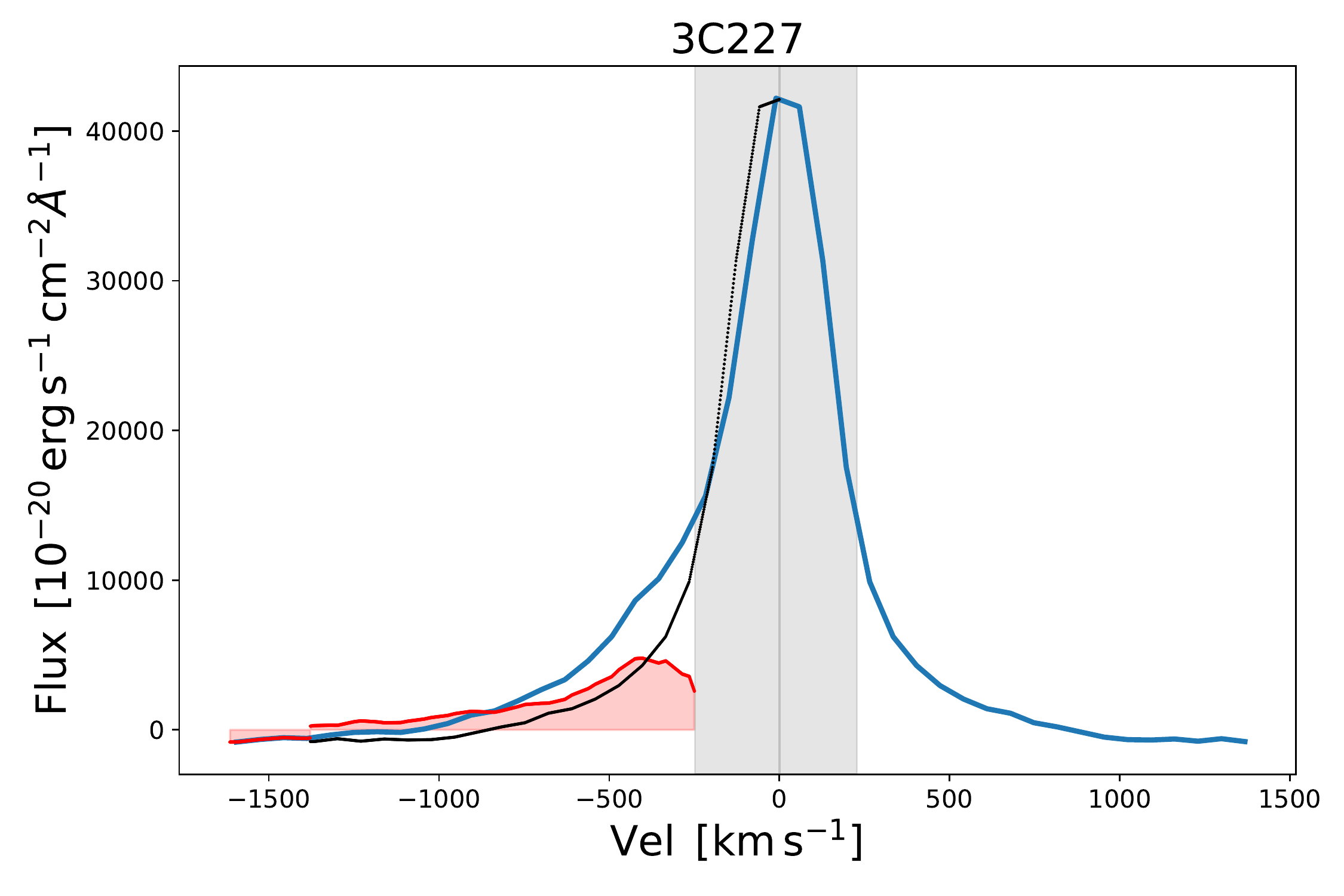}
\includegraphics[width=0.3\textwidth]{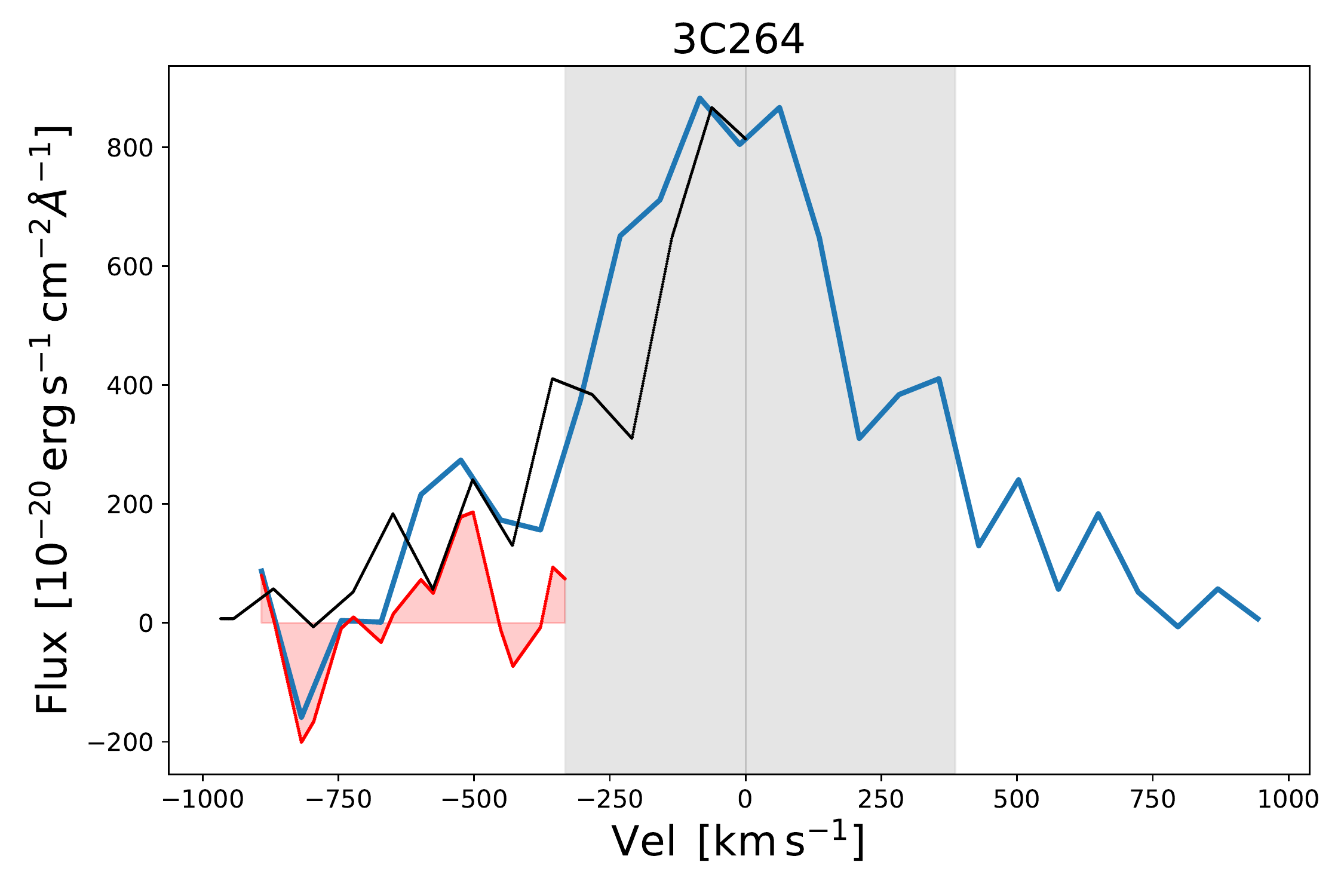}
\caption{Nuclear spectra of the 33 3C sources with a detection of the
  [O~III] emission line. Superposed on the [O~III] profile (blue) is the mirror image of the redshifted line (black). Their
  difference (red) reveals any line asymmetry. We
  estimated the median velocity ($V_{50}$) of the asymmetric line wing
  and its flux (F$_W$), after masking the spectral region at one-third of the
  maximum height of the profile (grey region).}
\end{figure*}

\addtocounter{figure}{-1}
\begin{figure*}
\centering
\includegraphics[width=0.3\textwidth]{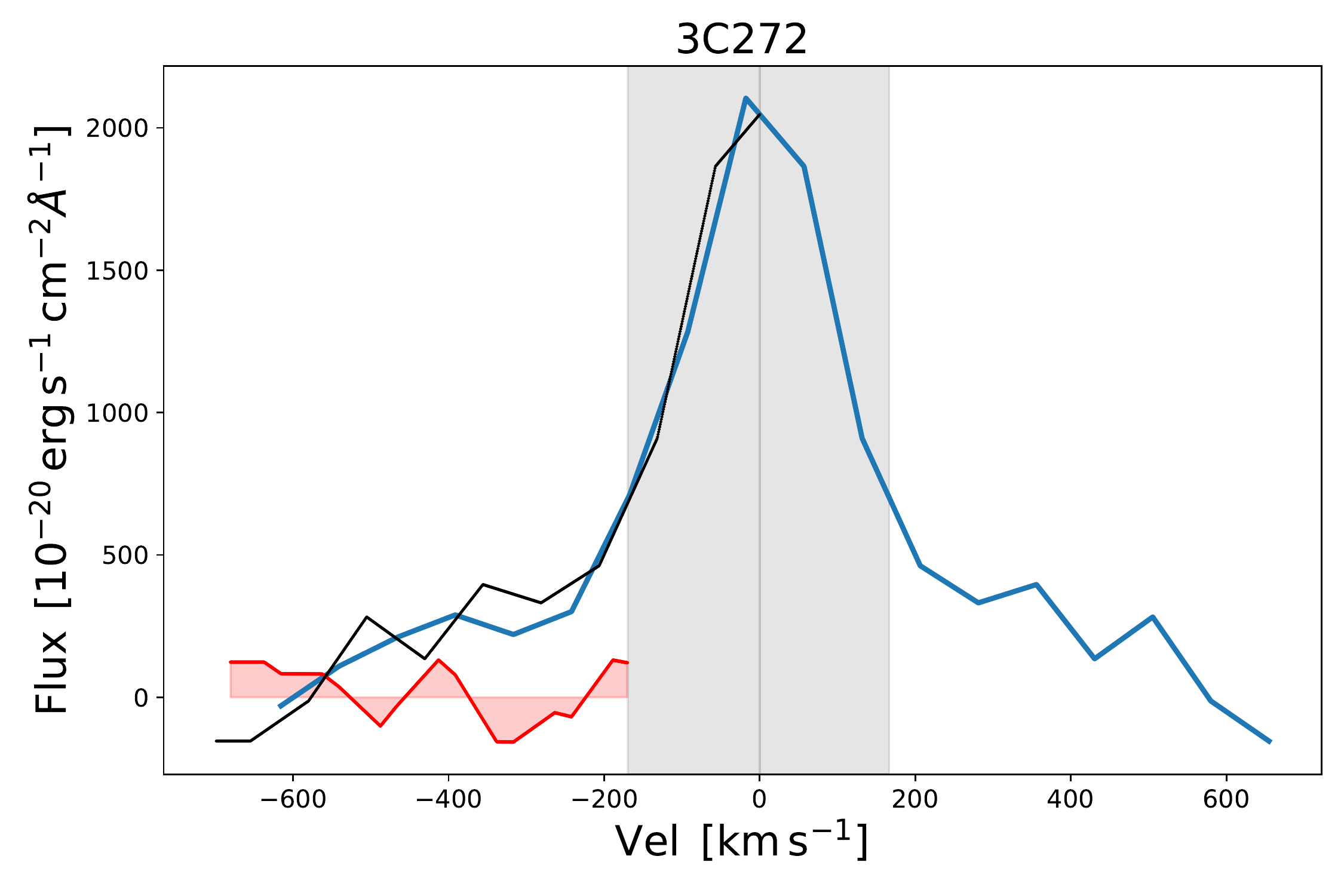}
\includegraphics[width=0.3\textwidth]{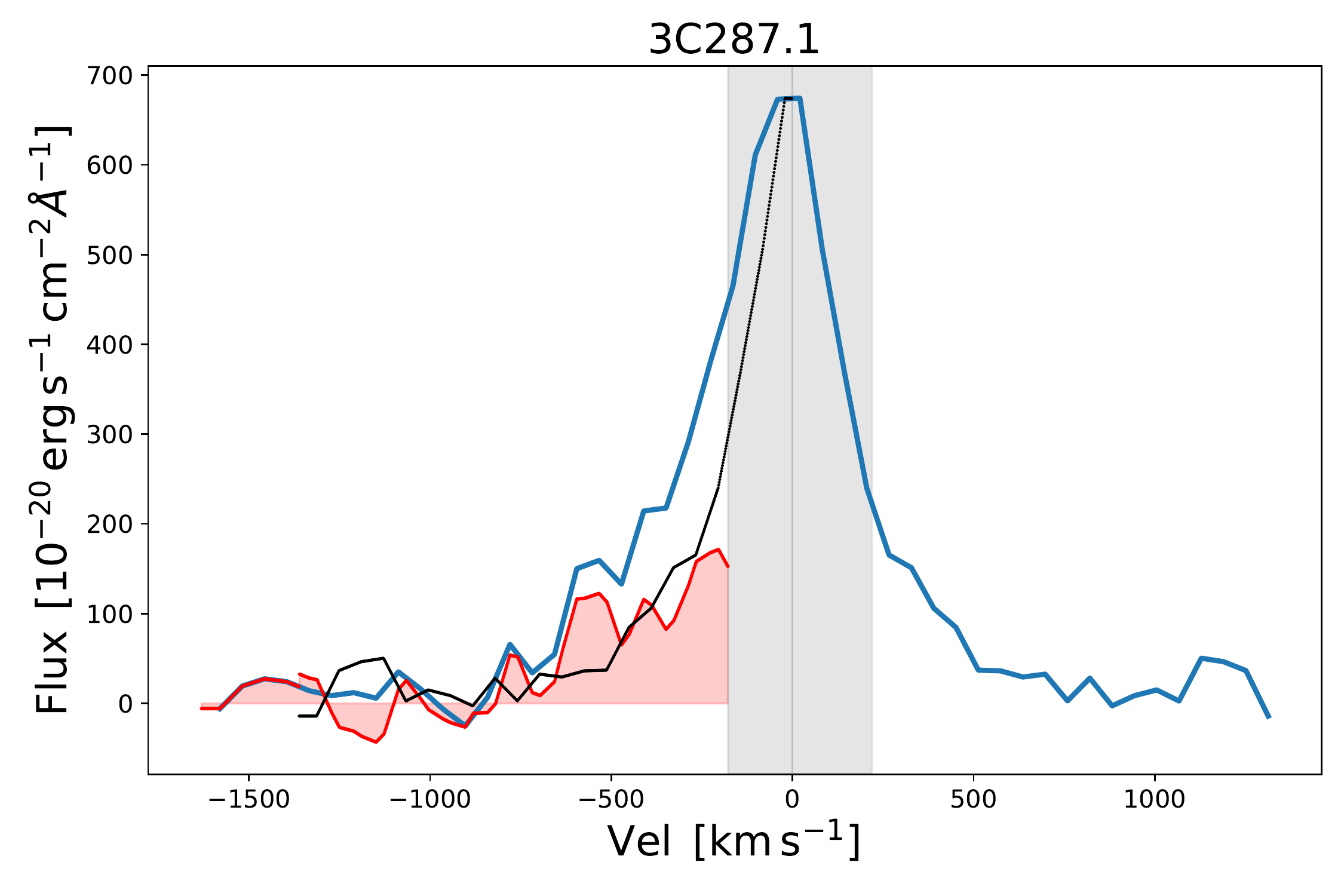}
\includegraphics[width=0.3\textwidth]{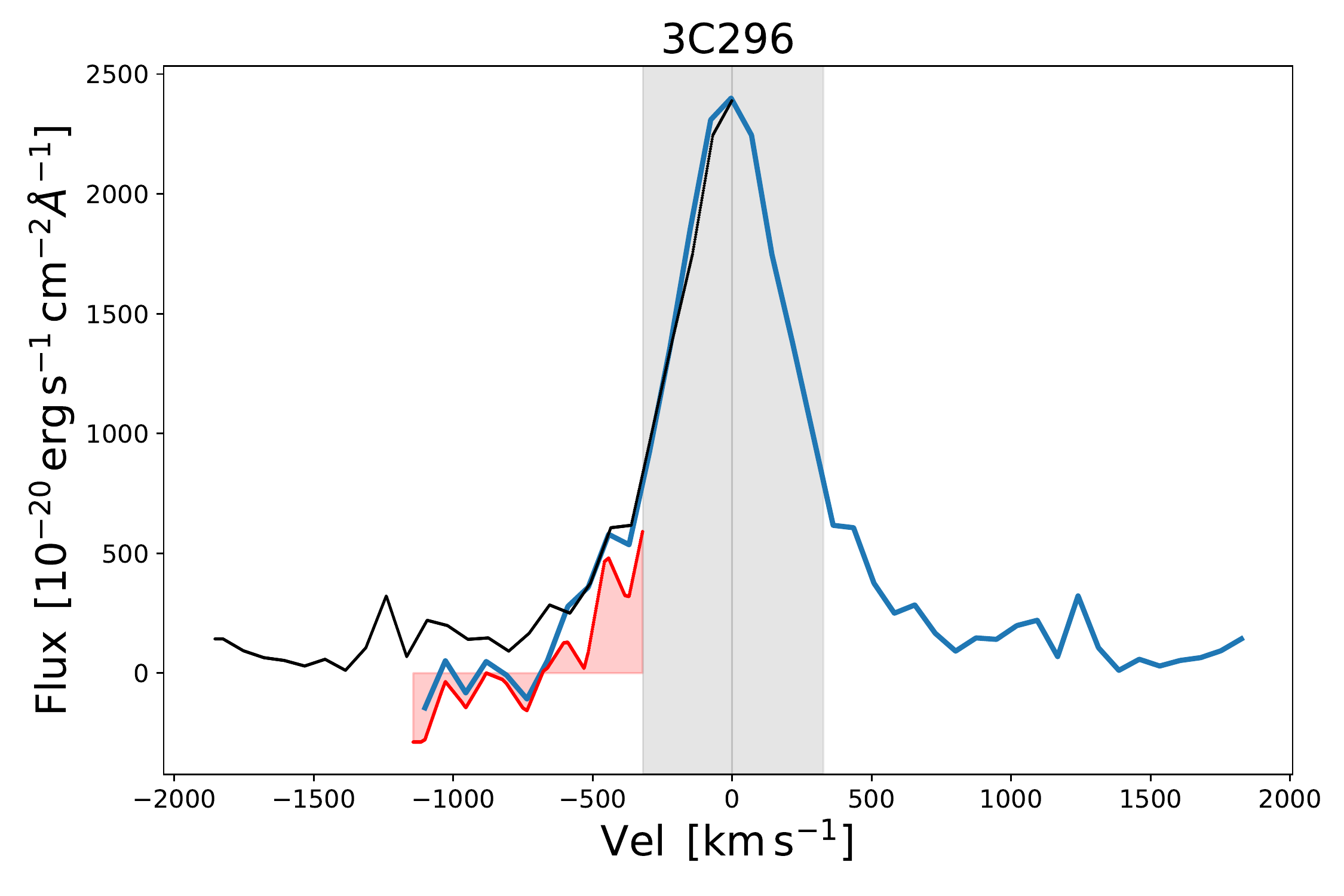}
\includegraphics[width=0.3\textwidth]{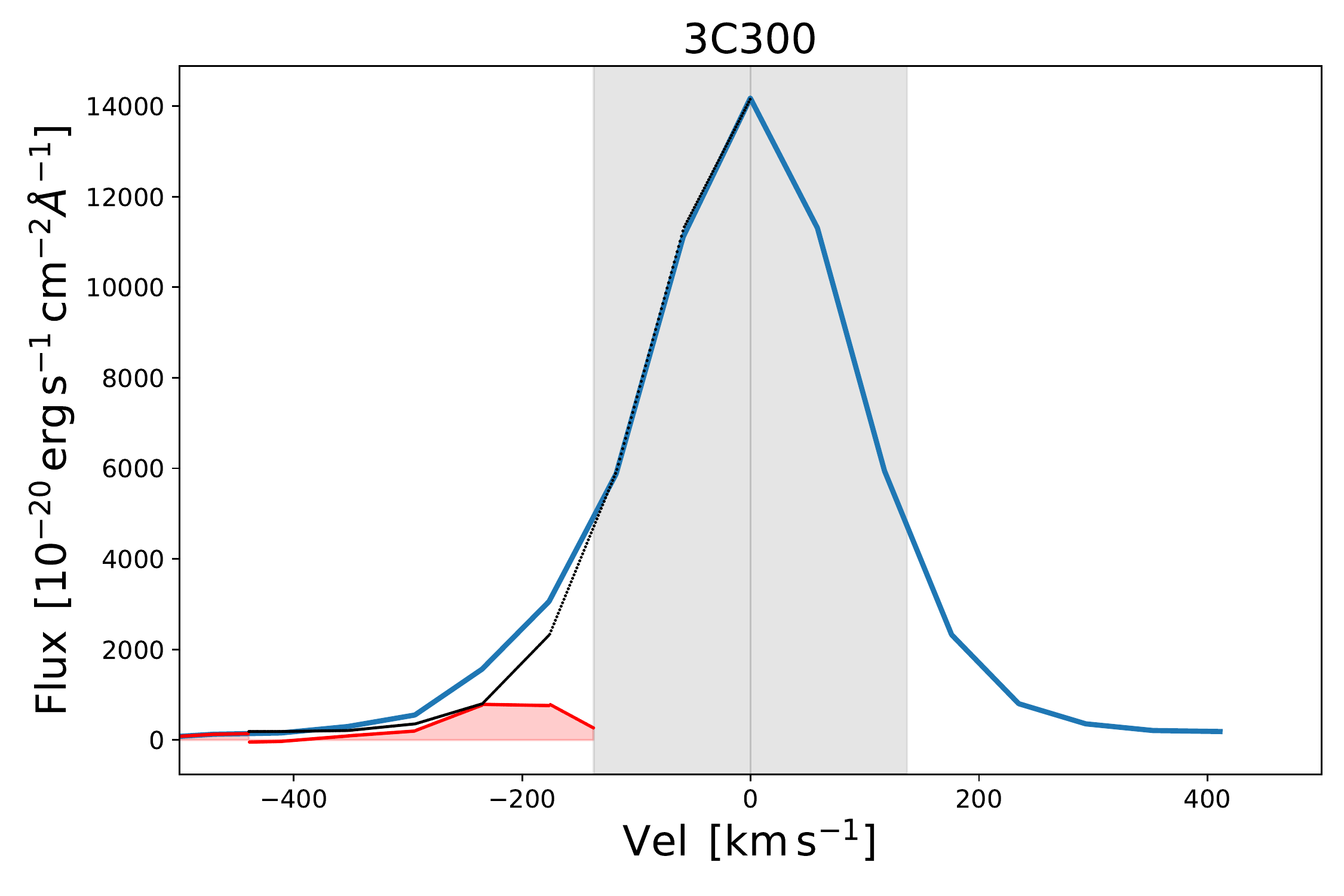}
\includegraphics[width=0.3\textwidth]{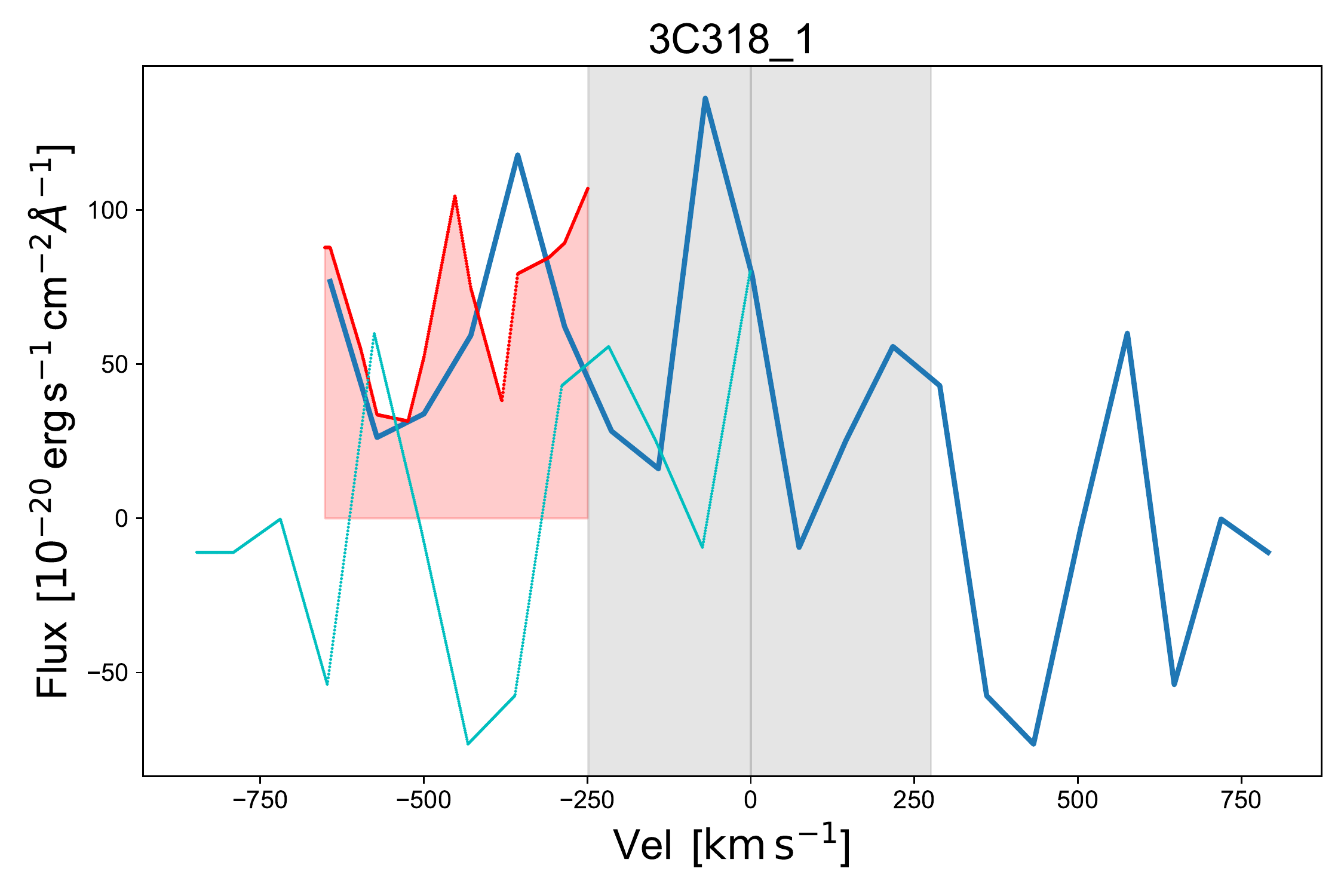}
\includegraphics[width=0.3\textwidth]{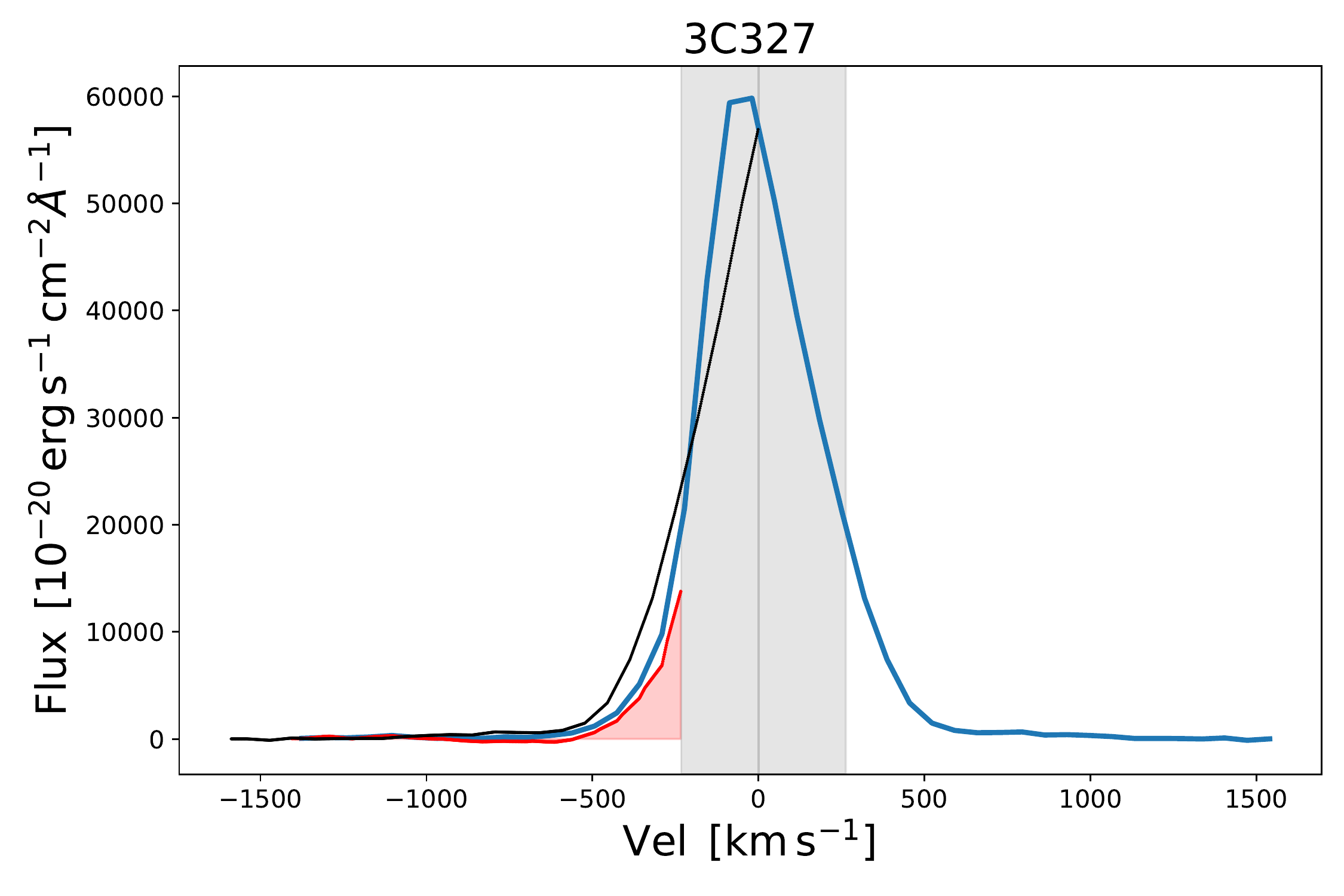}
\includegraphics[width=0.3\textwidth]{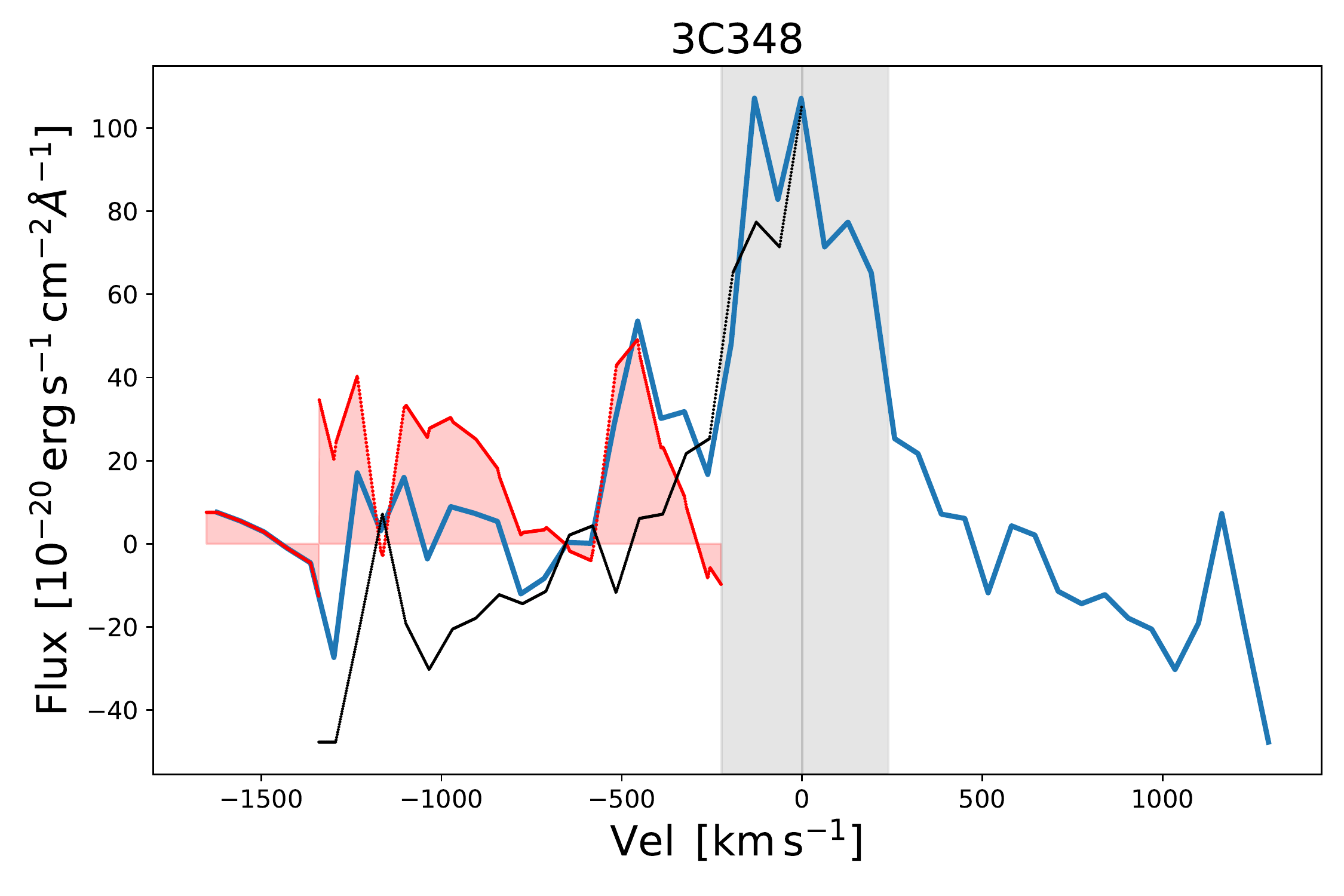}
\includegraphics[width=0.3\textwidth]{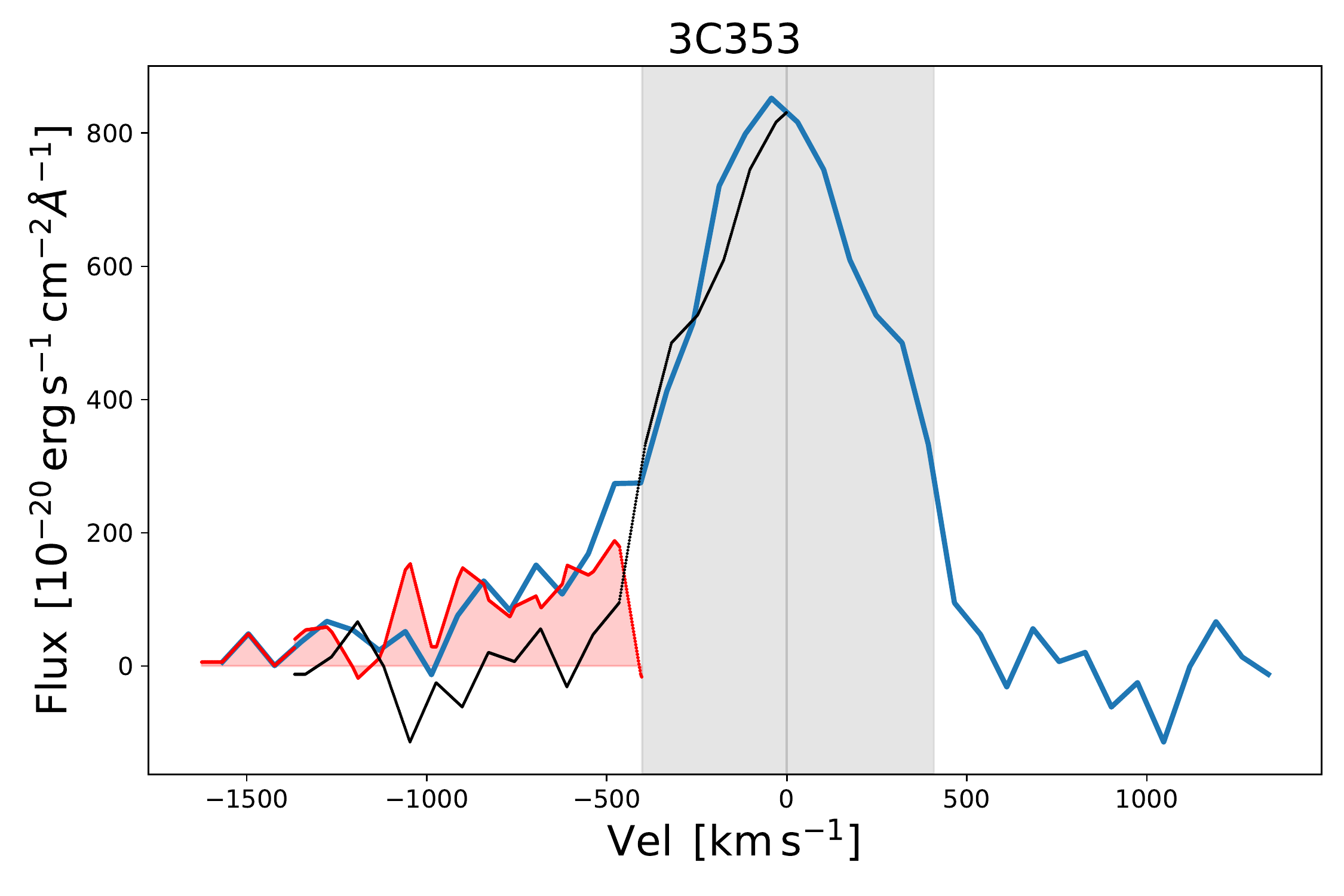}
\includegraphics[width=0.3\textwidth]{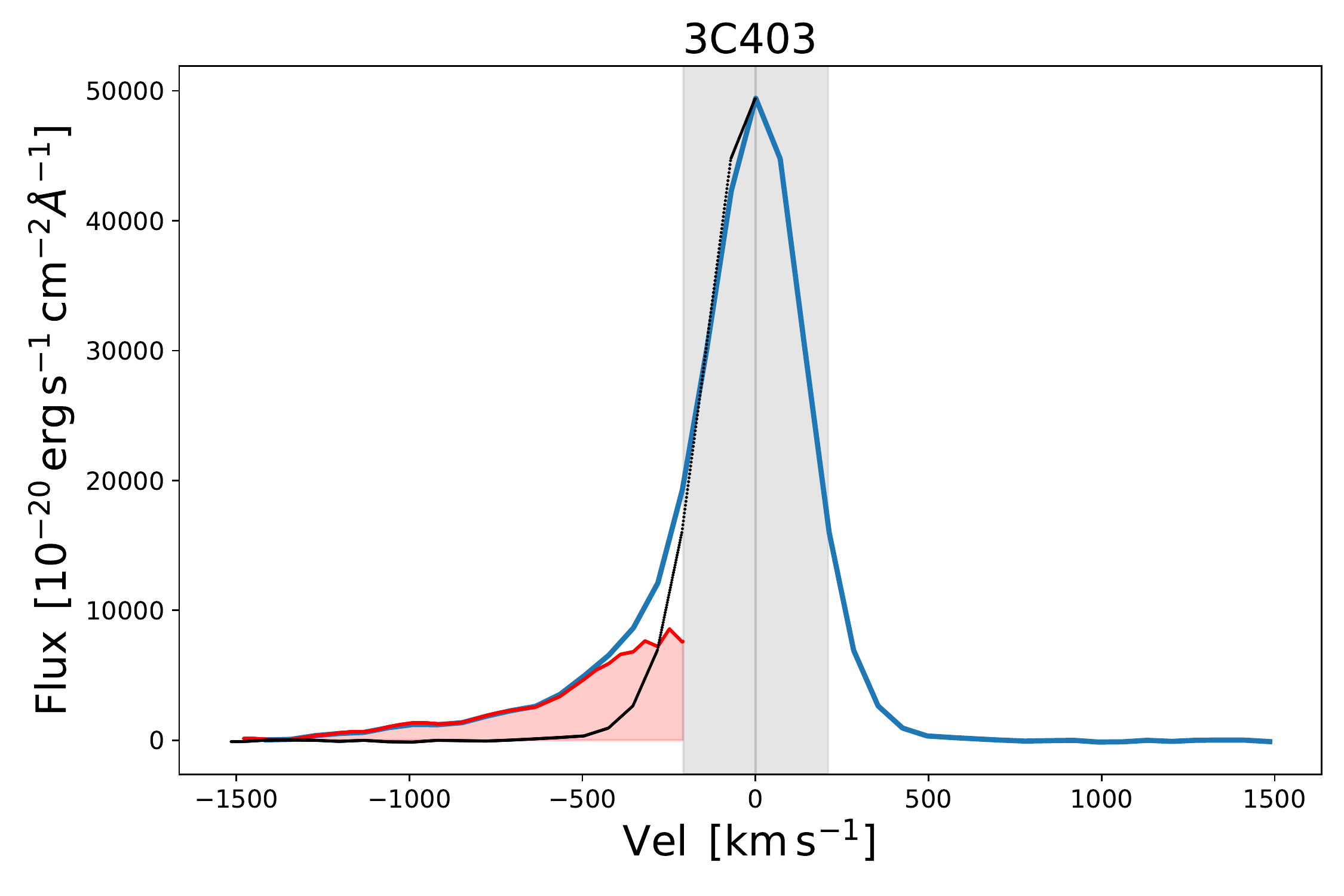}
\includegraphics[width=0.3\textwidth]{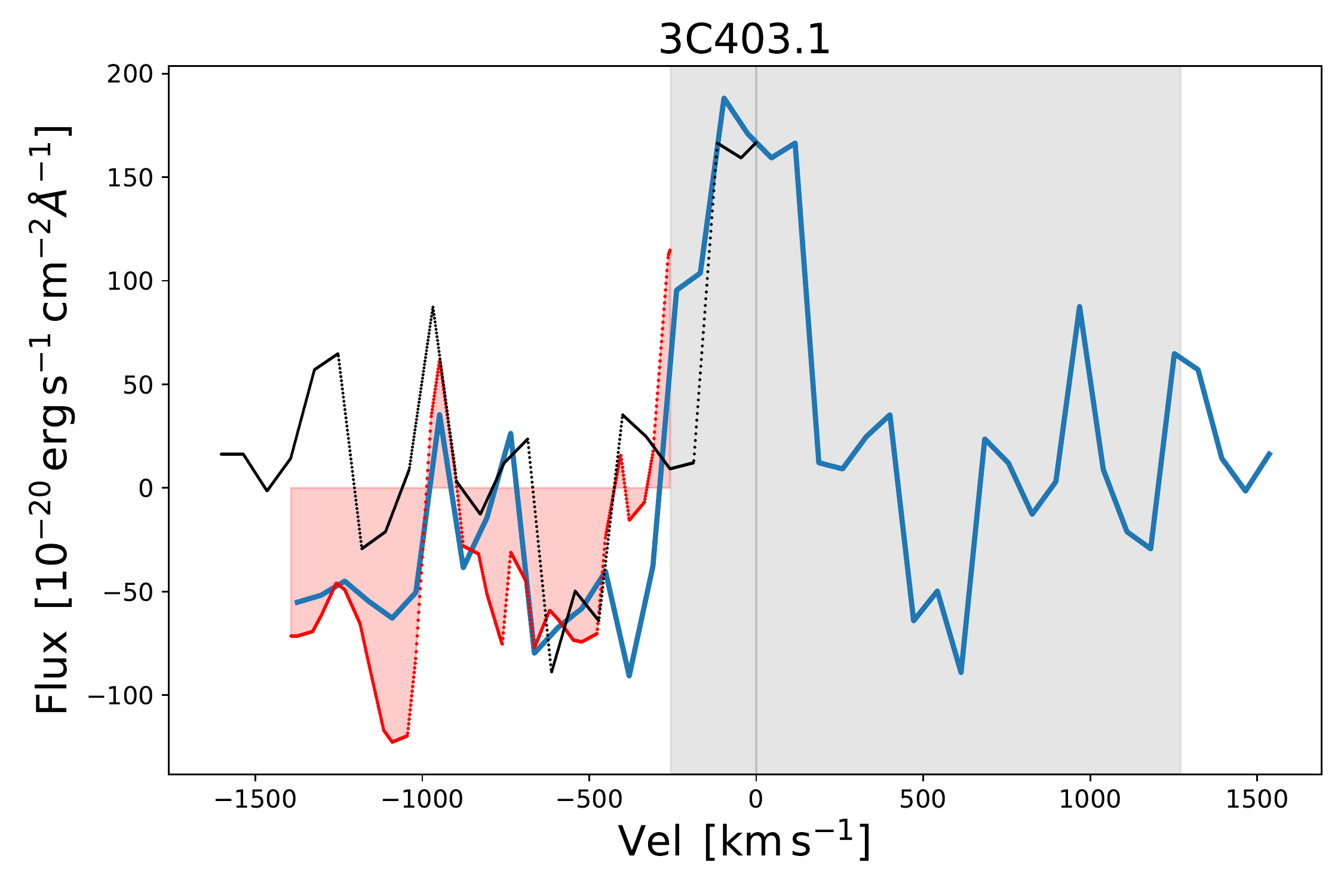}
\includegraphics[width=0.3\textwidth]{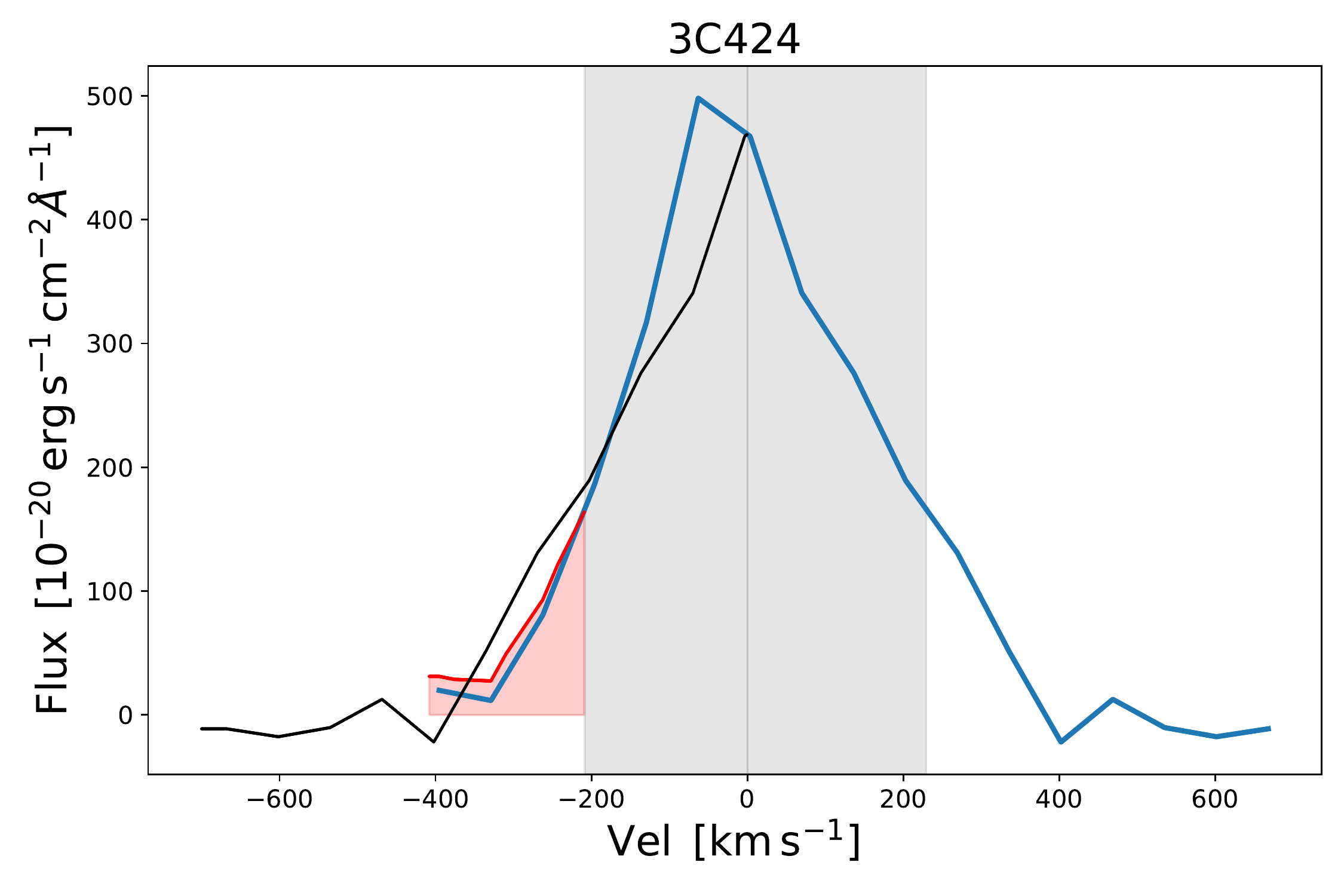}
\includegraphics[width=0.3\textwidth]{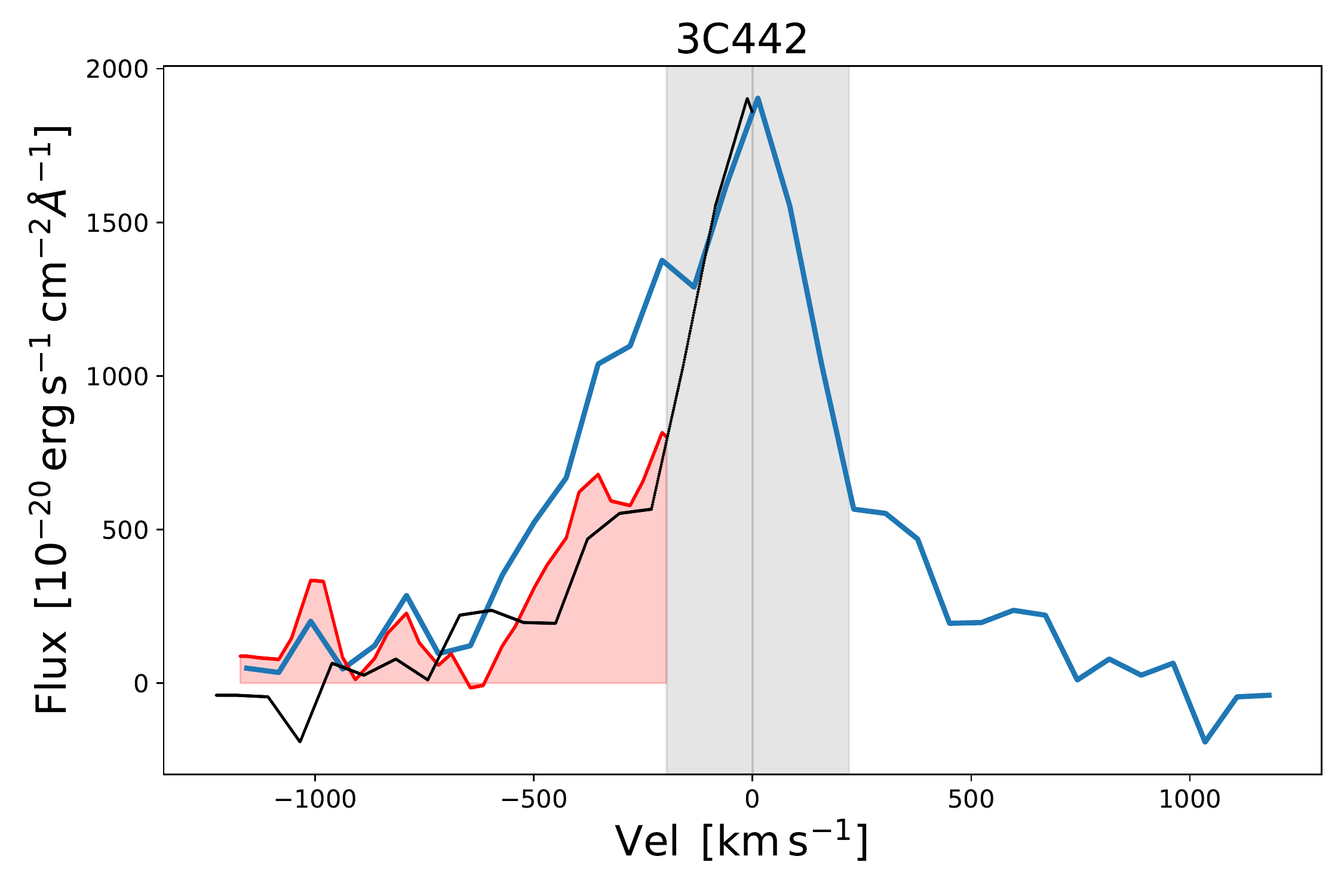}
\includegraphics[width=0.3\textwidth]{fignuc/3C445nuc.pdf}
\includegraphics[width=0.3\textwidth]{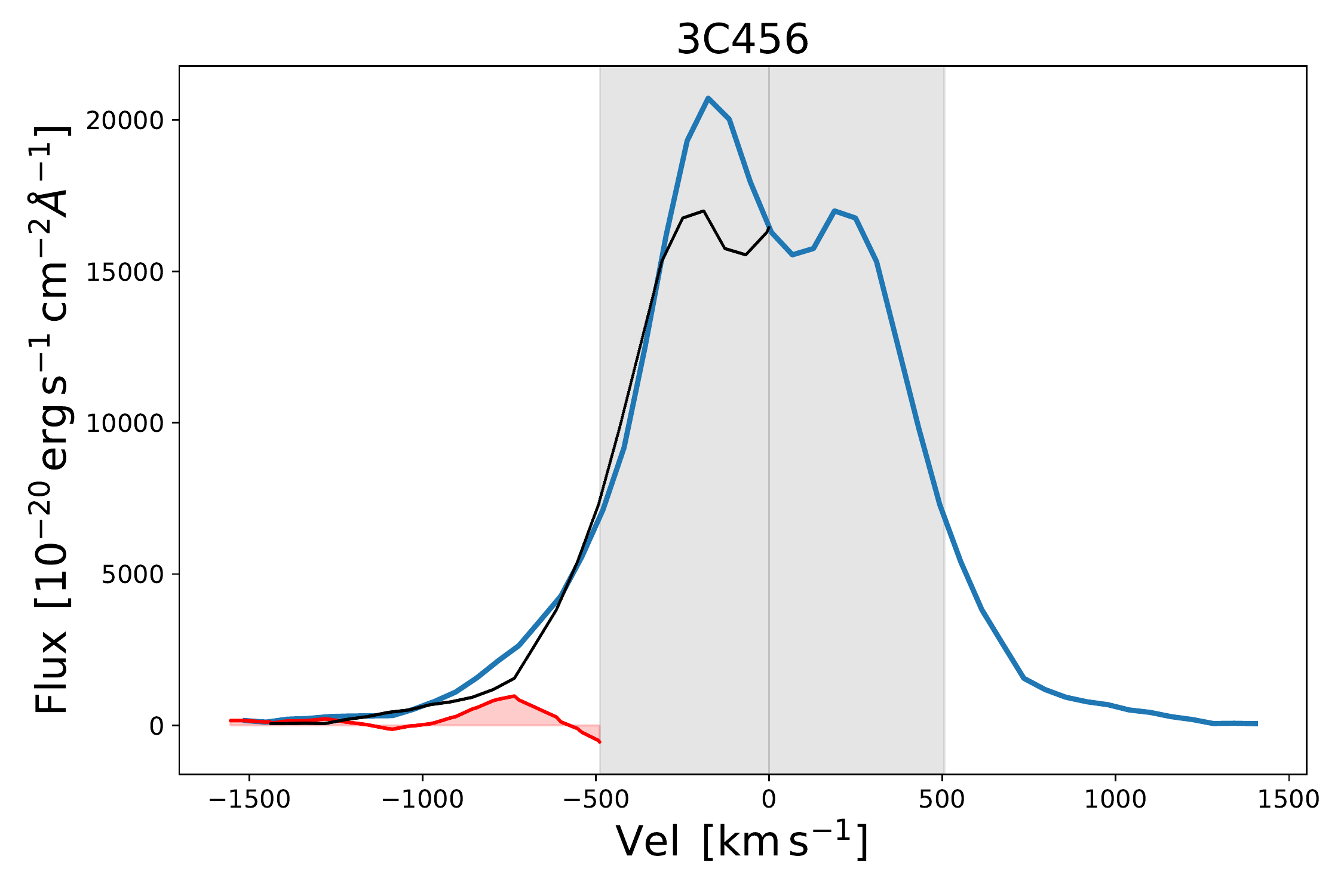}
\includegraphics[width=0.3\textwidth]{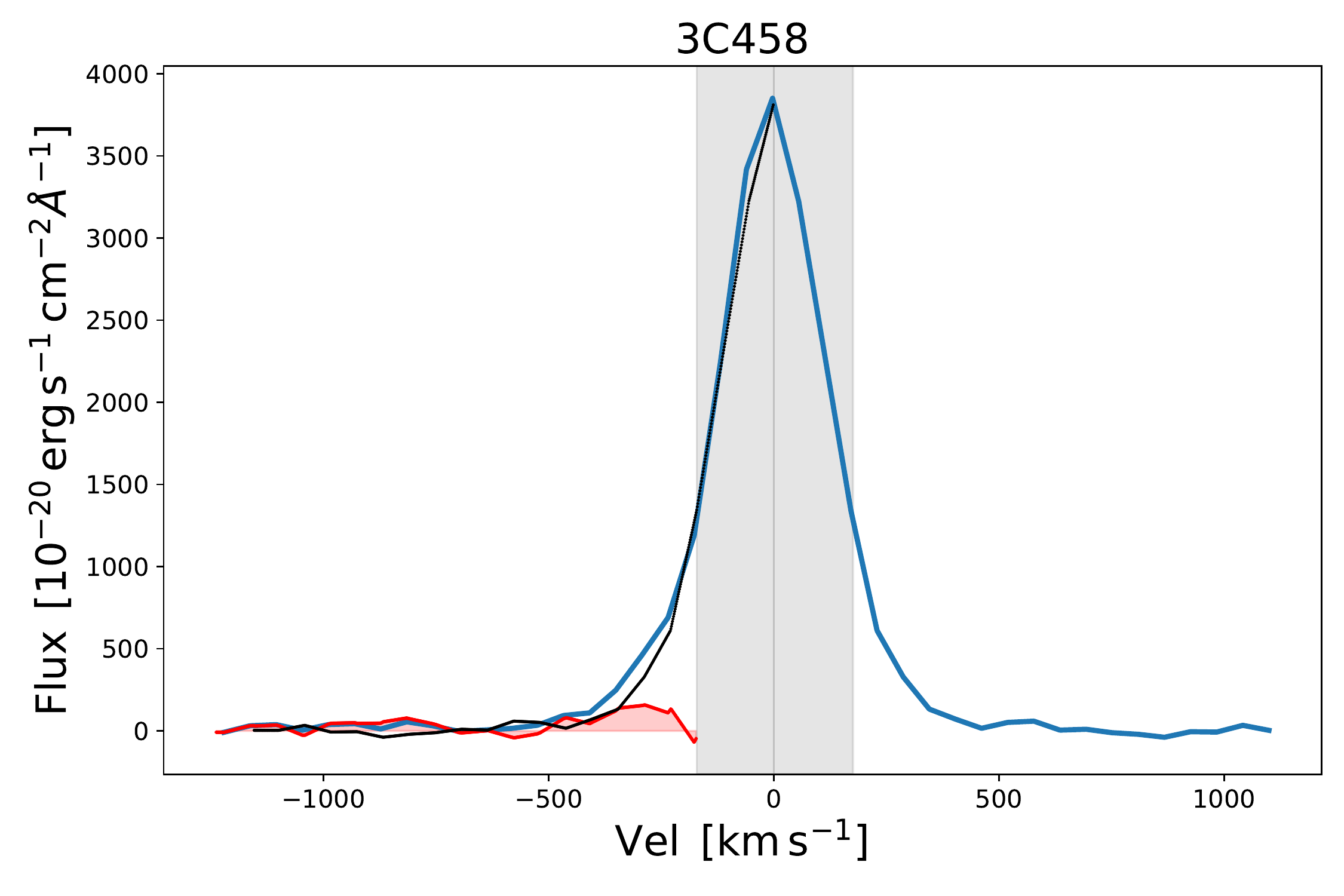}
\includegraphics[width=0.3\textwidth]{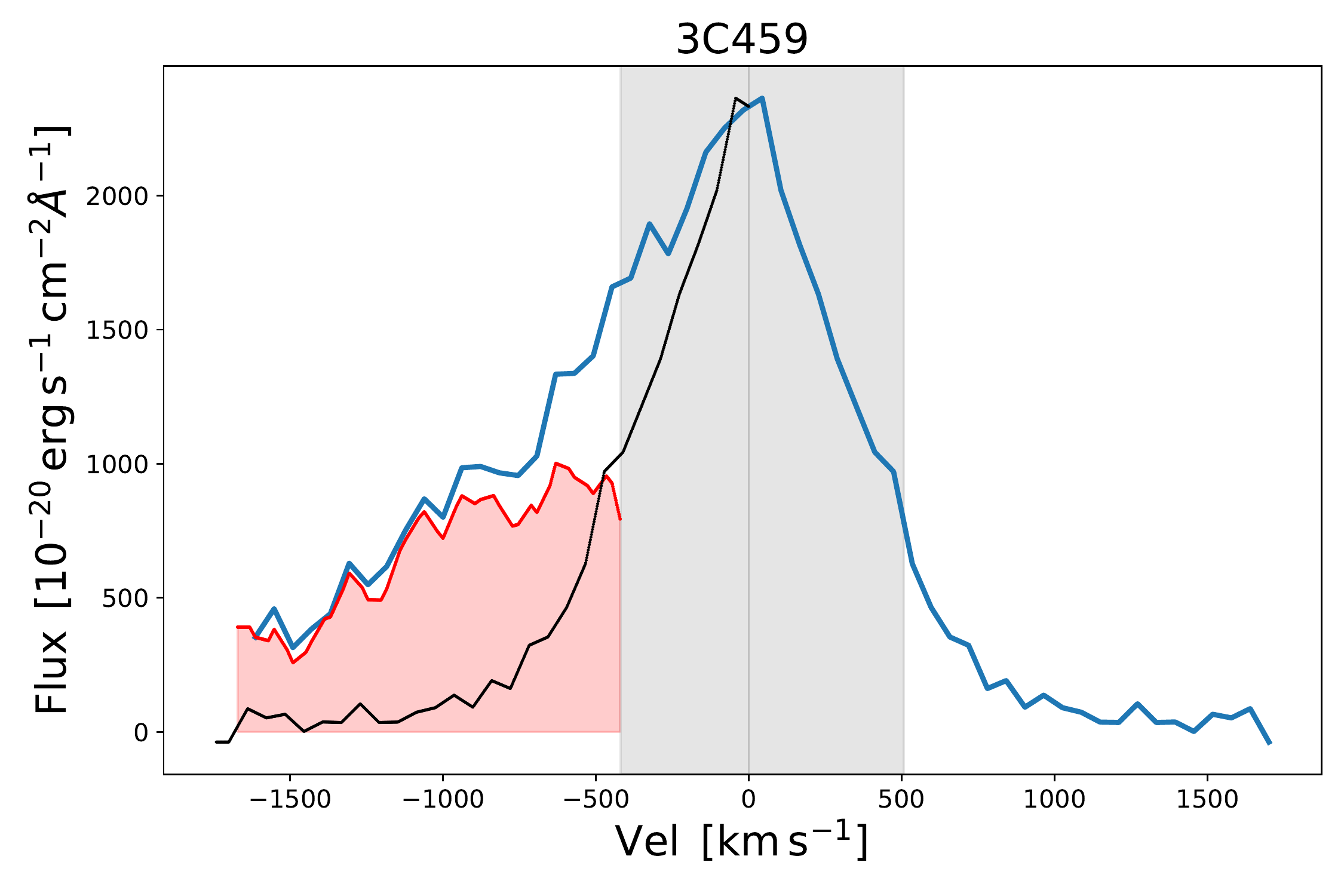}
\caption{- continued.}  
\label{fig:spectra}                                
\end{figure*}                                           

\onecolumn

\begin{figure}
\centering
\includegraphics[width=0.9\textwidth]{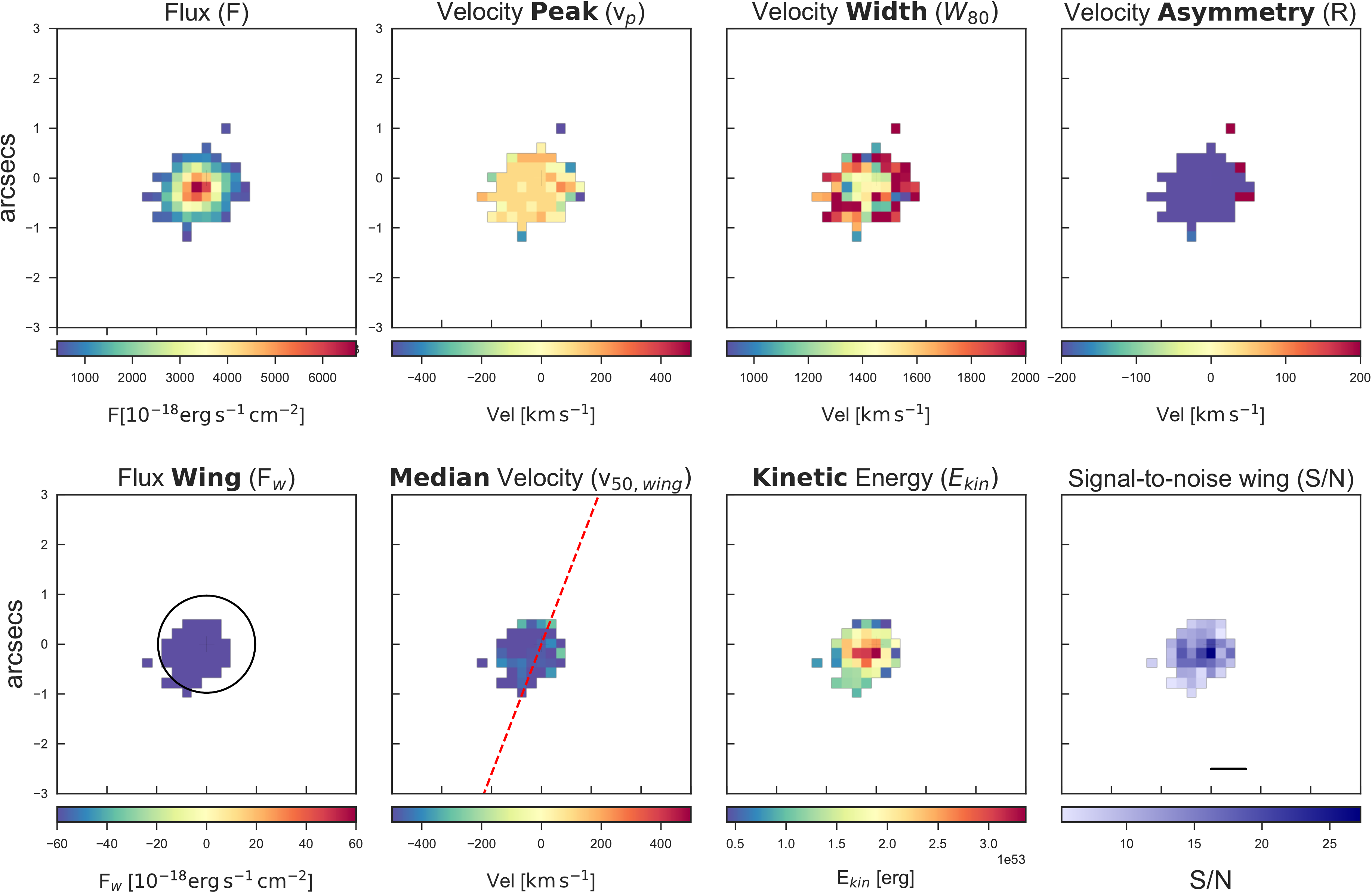}
\caption{3C~015, LEG, 1$\arcsec$ = 1.36 kpc.
The black circle in the first panel has a diameter of 3 times the seeing of
the observations; the dashed line in the second panel marks the radio position
angle. Its nuclear outflow appear to be confined in a
circular region of radius $\sim$ 0.98 kpc and having a blue-shifted
component peaking at $\sim$ -890 $\text {km}\,\text{s}^{-1}$.  The
outflow is well resolved enabling us to observe a kinetic energy
increasing towards the center.}
\label{fig:3C015}
\end{figure}

 \begin{figure}
\centering
\includegraphics[width=0.9\textwidth]{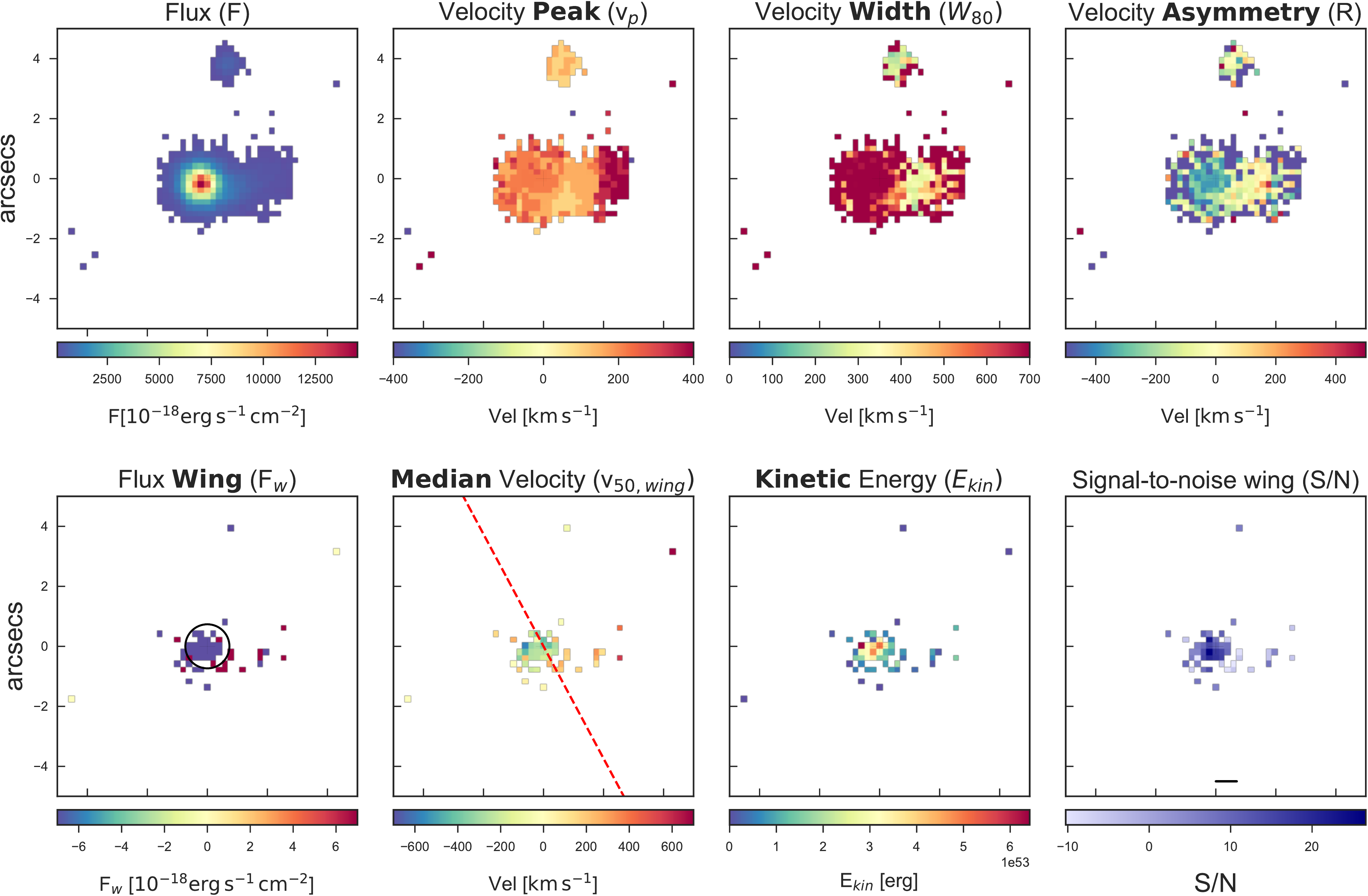}
\caption{3C~017, BLO, 1$\arcsec$ = 3.37 kpc. The black circle in the first panel has a diameter of 3 times the seeing of
the observations; the dashed line in the second panel marks the radio
position angle. the outflowing gas is moving toward us reaching
 negative values of velocity. Unfortunately its compactness (the upper
 limit of its extension from the center is $\sim$ 1.68 kpc) prevent us
 from producing a well resolved map of its kinetic energy.  The
 ionized gas achieves a velocity up to $\sim$ - 1100 $\text
 {km}\,\text{s}^{-1}$. 
}
\label{fig:3C017}
\end{figure}

 \clearpage
 \begin{figure}
\centering

\includegraphics[width=0.9\textwidth]{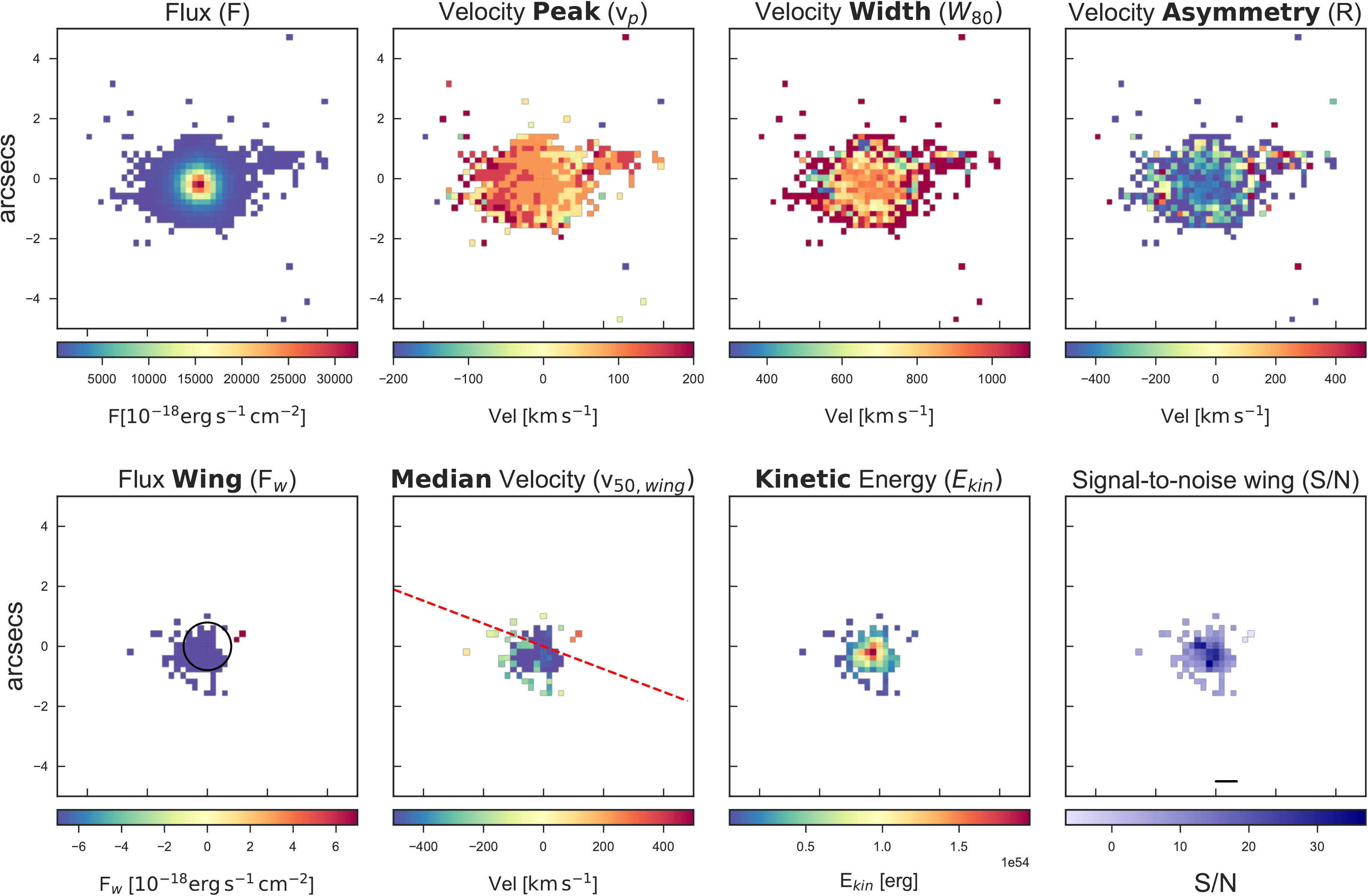}
\caption{3C~018, BLO, 1$\arcsec$ = 3.00 kpc. The black circle in the first panel has a diameter of 3 times the seeing of
the observations; the dashed line in the second panel marks the radio position
angle. The ionized gas appears well resolved, reaching high
 negative velocities up to $\sim$ -1120 $\text
 {km}\,\text{s}^{-1}$. The kinetic energy gradually decrease from the
 central nucleus to the edge ($\sim$ 2.52 kpc). 
}
\end{figure}

  \begin{figure}
\centering
\includegraphics[width=0.9\textwidth]{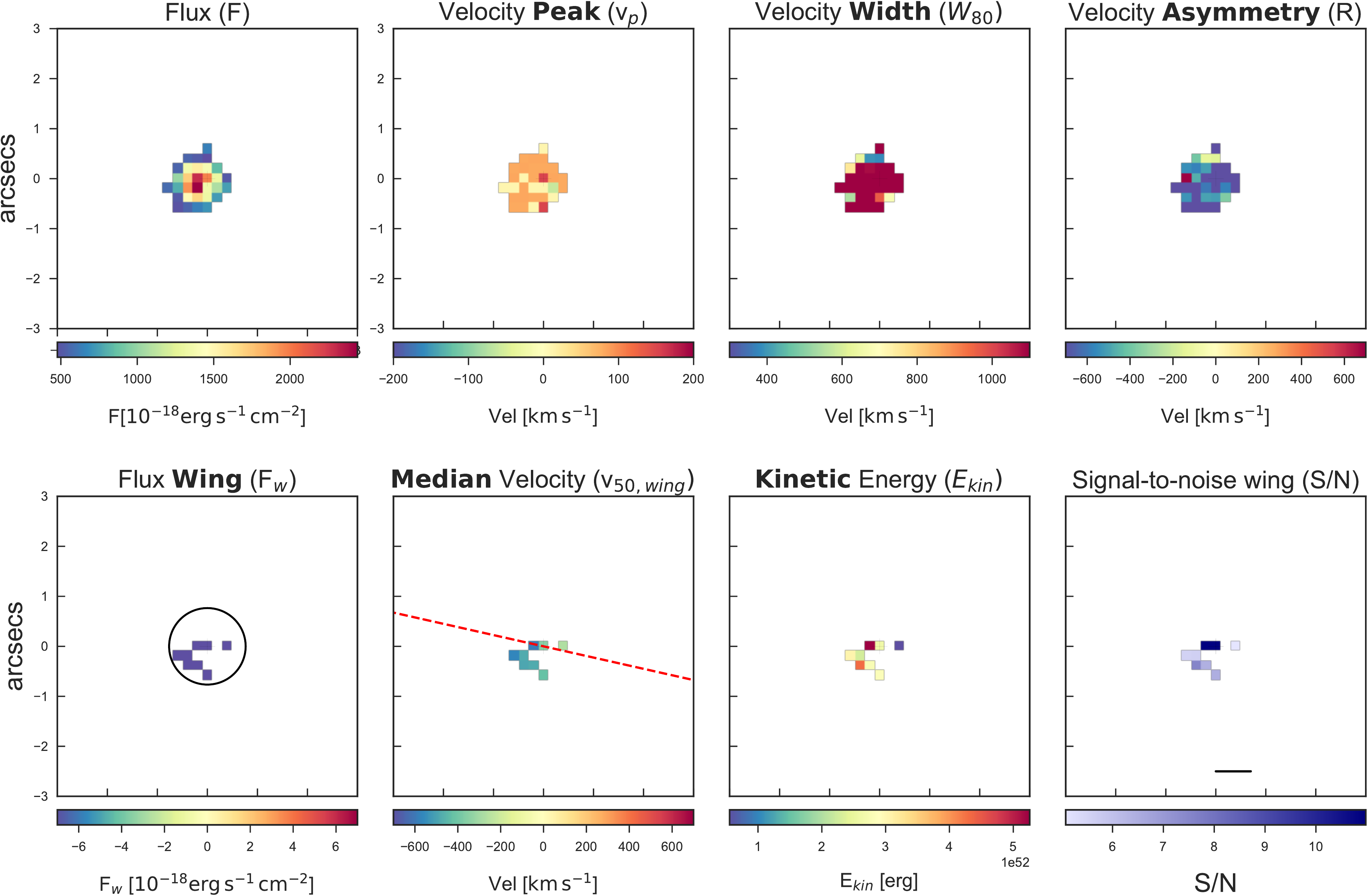}
\caption{3C~029, LEG, 1$\arcsec$ = 0.88 kpc. The black circle in the first panel has a diameter of 3 times the seeing of
the observations; the dashed line in the second panel marks the radio position
angle. Similarly to 3C~017 the emission line are only few
  extended and not well resolved.  The maximum extension is reached in
  the SW region at $\sim$ 0.20 kpc from the nucleus. We observe only
  the blue component of the outflow.}
\end{figure}


  \begin{figure}
\centering
\includegraphics[width=0.9\textwidth]{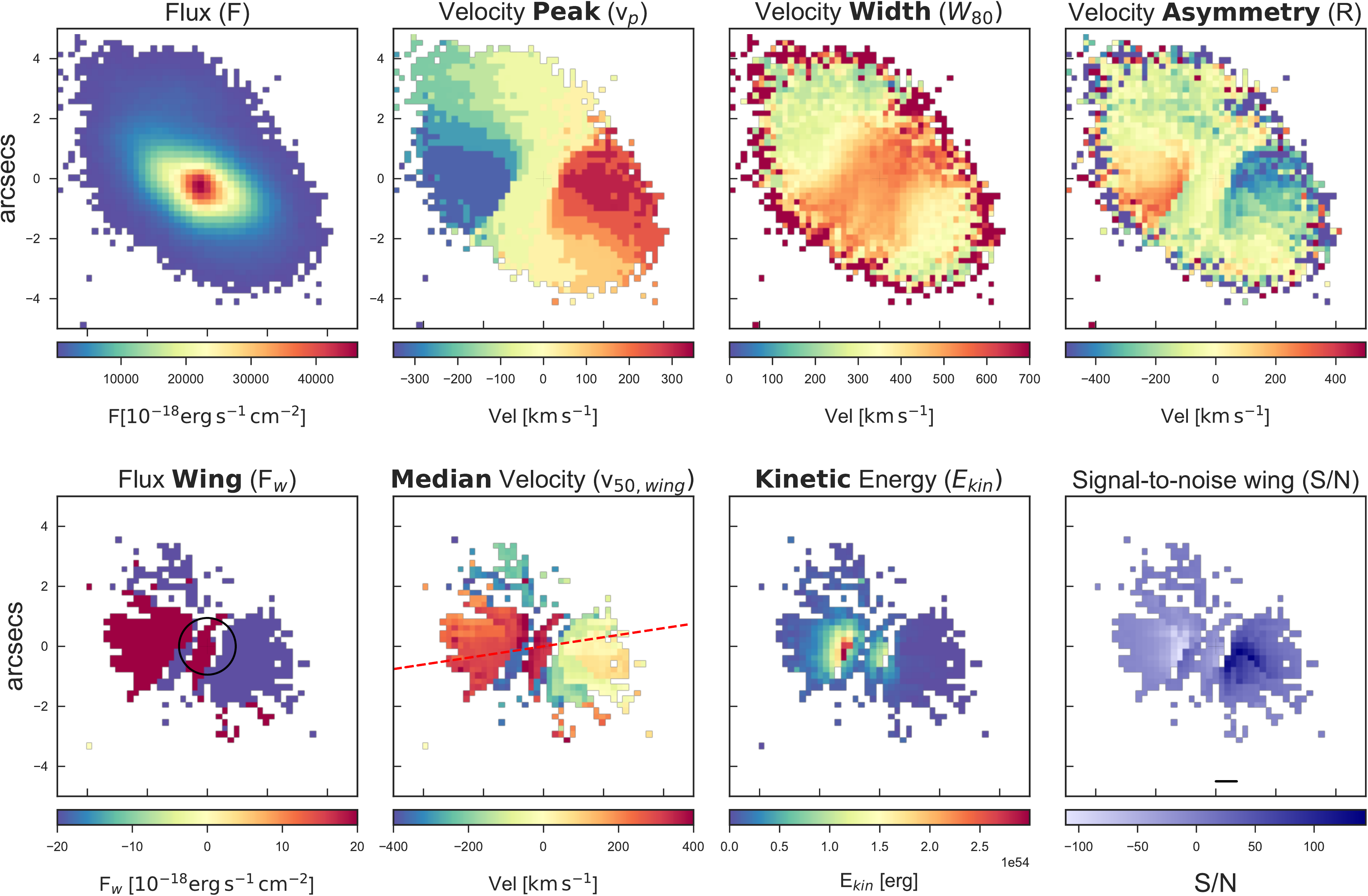}

\caption{3C~033, HEG, 1$\arcsec$ = 1.14 kpc. The black circle in the first panel has a diameter of 3 times the seeing of
the observations; the dashed line in the second panel marks the radio position
angle. We observe an outflow symmetrically extended
  towards the NW and the SE directions out to $\sim$ 3.4" ($\sim$ 3.88
  kpc) from the nucleus. In the NW region the gas assumes positive
  velocities (up to $\sim$ 1800 $\text {km}\,\text{s}^{-1}$) and in
  the SE negative values (up to $\sim$ -2000 $\text
  {km}\,\text{s}^{-1}$), in both cases they slowly decrease along the
  maximum extension. The kinetic energy grows from the most external
  regions close to the center, reaching the maximum values at $\sim$
  1.00 kpc from the nucleus and decreasing again approaching the
  center. This is due to our field of view in which the nucleus is
  obscured by the torus surrounding the central engine. In fact the
  highest values of S/N are measured in the same area in which the
  energy gets the greatest values.}
\end{figure}

\begin{figure}
\centering
\includegraphics[width=0.9\textwidth]{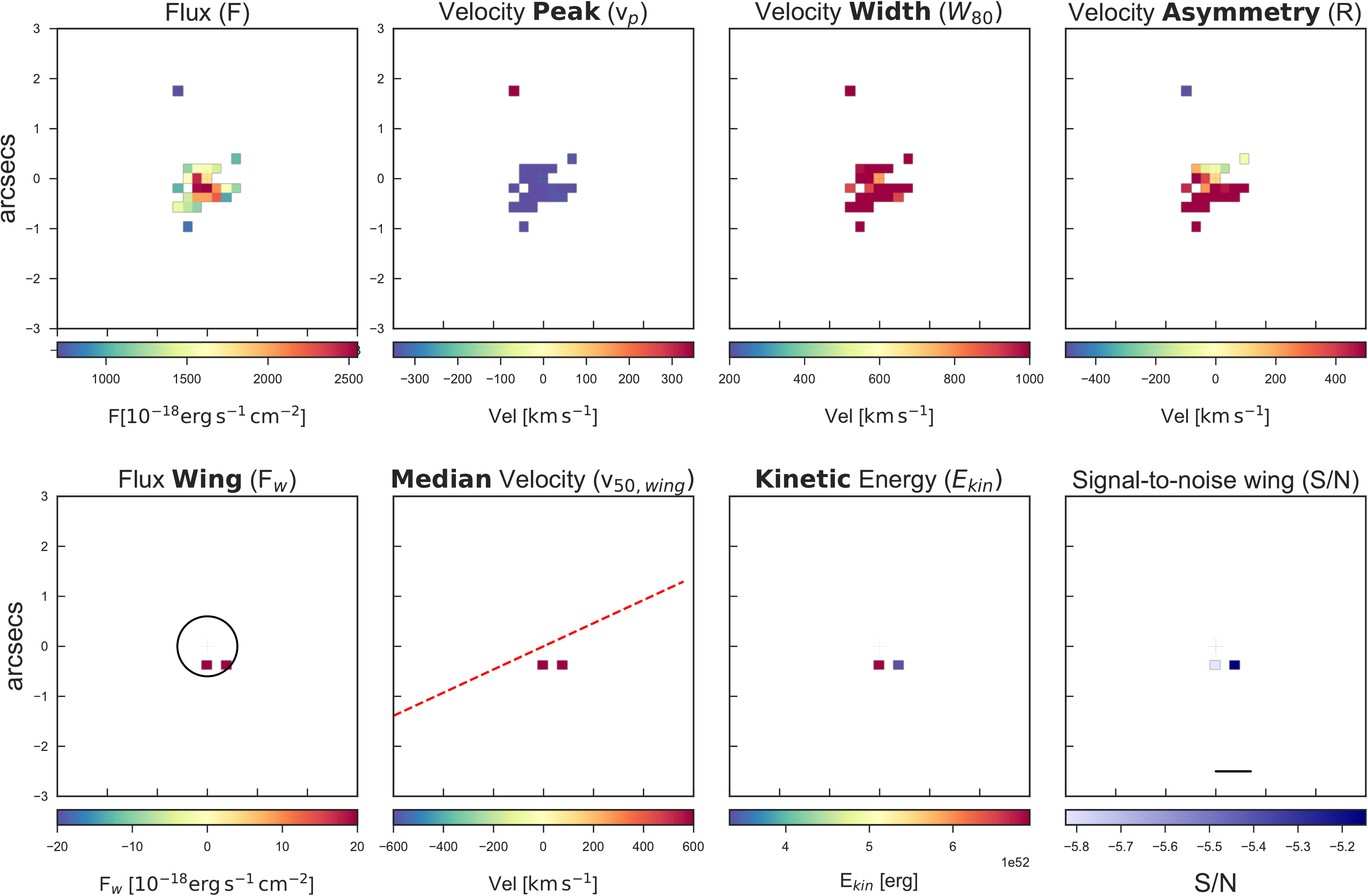}
\caption{3C040, LEG, 1$\arcsec$ = 0.36 kpc. The black circle in the first panel has a diameter of 3 times the seeing of
the observations; the dashed line in the second panel marks the radio position
angle.}
\end{figure}

  \begin{figure}
\centering
\includegraphics[width=0.9\textwidth]{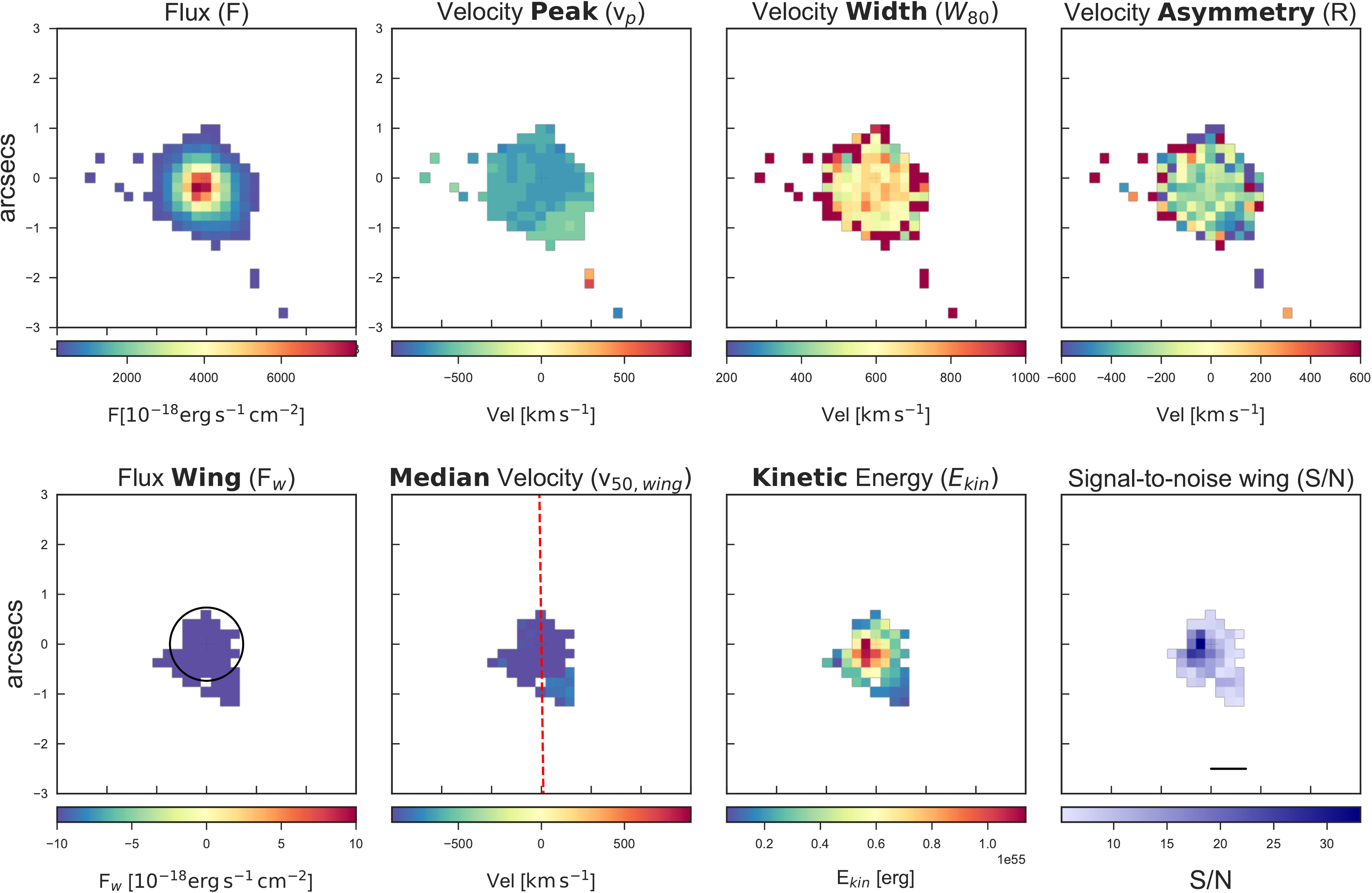}
\caption{3C~063, HEG, 1$\arcsec$ = 1.36 kpc. The black circle in the first panel has a diameter of 3 times the seeing of
the observations; the dashed line in the second panel marks the radio position
angle. The gas in outflow shows a quite circular shape
  with a radius of $\sim$ 1.14 kpc with a elongation in the SE
  direction ($\sim$ out to 2.12 kpc). We observe just the blue
  component of the [O~III] with a velocity down to $\sim$ 1900 $\text
  {km}\,\text{s}^{-1}$.}
  \end{figure}


    \begin{figure}
\centering
\includegraphics[width=0.9\textwidth]{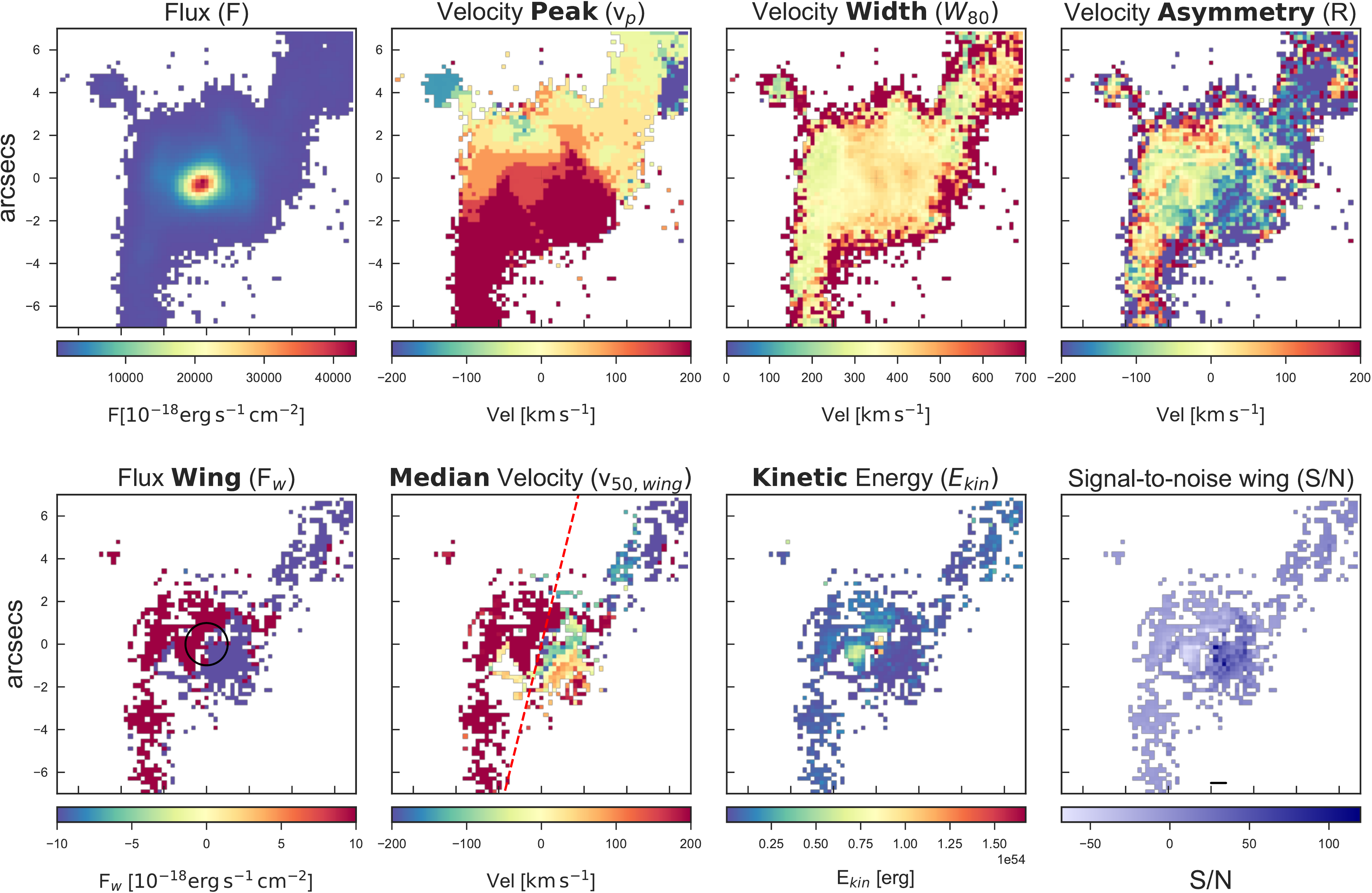}
\caption{3C~079, HEG, 1$\arcsec$ = 3.75 kpc. The black circle in the first panel has a diameter of 3 times the seeing of
the observations; the dashed line in the second panel marks the radio position
angle. The outflow gas is confined in a circular region
    of radius $\sim$ 7.5 kpc, from here two filamentary structures
    extend beyond $\sim$ 36 kpc and $\sim$ 26 kpc respectively towards
    NE and SW. The filaments appears as the division axis between the
    gas launched in our direction (upper area) and the one in the
    opposite side (bottom area). The central region of the source is
    obscured, prevent us to observe the energetic of the nucleus. The
    maximum velocity achieved by the [O~III] is $\sim$ 2450 $\text
    {km}\,\text{s}^{-1}$, that represents the maximum velocity reached
    by the outflow in the whole sample and also by the 3C~180.}
    \end{figure}

   \begin{figure}
\centering
\includegraphics[width=0.9\textwidth]{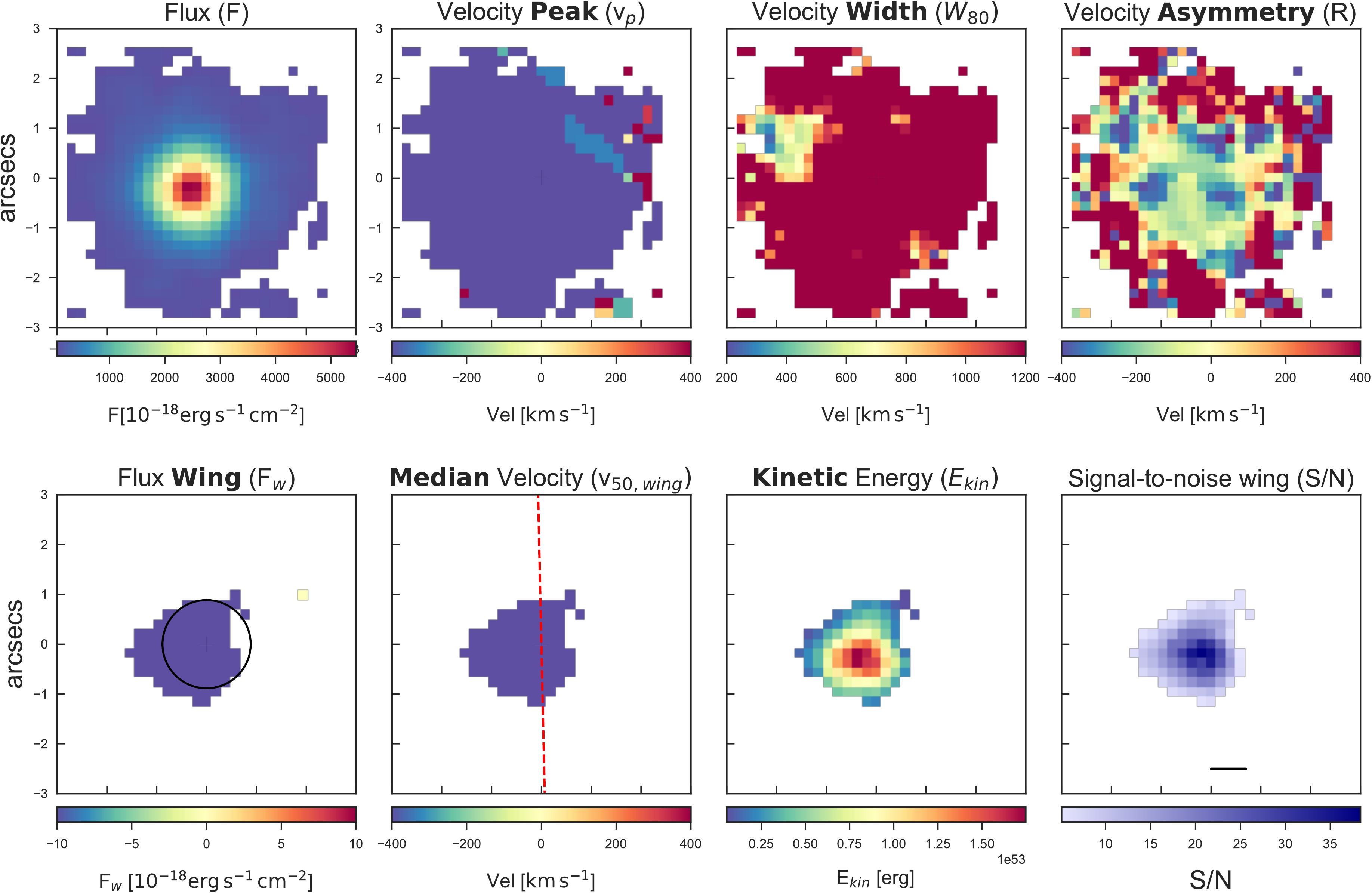}
\caption{3C~088, LEG, 1$\arcsec$ = 0.60 kpc. The black circle in the first panel has a diameter of 3 times the seeing of
the observations; the dashed line in the second panel marks the radio position
angle. The source is well resolved, the kinetic energy of
   the gas increases in the inner part and its velocities are negative
   out to $\sim$ -1360 $\text {km}\,\text{s}^{-1}$. The outflow not
   expands beyond 0.72 kpc from the nucleus.}
\end{figure}

   \begin{figure}
\centering
\includegraphics[width=0.9\textwidth]{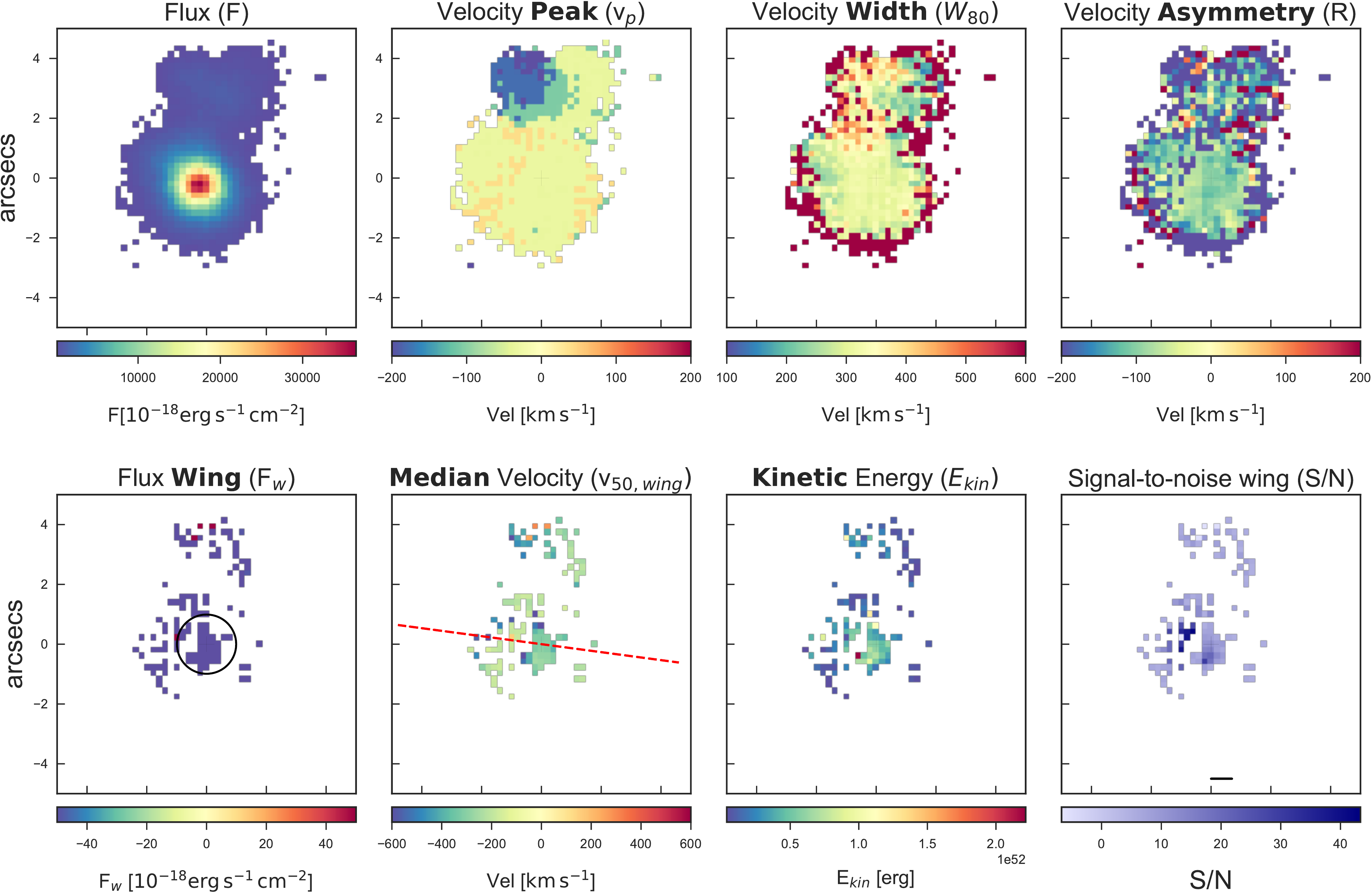}
\caption{3C~098, LEG, 1$\arcsec$ = 0.60 kpc. The black circle in the first panel has a diameter of 3 times the seeing of
the observations; the dashed line in the second panel marks the radio position
angle.  The MUSE field-of-view covers only $\sim$ 1/4 of
   the extension of the radio source. Even if the map is not well
   resolved we can see that the ionized gas of the outflow reaching
   negative velocities beyond $\sim$ 2000 $\text {km}\,\text{s}^{-1}$
   and it expands into an irregular structure.}
\end{figure}

   \begin{figure}
\centering
\includegraphics[width=0.9\textwidth]{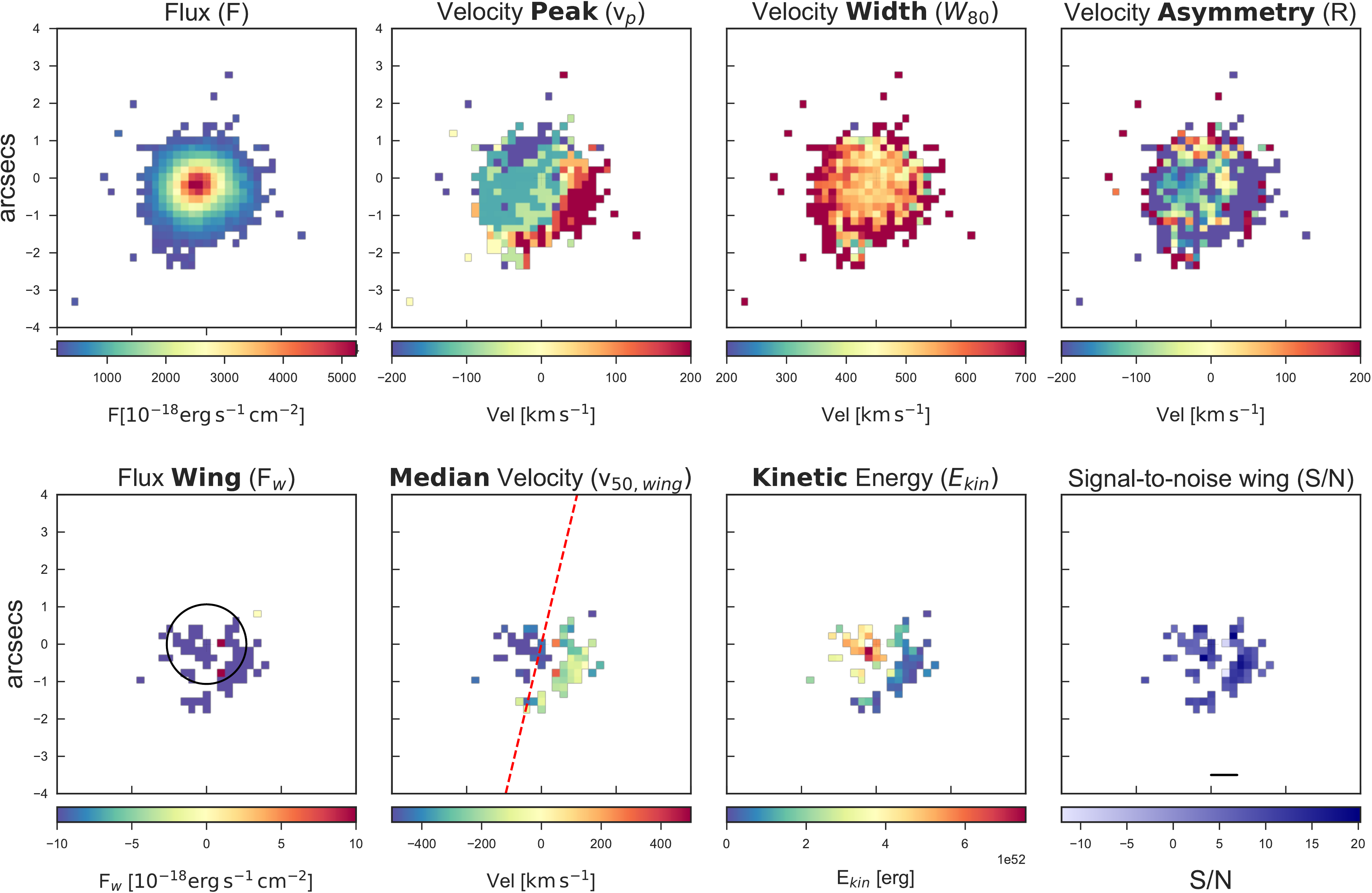}
\caption{3C~105, BLO, 1$\arcsec$ = 1.63 kpc. The black circle in the first panel has a diameter of 3 times the seeing of
the observations; the dashed line in the second panel marks the radio position
angle.}
\end{figure}

 \begin{figure}
\centering
\includegraphics[width=0.9\textwidth]{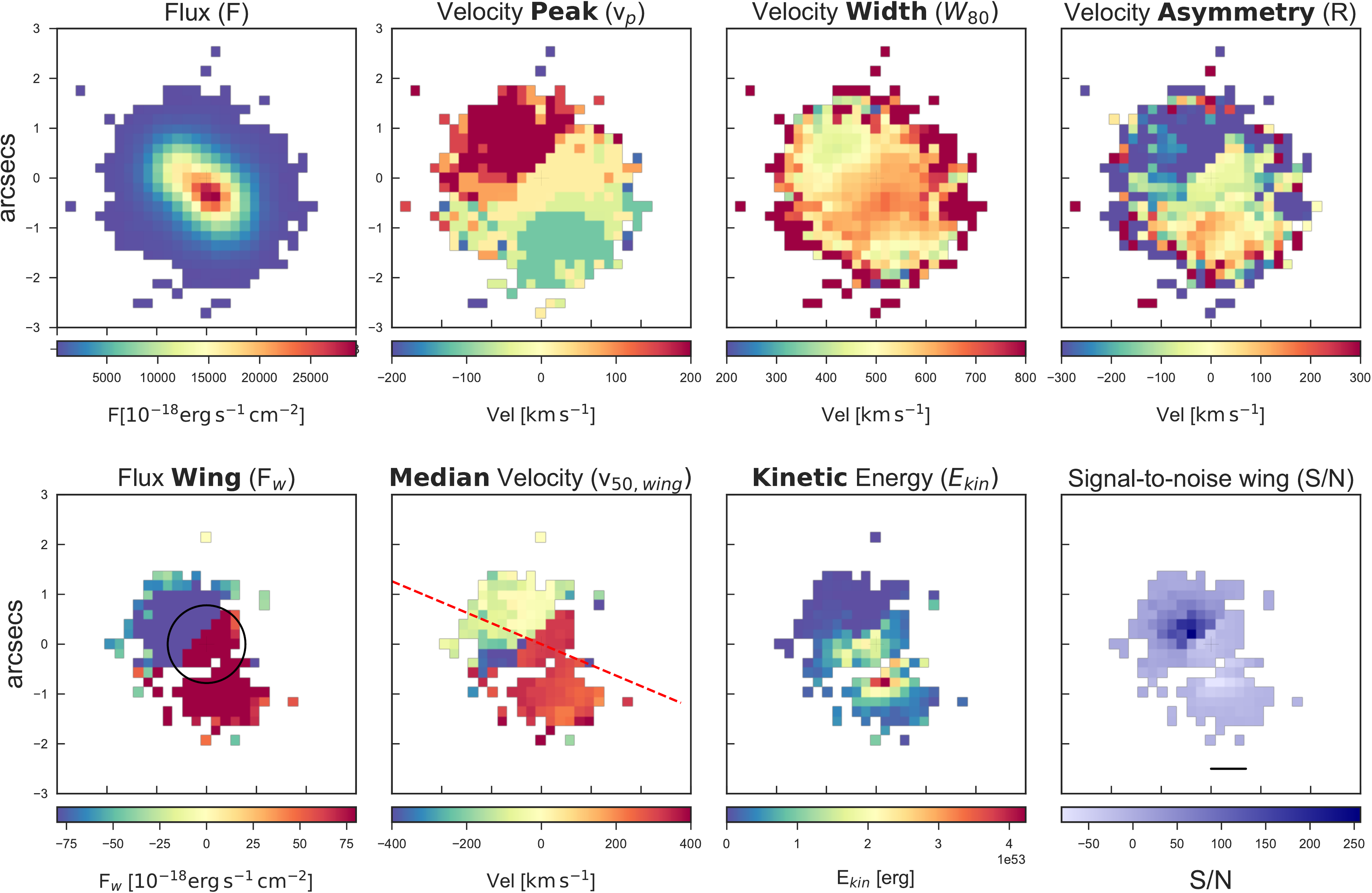}
\caption{3C~135, HEG, 1$\arcsec$ = 3.37 kpc. The black circle in the first panel has a diameter of 3 times the seeing of
the observations; the dashed line in the second panel marks the radio position
angle. We find a shape already observed in the 3C~033, an
 extended outflow characterized by two geometrically symmetrical lobes
 that cover $\sim$ 1.8" ($\sim$ 4 kpc). The lobe that extends in the
 NW is characterized by high negative velocities of the gas, with a
 maximum value that amount to $\sim$ -1940 $\text
 {km}\,\text{s}^{-1}$. In the other lobe that evolves towards SE the
 gas assumes lower positive velocities down to $\sim$ 570 $\text
 {km}\,\text{s}^{-1}$. In the nucleus we measured a value of $\sim$
 290 $\text {km}\,\text{s}^{-1}$ for the median velocity of the gas,
 anyway it is obscured preventing us to explore its energetic. Also
 for this source the radio image exceed the MUSE field of view.}

 \end{figure}

 \begin{figure}
\centering
\includegraphics[width=0.9\textwidth]{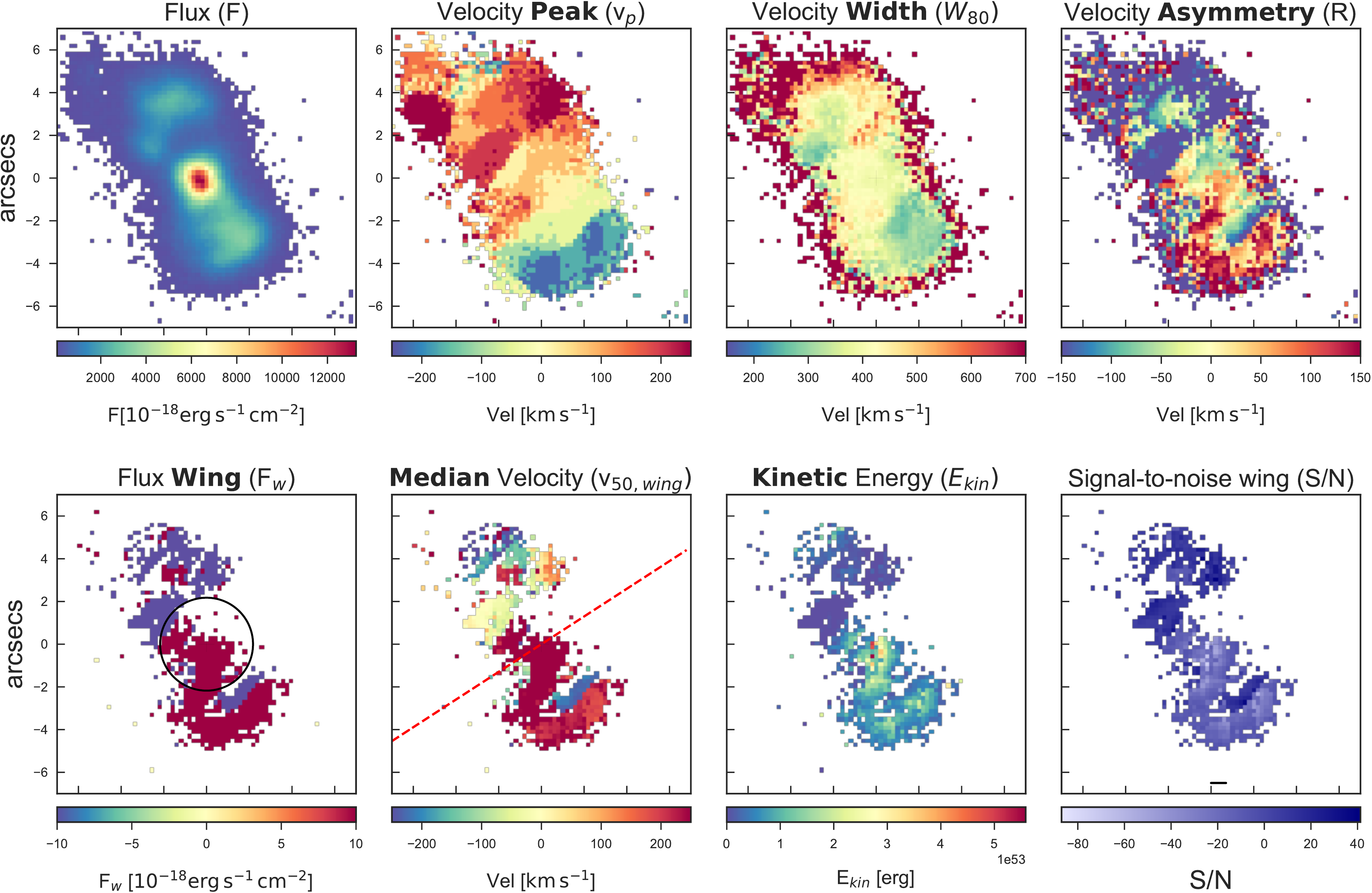}
\caption{3C~180, HEG, 1$\arcsec$ = 3.37 kpc. The black circle in the first panel has a diameter of 3 times the seeing of
the observations; the dashed line in the second panel marks the radio position
angle. In the inner region the ionized gas is quite
 compact (up to $\sim$ 8.5 kpc from the nucleus), outside the central
 region (from $\sim$ 8.5 kpc to $\sim$ 15 kpc) the most distant gas is
 located in a larger area especially in the SE section, here the gas
 is dominated by high positive velocities that reach $\sim$ 2450
 $\text {km}\,\text{s}^{-1}$. This represents the maximum velocity
 reached by the outflow in the whole sample and also by the 3C~79. In
 this area we observe a thin segment of gas characterized by negative
 velocities. The NW area is dominated by the blue component.}
\end{figure}

   \begin{figure}
\centering

\includegraphics[width=0.9\textwidth]{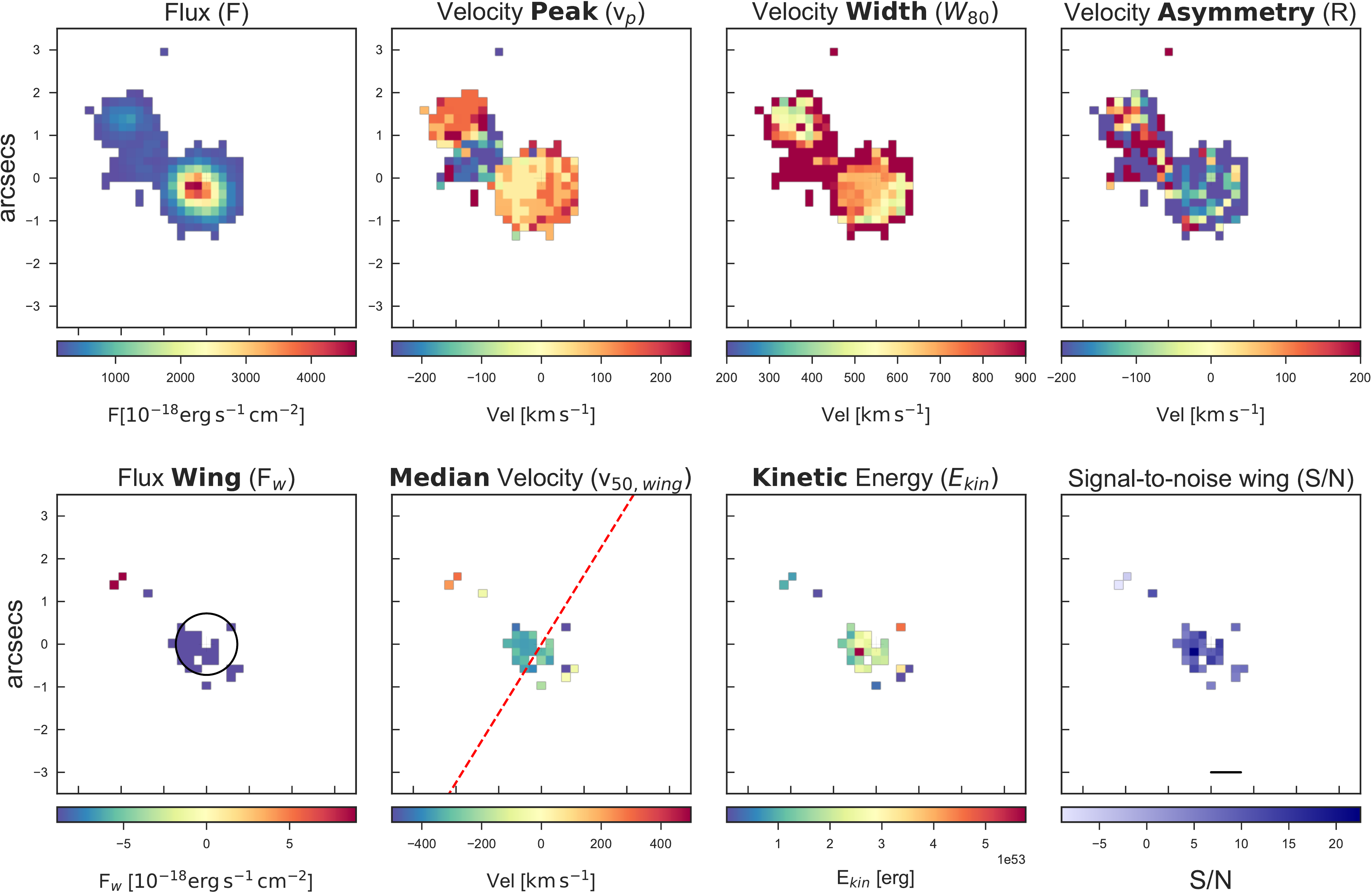}
\caption{3C~196.1, BLO, 1$\arcsec$ = 3.12 kpc. The black circle in the first panel has a diameter of 3 times the seeing of
the observations; the dashed line in the second panel marks the radio position
angle. The outflowing gas of this source is very compact,
 it is enclose in a region of radius $\sim$ 1.8 kpc. We observe the
 blue component of the [O~III] with a maximum velocity of $\sim$ -640
 $\text {km}\,\text{s}^{-1}$, unfortunately the image is not well
 resolved. In this case the MUSE field of view is larger than the
 radio image.}
\end{figure}


\begin{figure}
\centering
\includegraphics[width=0.9\textwidth]{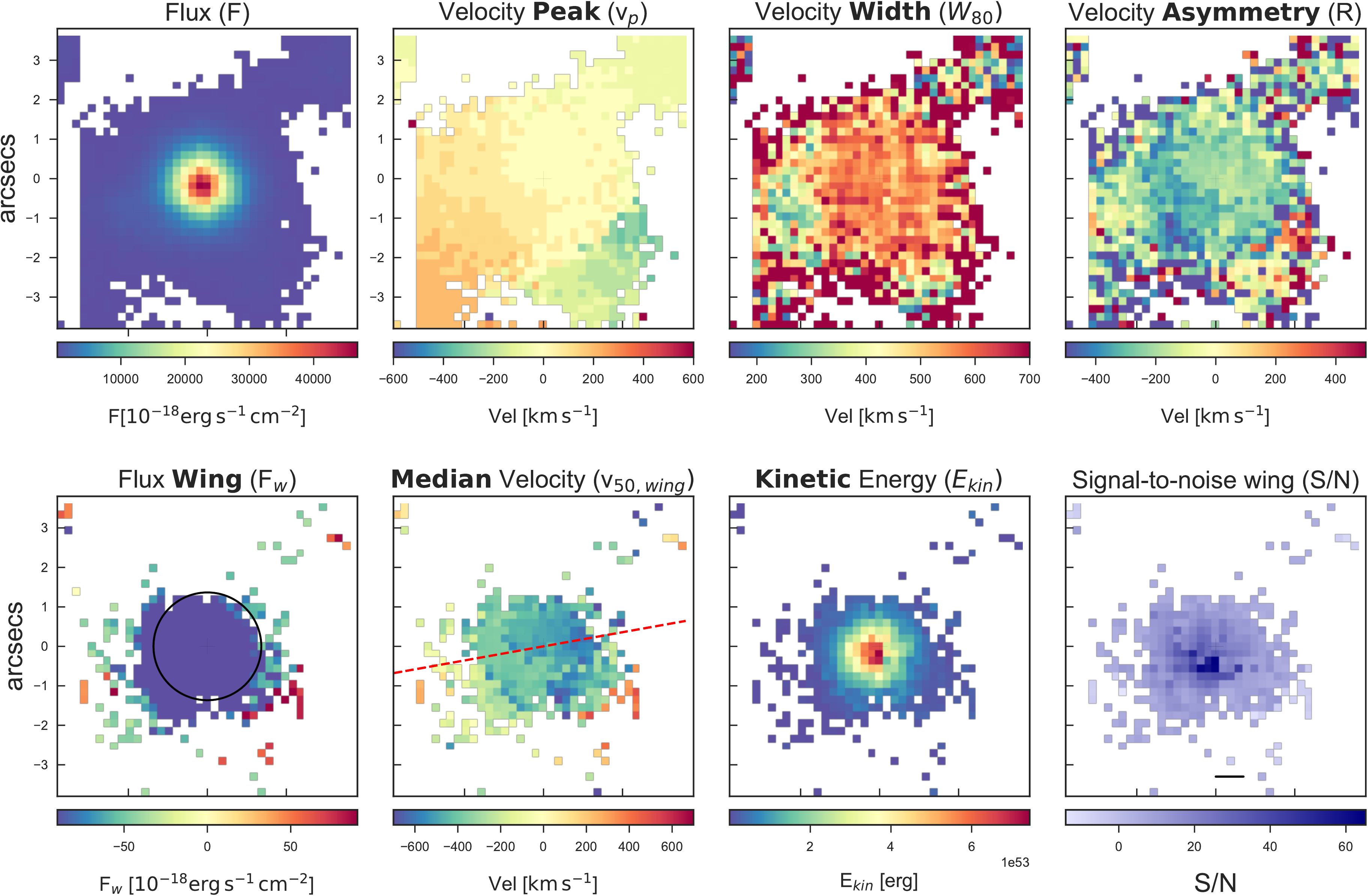}
\caption{3C~227, BLO, 1$\arcsec$ = 1.58 kpc. The black circle in the first panel has a diameter of 3 times the seeing of
the observations; the dashed line in the second panel marks the radio position
angle.  the emission lines are detected in a compact but
  resolved region extending by $\sim$ 2.53 kpc from the nucleus. The
  ionized gas assumes negative velocities with the greatest value at
  -1920 $\text {km}\,\text{s}^{-1}$ and the kinetic energy increases
  toward the center. The radio image terminate with two symmetrical
  hotspots exceeding the MUSE field of view.}
\end{figure}

\begin{figure}
\centering
\includegraphics[width=0.9\textwidth]{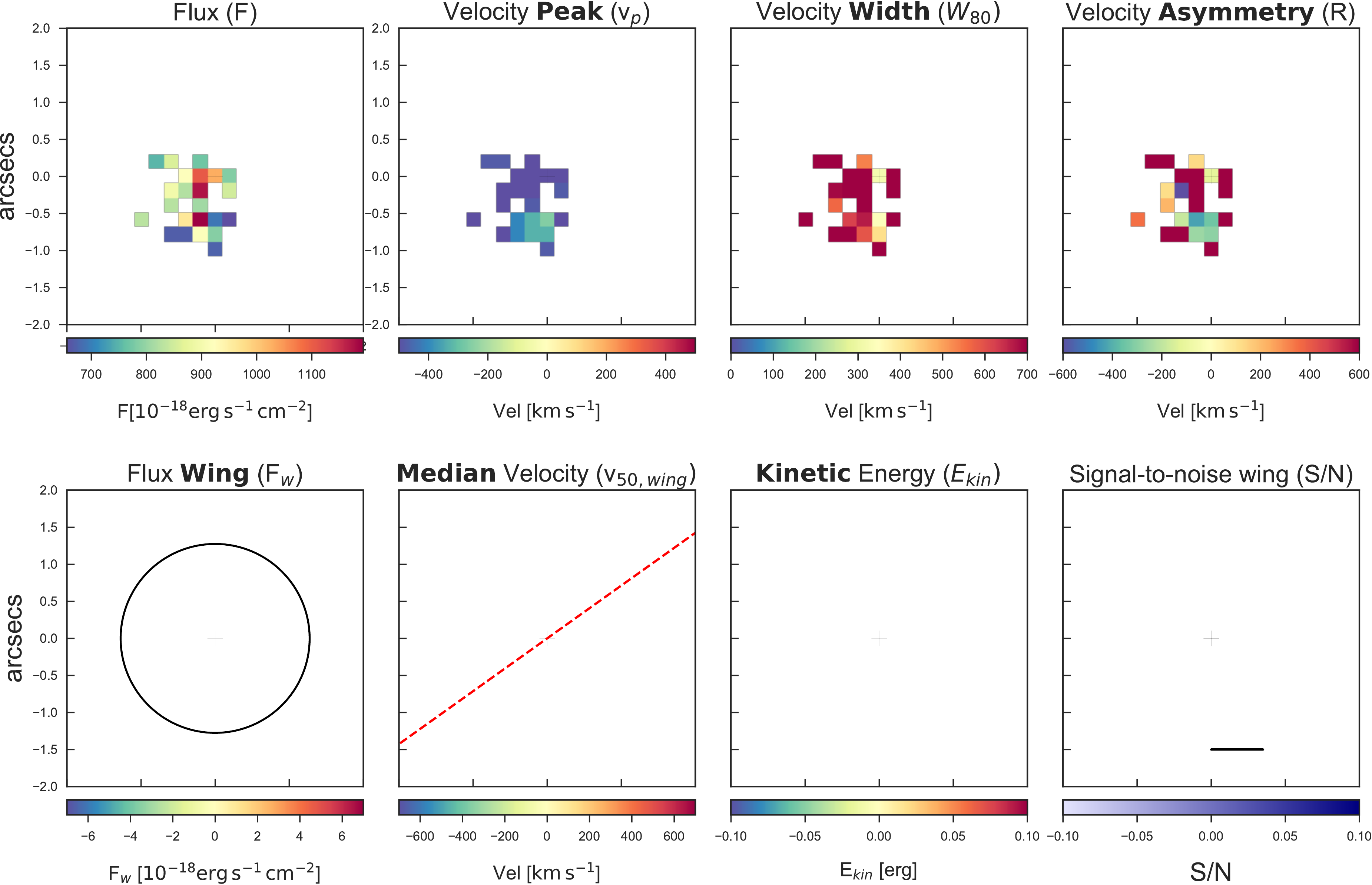}
\caption{3C264, LEG, 1$\arcsec$ = 0.44 kpc. The black circle in the first panel has a diameter of 3 times the seeing of
the observations; the dashed line in the second panel marks the radio position
angle.}
\end{figure}

\begin{figure}
\centering
\includegraphics[width=0.9\textwidth]{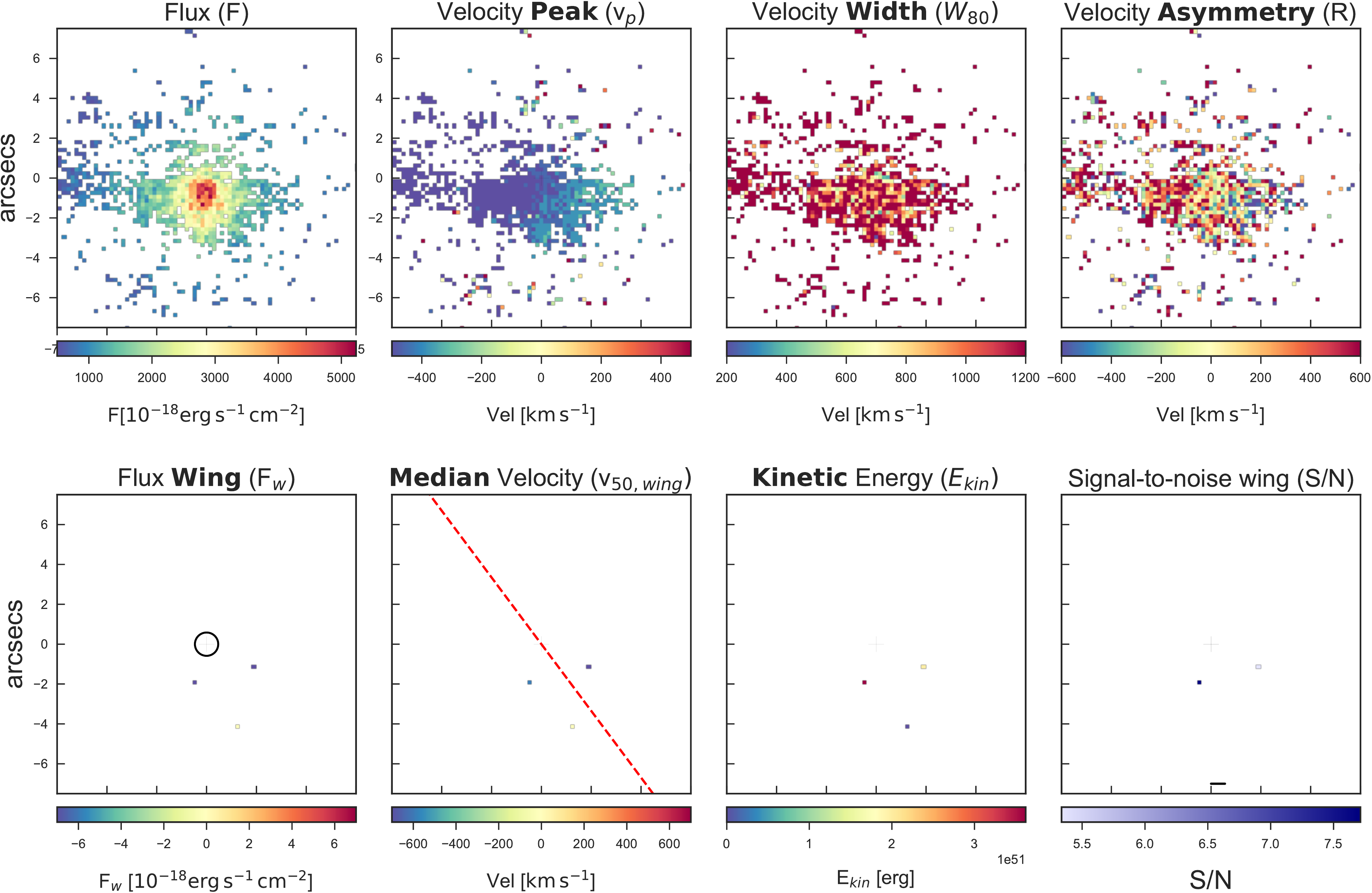}
\caption{3C272.1, LEG, 1$\arcsec$ = 0.07 kpc. The black circle in the first panel has a diameter of 3 times the seeing of
the observations; the dashed line in the second panel marks the radio position
angle.}
\end{figure}

  

\begin{figure}
\centering
\includegraphics[width=0.9\textwidth]{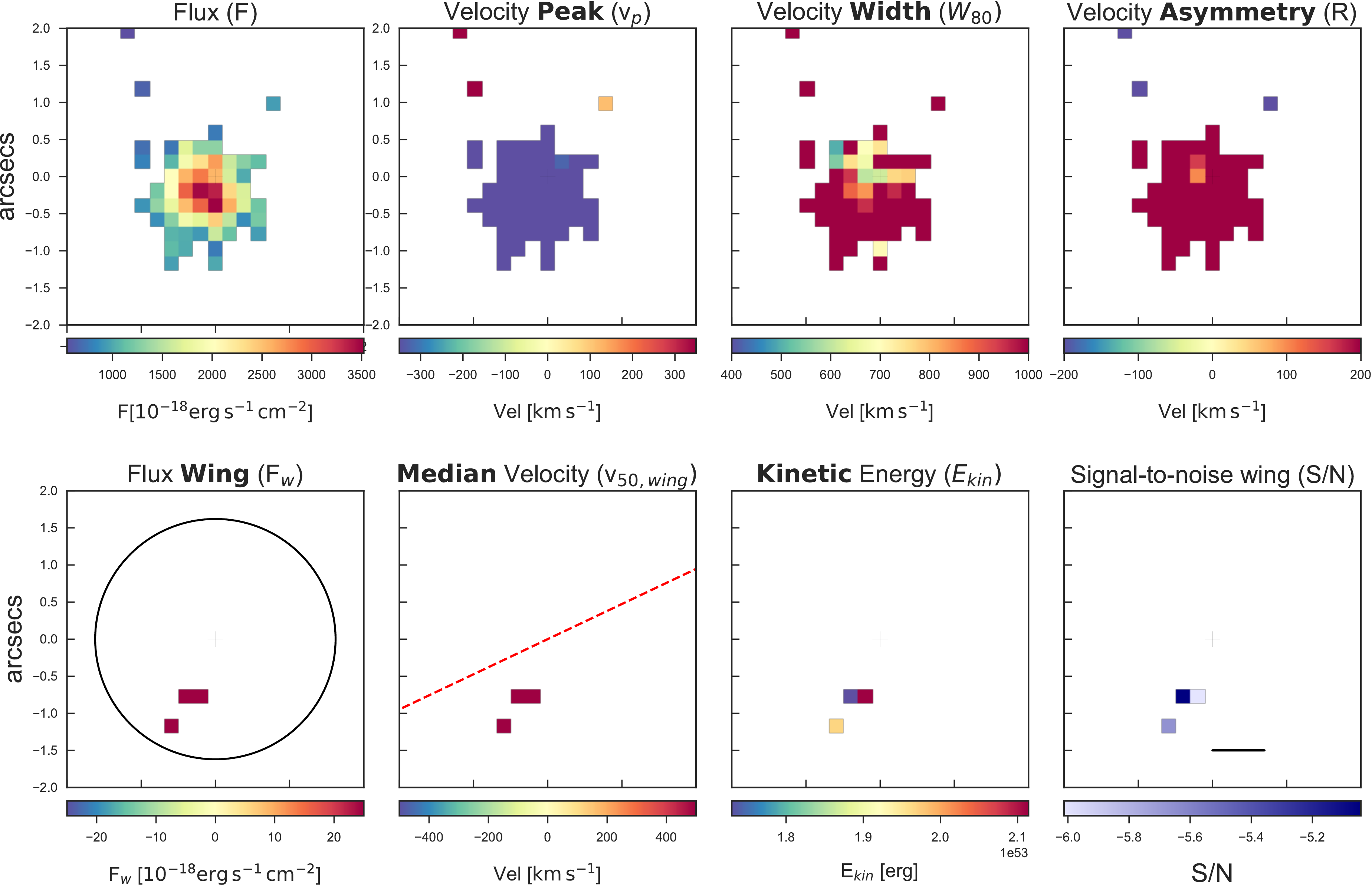}
\caption{3C296, LEG, 1$\arcsec$ = 0.49 kpc. The black circle in the first panel has a diameter of 3 times the seeing of
the observations; the dashed line in the second panel marks the radio position
angle.}
\end{figure}

\begin{figure}
\centering
\includegraphics[width=0.9\textwidth]{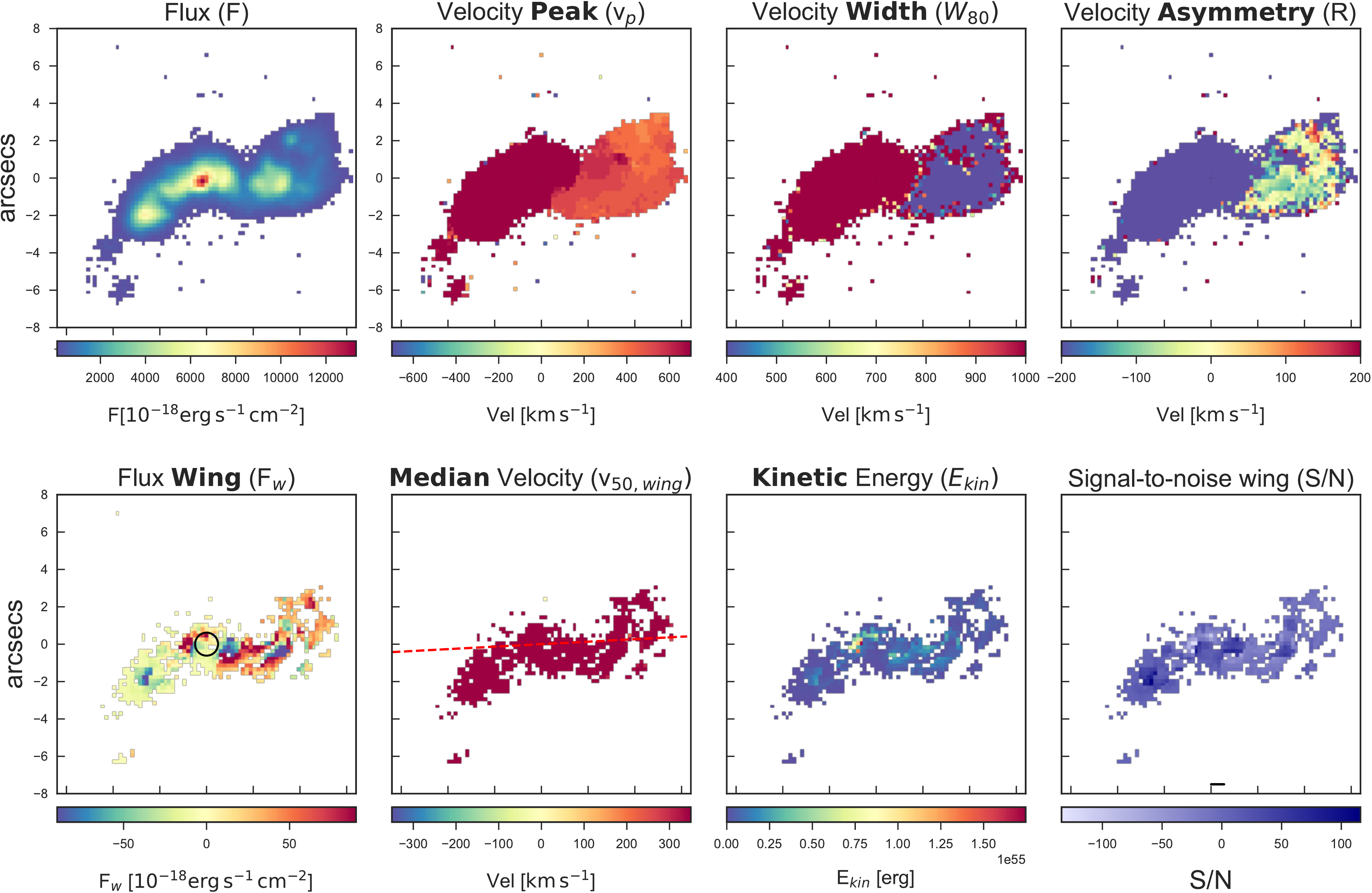}
\caption{3C~300, HEG, 1$\arcsec$ = 3.88 kpc. The black circle in the first panel has a diameter of 3 times the seeing of
the observations; the dashed line in the second panel marks the radio position
angle. This source shows an elliptical shape (the most
extended f our sample) with a residual filamentary structure in the NE
direction. The whole major axis of the ellipse has a length of $\sim$
7.2" ($\sim$ 27.94 kpc), from its east end the emission lines extend
beyond $\sim$ 12" ($\sim$ 9.31 kpc). The flux of the gas is major in
correspondence with the nucleus and it peaks again at $\sim$ 14.25 kpc
toward SW from the center. The outflow is mainly dominated by the blue
component which maximum velocity is $\sim$ -2160 $\text
{km}\,\text{s}^{-1}$. The red component of the gas is detected in the
NE filaments with a velocity that arrives up to $\sim$ 1450 $\text
{km}\,\text{s}^{-1}$. The kinetic energy presents two peaks located in
the same position as those of the flux. This double peak is probably a
sign of two sources from which the outflow originates.}
\end{figure}

\begin{figure}
\centering
\includegraphics[width=0.9\textwidth]{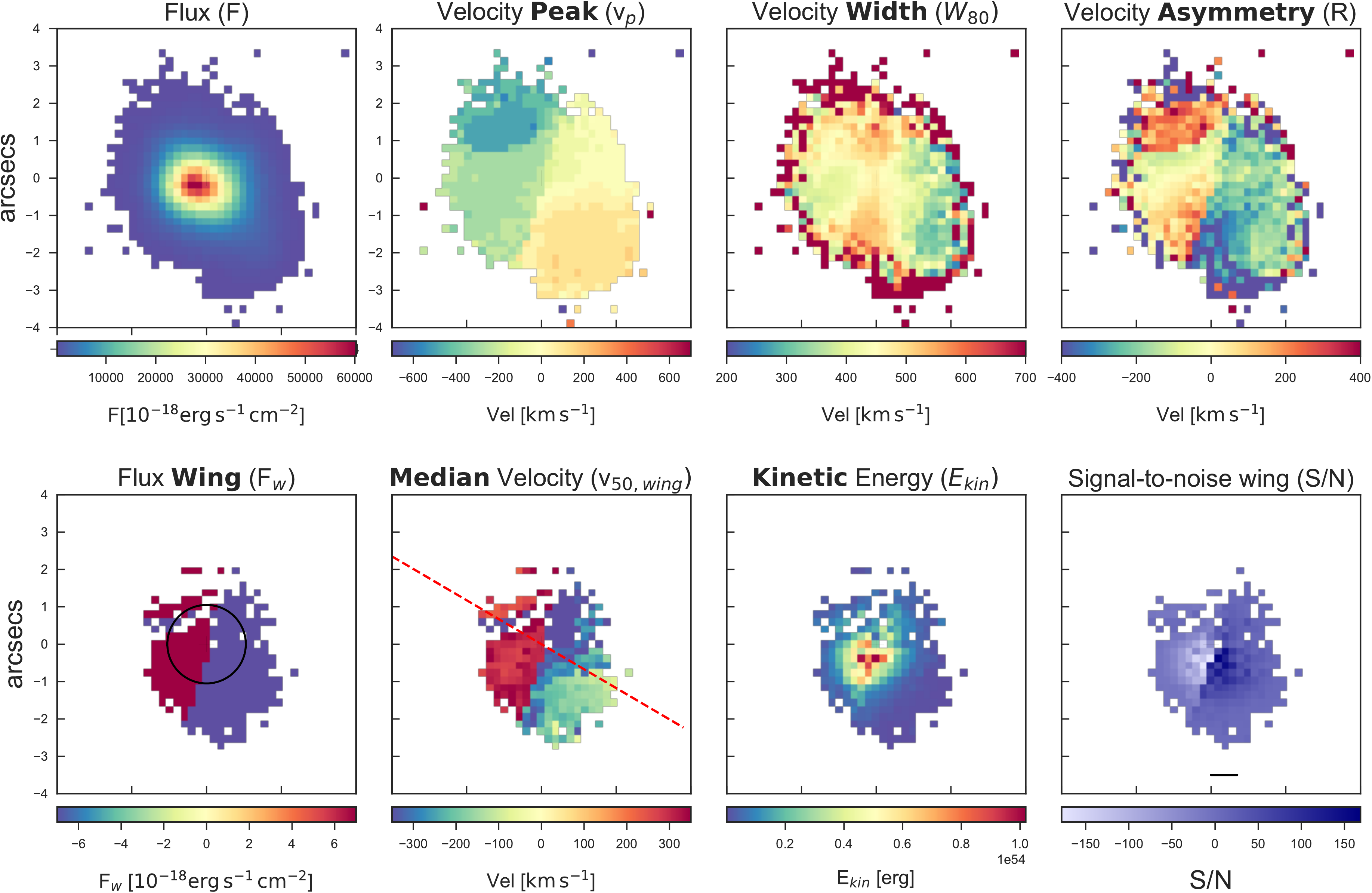}
\caption{3C~327, HEG, 1$\arcsec$ = 1.87 kpc. The black circle in the first panel has a diameter of 3 times the seeing of
the observations; the dashed line in the second panel marks the radio position
angle. Both component of the outflow are detected with a S/N
above 100. To the west the ionized gas reaches positive velocities and
in the opposite side negatives, achieving a maximum values by $\sim$
1440 $\text {km}\,\text{s}^{-1}$. The maps are well resolved, the
energy peaks at the center then decreases more and more towards the
external region (up to $\sim$ 3.74 kpc). }
\end{figure}

\begin{figure}
\centering
\includegraphics[width=0.9\textwidth]{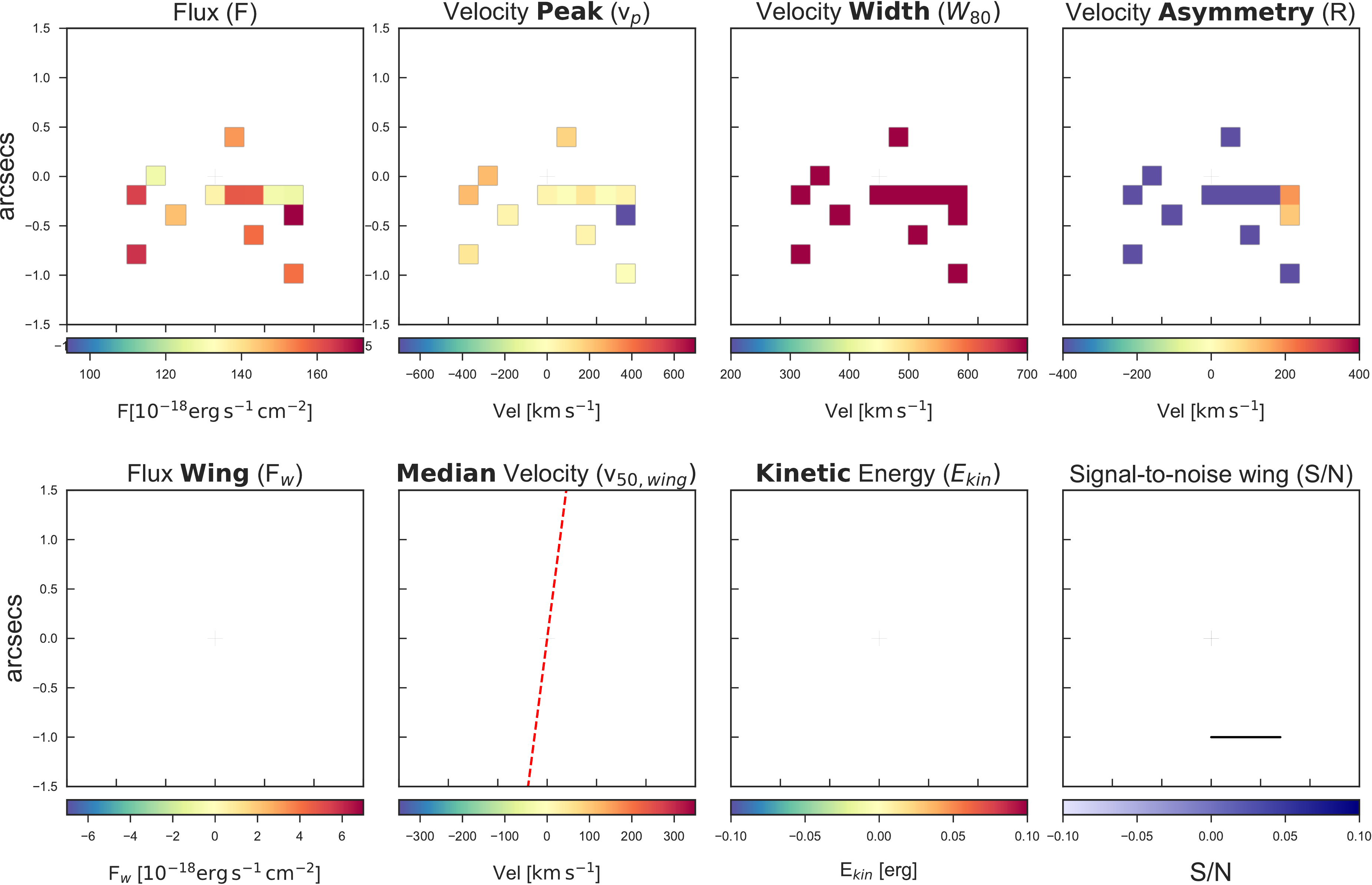}
\caption{3C348, LEG, 1$\arcsec$ = 2.56 kpc. The black circle in the first panel has a diameter of 3 times the seeing of
the observations; the dashed line in the second panel marks the radio position
angle.}
\end{figure}

\begin{figure}
\centering
\includegraphics[width=0.9\textwidth]{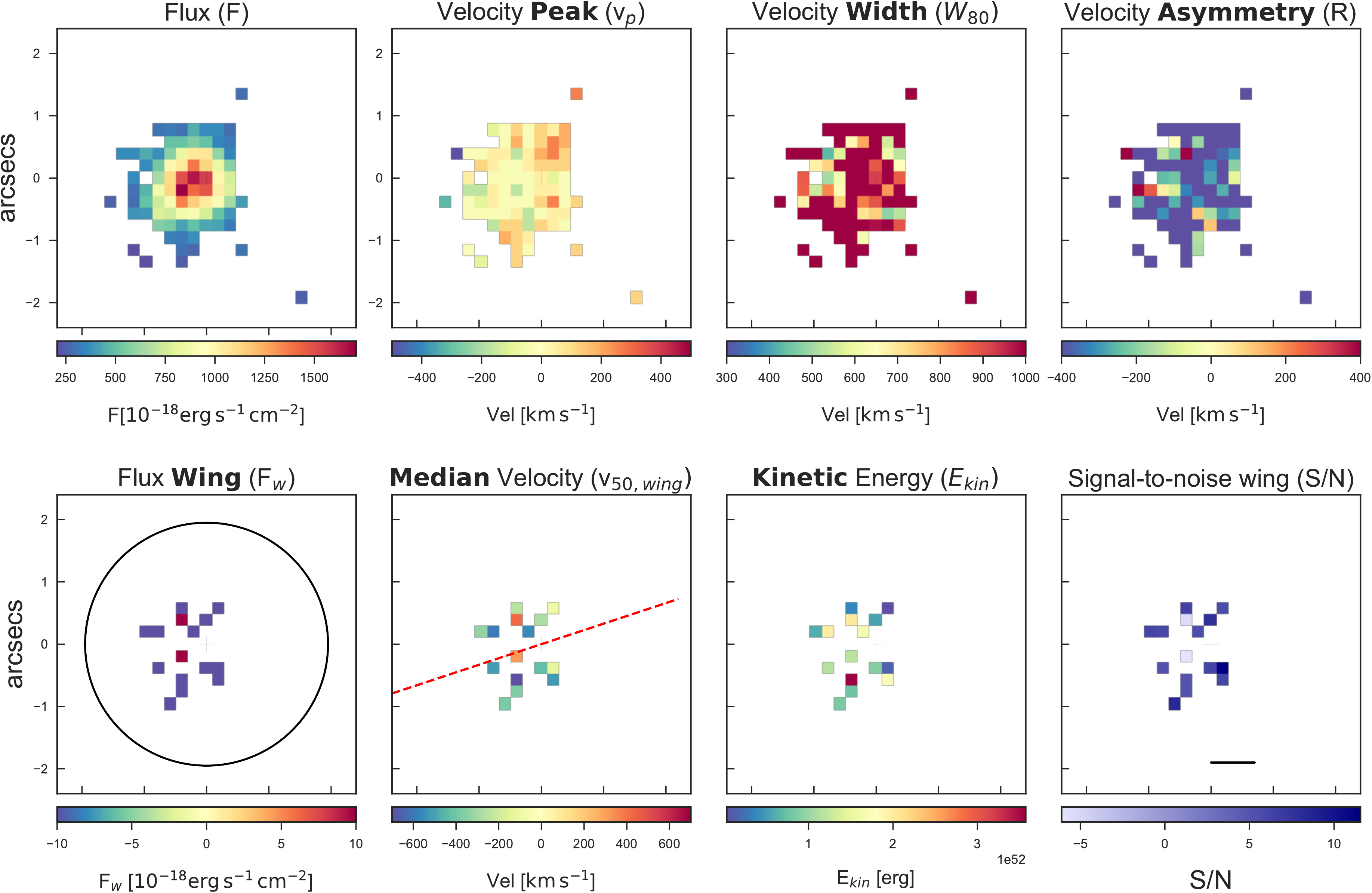}
\caption{3C353, LEG, 1$\arcsec$ = 0.59 kpc. The black circle in the first panel has a diameter of 3 times the seeing of
the observations; the dashed line in the second panel marks the radio position
angle.}
\end{figure}

\begin{figure}
  \centering
  \includegraphics[width=0.9\textwidth]{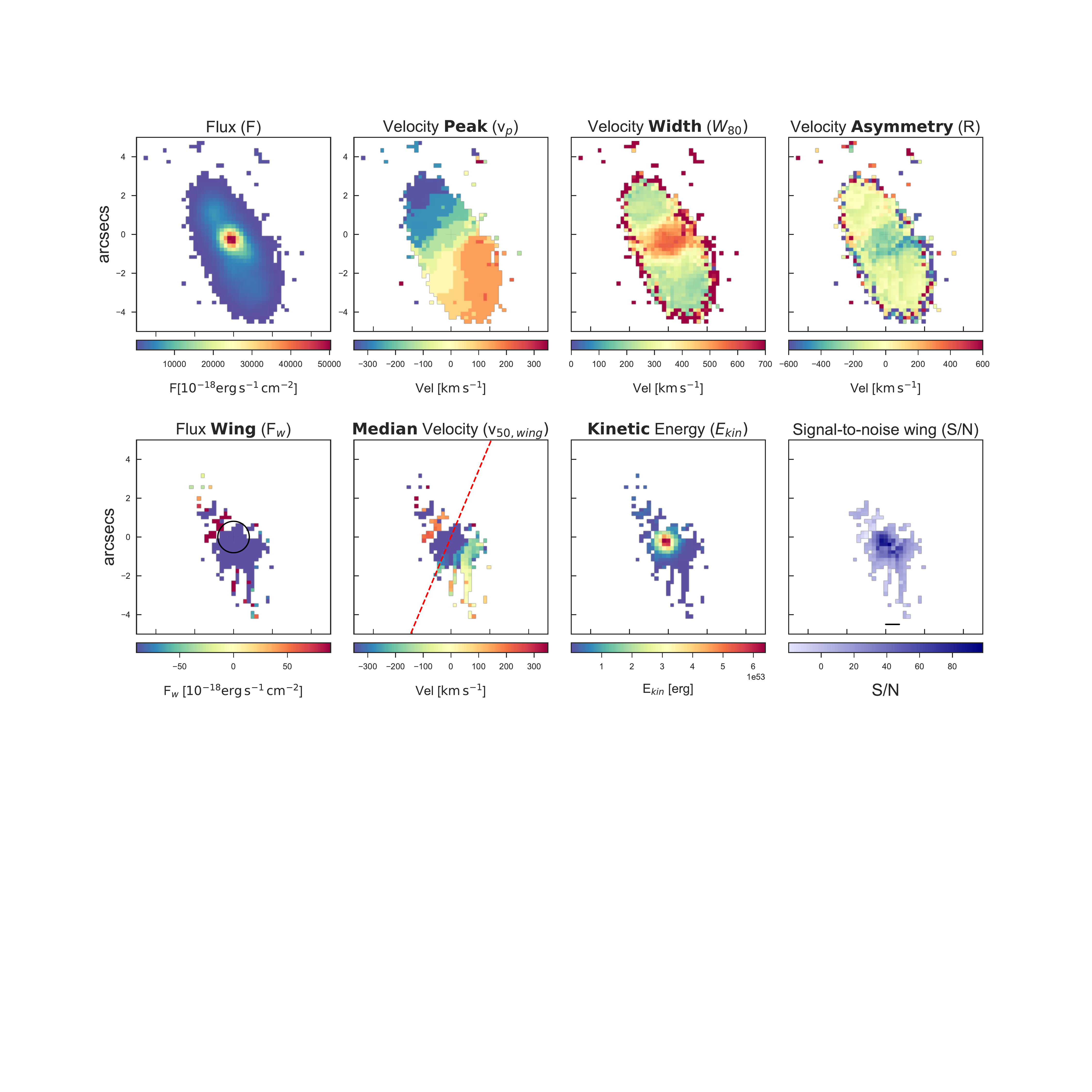}
\caption{3C~403, HEG, 1$\arcsec$ = 1.13 kpc. The black circle in the first panel has a diameter of 3 times the seeing of
the observations; the dashed line in the second panel marks the radio position
angle. The gas in outflow is mainly observed in a compact region. Its flux is concentrated into a radius of $\sim$ 1$\arcsec$. However, the gas also weakly extends in the NW-SE directions. The [O~III] is blueshifted, reaching $\sim$ -1940 $\text {km}\,\text{s}^{-1}$. Its energy peaks in the nucleus where the median velocity of the gas is by $\sim$ -430 $\text {km}\,\text{s}^{-1}$.}
\end{figure}



\begin{figure}
\centering
\includegraphics[width=0.9\textwidth]{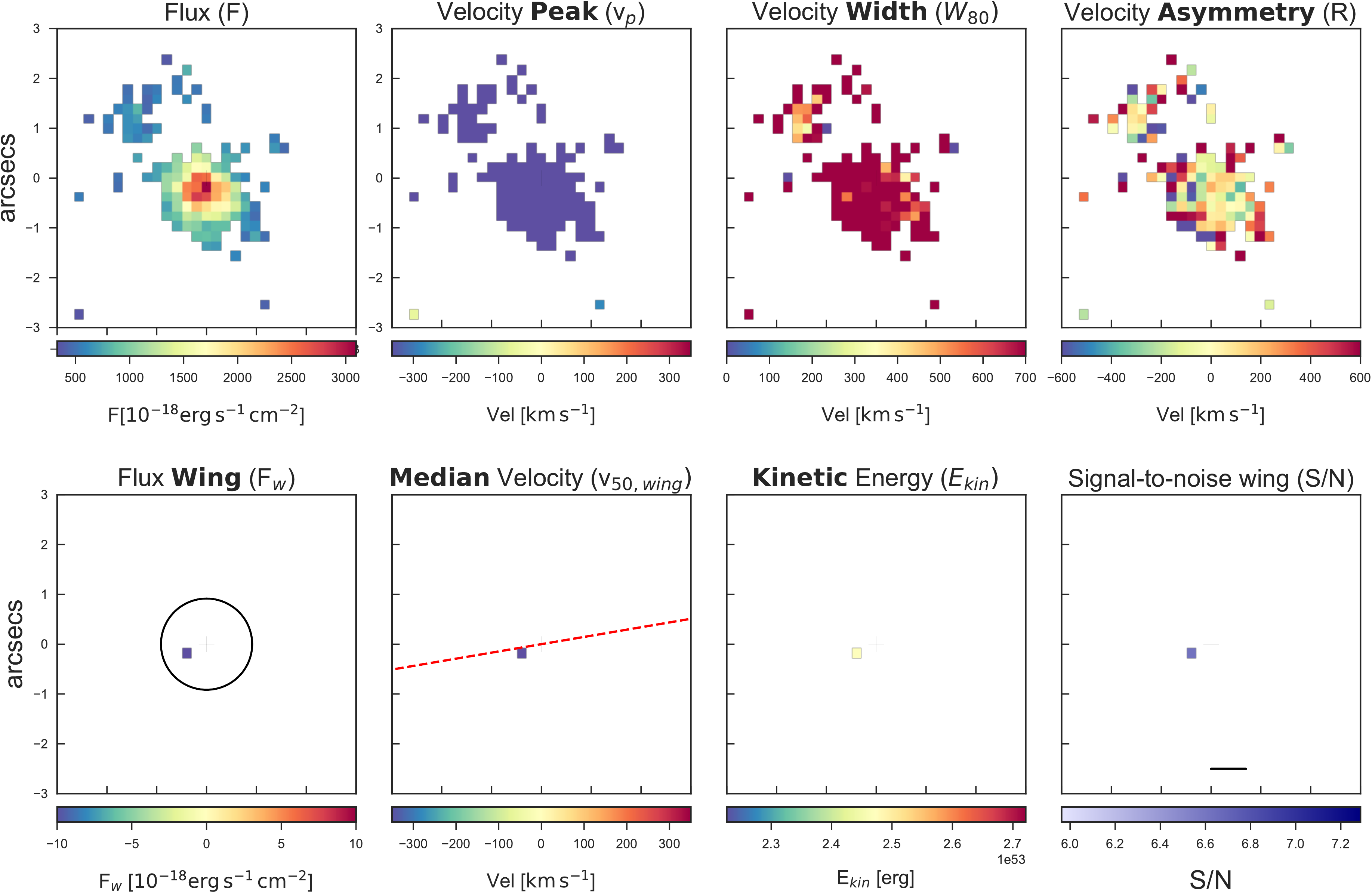}
\caption{3C442, LEG, 1$\arcsec$ = 0.51 kpc. The black circle in the first panel has a diameter of 3 times the seeing of
the observations; the dashed line in the second panel marks the radio position
angle.}
\end{figure}

\begin{figure}
\centering
\includegraphics[width=0.9\textwidth]{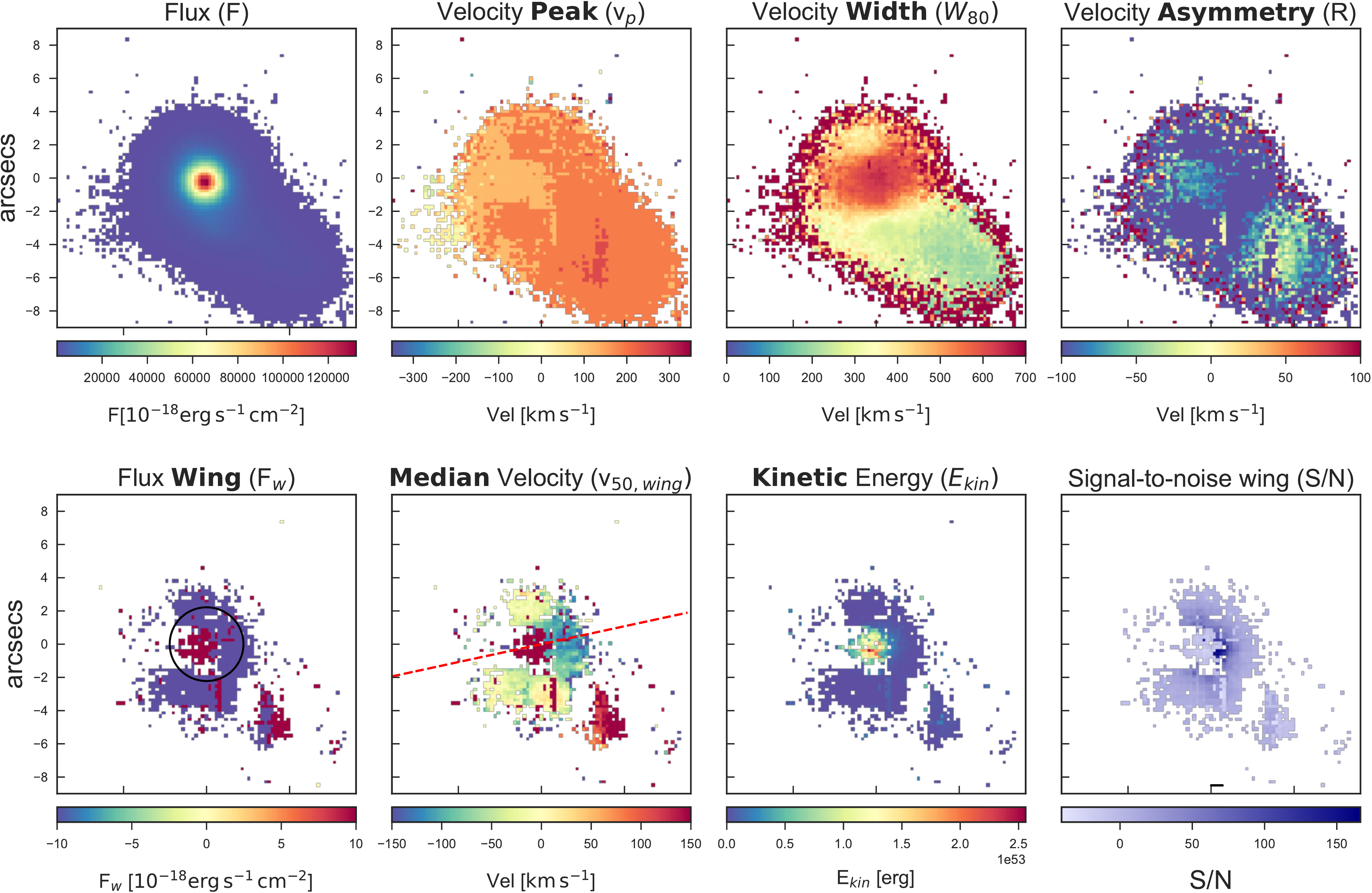}
\caption{3C~445, BLO, 1$\arcsec$ = 1.07 kpc. The black circle in the first panel has a diameter of 3 times the seeing of
the observations; the dashed line in the second panel marks the radio position
angle. In this source the outflowing gas assumes a flipped C
shape extended by $\sim$ 8.60 kpc. The flux is approximately constant
throughout the emission region while the velocity of the gas is major
in the central area and decrease towards the two extremities. Here the
emission lines are blueshifted showing velocities up to $\sim$ -2160
$\text {km}\,\text{s}^{-1}$ (the maximum velocity reached between the
BLOs). At the distance of $\sim$ 1.50 kpc from this region a smaller
redshifted area is detected with dimensions of $\sim$ 3 kpc, here the
gas reaches velocities down to $\sim$ 1100 $\text
{km}\,\text{s}^{-1}$. The energy is higher in correspondence of the
highest blueshifted velocities.}
\end{figure}

\begin{figure}
\centering
\includegraphics[width=0.9\textwidth]{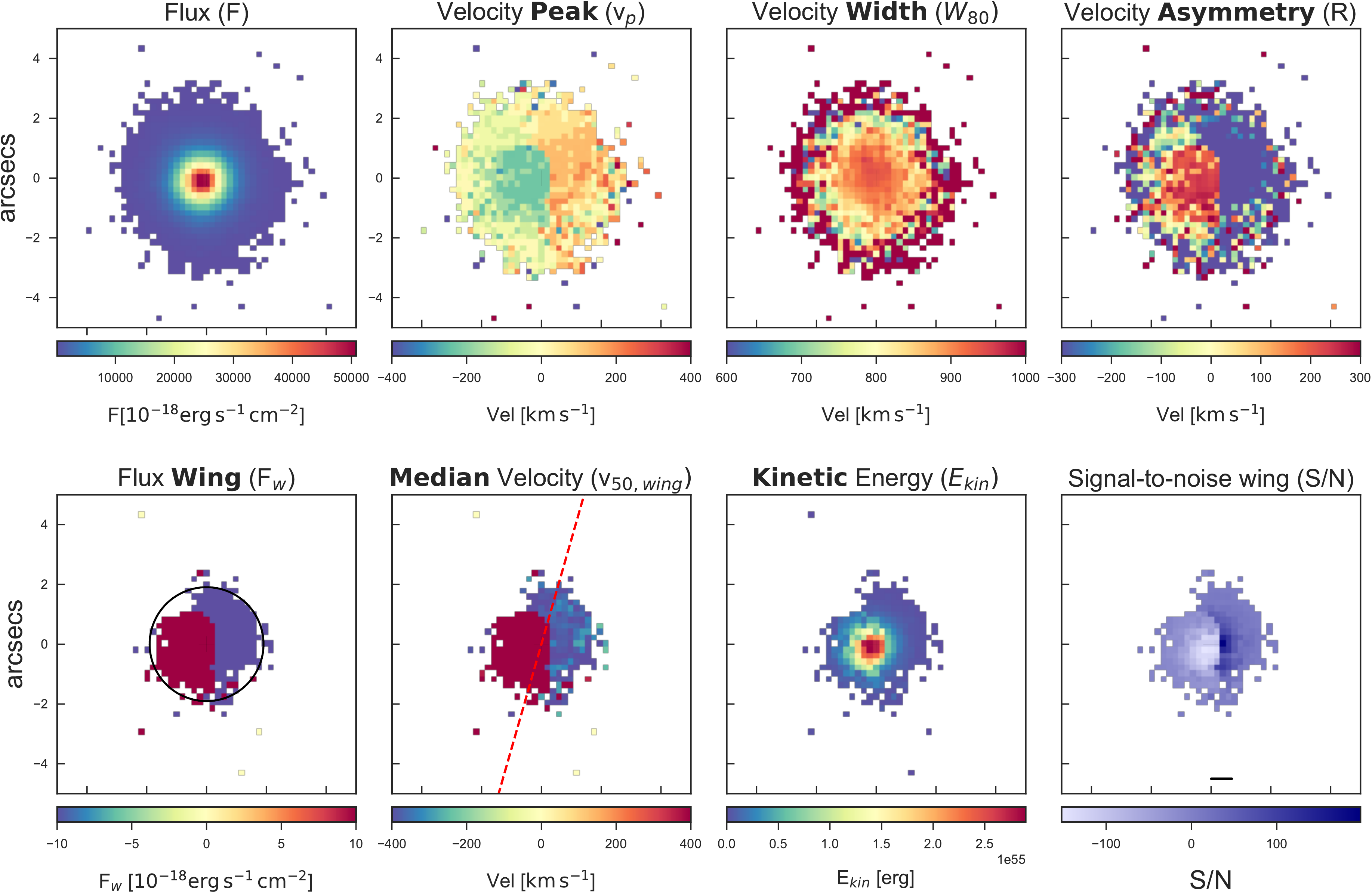}
\caption{3C~456, HEG, 1$\arcsec$ = 3.51 kpc. The black circle in the first panel has a diameter of 3 times the seeing of
the observations; the dashed line in the second panel marks the radio position
angle. The outflow maps of this source are very similar to
the ones obtained for 3C~327. In this case the emission lines are just
more extended: the outflow is enclosed into an area of radius $\sim$
7.02 kpc. In the west region the redshifted component is detected (up
to $\sim$ 1100 $\text {km}\,\text{s}^{-1}$), in the east the
blueshifted one (up to $\sim$ -1900 $\text {km}\,\text{s}^{-1}$). The
image is clearly resolved, the energy peaks on the nucleus and
decreases towards the border.}
\end{figure}

\begin{figure}
\centering
\includegraphics[width=0.9\textwidth]{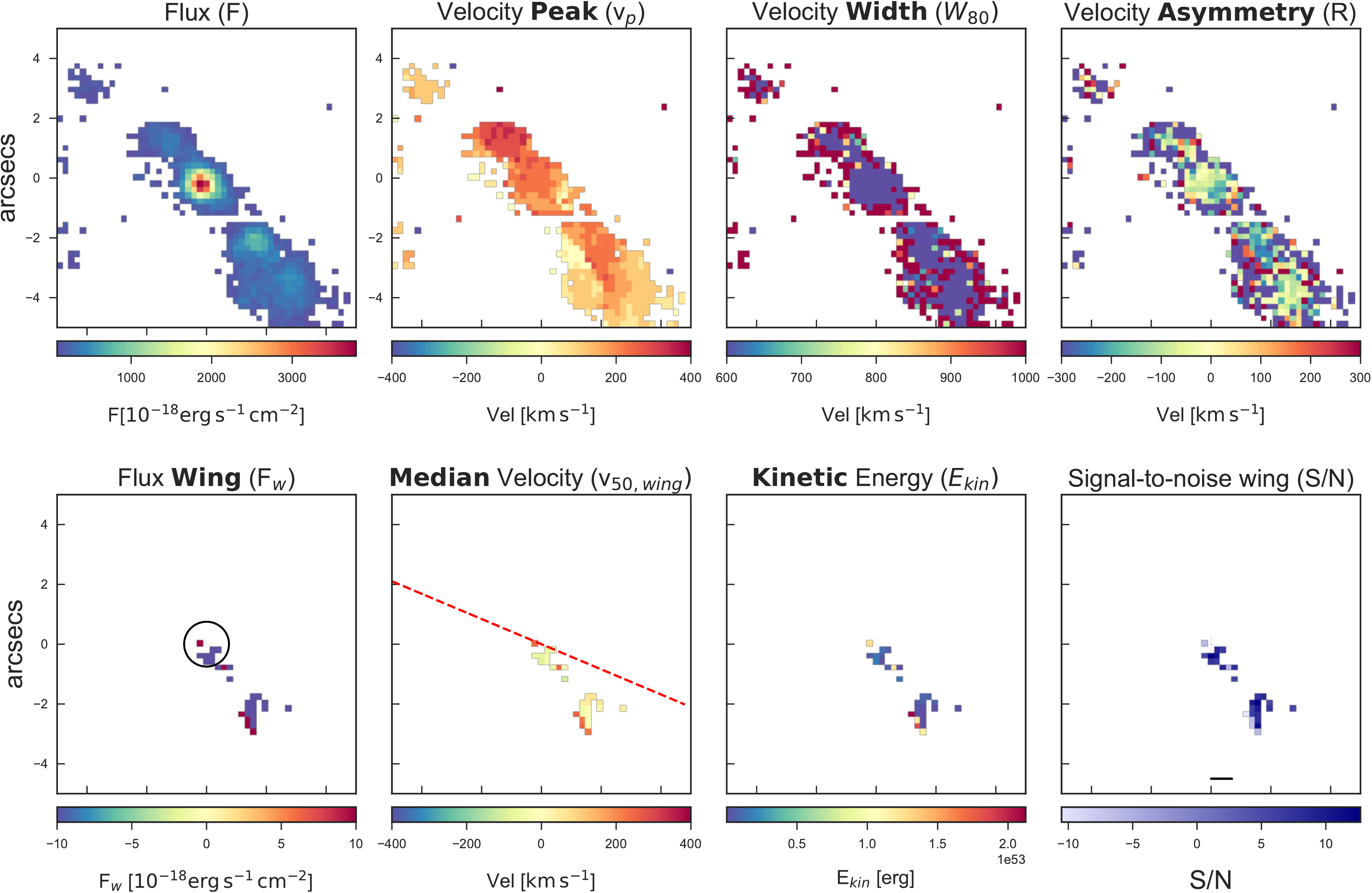}
\caption{3C458, HEG, 1$\arcsec$ = 4.07 kpc. The black circle in the first panel has a diameter of 3 times the seeing of
the observations; the dashed line in the second panel marks the radio position
angle.}
\end{figure}

\begin{figure}
\centering
\includegraphics[width=0.9\textwidth]{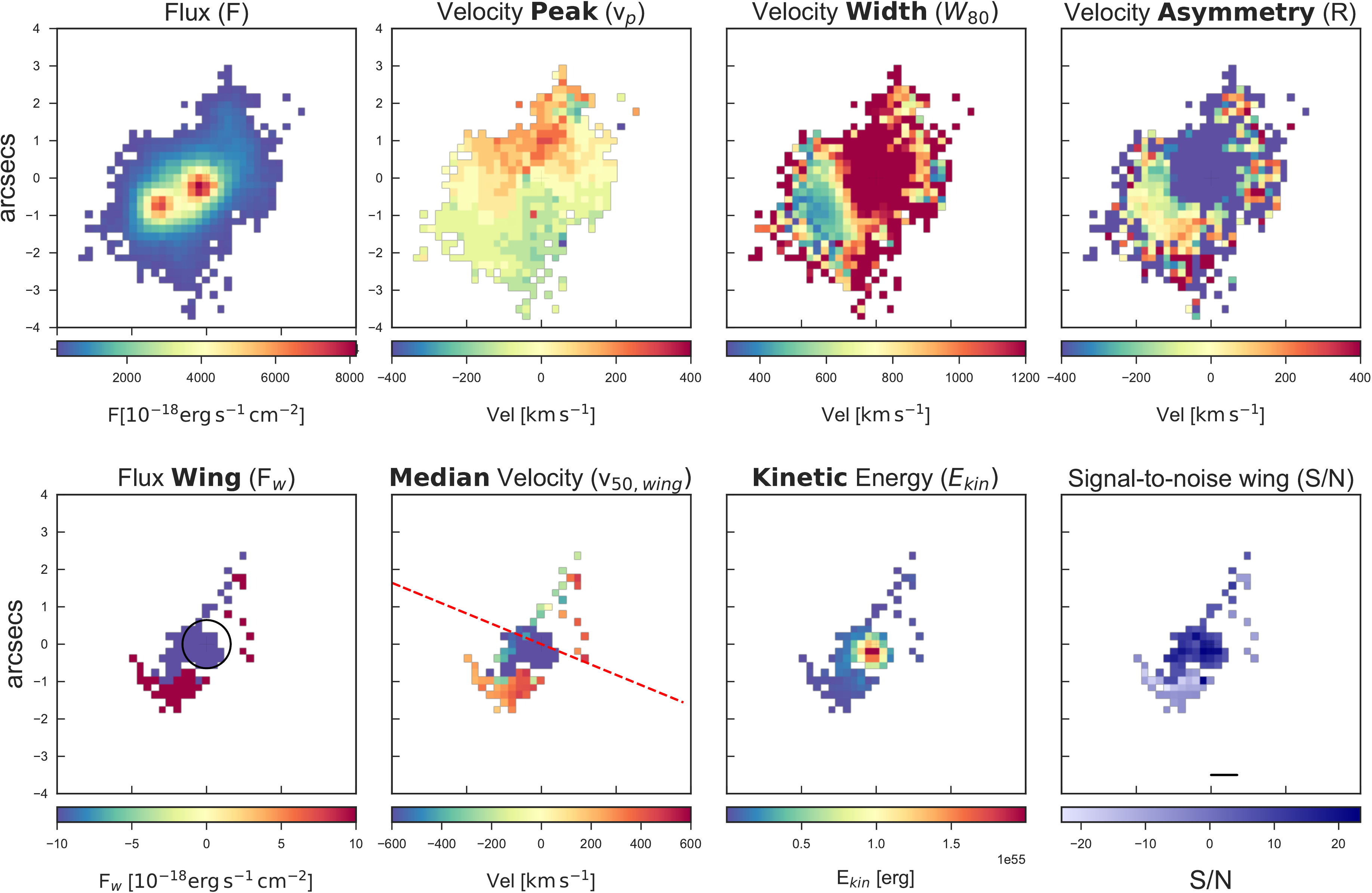}
\caption{3C~459, BLO, 1$\arcsec$ = 3.3 kpc. The black circle in the first panel of the bottom has a diameter of 3 times the seeing of
the observations; the dashed red line in the second panel of the bottom marks the radio position
angle. We presented detailed results obtained for this
source in \citet{Balmaverde2018}. The outflowing gas is mainly
confined in a region of radius $\sim$ 2.86 kpc. Here the gas has
high negative velocities up to $\sim$ -2040 $\text
{km}\,\text{s}^{-1}$. In the south there is a thin strip of
redshifted gas. The maximum of the energy is found in
correspondence with the blueshifted component of the ionized gas.}
\end{figure}

\end{appendix}

\end{document}